\documentclass[a4paper,11pt]{article}
\usepackage{jheppub}
\usepackage{lineno}

\usepackage{amsmath, amssymb, graphics,mathtools}
\usepackage{commath}
\usepackage{amsfonts,euscript,stmaryrd,braket}
\usepackage{graphics,tikz}
\usetikzlibrary{arrows,decorations.markings,patterns}
\usepackage{comment}
\usepackage{caption}
\usepackage{subcaption}
\usepackage{slashed}
\usepackage{wrapfig}
 \usepackage{booktabs}
 \usepackage{multirow}
\usepackage{tabularx}
\usepackage{longtable}
\usepackage{bbold,bbding}
\usepackage{scalerel}
\usepackage{amsthm} 
\usepackage{esint}
\newcommand{\bal}{\begin{equation}\begin{aligned}}
\newcommand{\eal}{\end{aligned} \end{equation}}

\newcommand{\htheta}{\hat{\theta}}
\newcommand{\hL}{\hat{\mathcal{L}}}

\newcommand{\eL}{\epsilon_{\text{\tiny L}}}
\usepackage[utf8]{inputenc}
\usepackage{epsfig}
\usepackage{graphicx}
\usepackage[multiple]{footmisc}
\usepackage{amssymb,amsmath}
\usepackage{fancybox,framed,tikz}
\usetikzlibrary{decorations.pathmorphing,patterns}
\usepackage{dsfont}
\usepackage{braket}
\usepackage{rotating}
\usepackage{multirow}
\usepackage{upgreek}
\newcommand{\be}{\begin{equation}}
\newcommand{\ee}{\end{equation}}
\newcommand{\bea}{\begin{eqnarray}}
\newcommand{\eea}{\end{eqnarray}}
\newcommand{\nn}{\nonumber}

\newcommand{\nL}{n_{\text{\tiny $L$}}}
\newcommand{\mL}{m_{\text{\tiny $L$}}}

\renewcommand{\a}{\alpha}
\renewcommand{\b}{\beta}

\newcommand{\pa}{\partial}
\newcommand{\g}{\gamma}
\newcommand{\G}{\Gamma}

\renewcommand{\L}{\hL}

\newcommand{\R}{\Rho}
\newcommand{\s}{\sigma}

\newcommand{\Tr}{\textup{Tr}}

\newcommand\Theff{\Theta_\text{\tiny eff}}
 
\usepackage{color}

\definecolor{mypink1}{rgb}{0.958, 0.188, 0.478}

\newcommand{\ba}{\begin{eqnarray}}
\newcommand{\ea}{\end{eqnarray}}

 \def\r{\rho}

\def\mbar{\overline{m}}

\newcommand{\grp}[1]{\mathrm{#1}}

\newcommand{\grSO}{\grp{SO}}
\newcommand{\grSL}{\grp{SL}}

\newcommand{\XX}{\mathcal{X}}
\newcommand{\YY}{\mathcal{Y}}

 \def\L{\mathcal{L}}
\hypersetup{}

\tikzset{Witten diagram/.style={execute at begin picture={%
\draw[blue ,fill=blue!05] circle[radius=\pgfkeysvalueof{/tikz/Witten/radius}];
\path node (X){\phantom{X}};
},baseline={(X.base)}},vertex/.style={circle,fill,inner sep=1.414pt,node
contents={}},
Witten/.cd,radius/.initial=1.414cm}

\usepackage{color}
\newif\ifcomments    
\commentstrue

\newcommand*{\Scale}[2][4]{\scalebox{#1}{$#2$}}%
\usetikzlibrary{arrows.meta, positioning, calc}

\usepackage{hyperref}

\def\sn{{\rm sn}}
\def\cn{{\rm cn}}
\def\dn{{\rm dn}}

\arxivnumber{2512.15698} 

\title{\boldmath {Probing the dynamics of stringy flux tubes \\ with large $R$-charge}}

\author[a]{Davide~Bonomi,}
 \author[b]{Valentina~Forini,}
 \author[c]{Valentina Giangreco M. Puletti,}
 \author[d]{Luca~Griguolo,}
 \author[e]{Domenico~Seminara}
 \affiliation[a]{Deutsches Elektronen-Synchrotron DESY, \\
Notkestr. 85, 22607 Hamburg, Germany}
 \affiliation[b]{Institut f\"ur Physik, Humboldt-Universit\"at zu Berlin and IRIS Adlershof, \\
Zum Gro\ss en Windkanal 2, 12489 Berlin, Germany}
 \affiliation[c]{Science Institute, University of Iceland, \\  Dunhaga 3, 107 Reykjavík, Iceland}
 \affiliation[d]{Dipartimento SMFI, Università di Parma and INFN, Gruppo Collegato di Parma,\\
Viale G.P. Usberti 7/A, I-43100 Parma, Italy}
 \affiliation[e]{Dipartimento di Fisica, Universit\'a di Firenze and INFN Sezione di Firenze, \\
via G. Sansone 1, 50019 Sesto Fiorentino, Italy}

\emailAdd{davide.bonomi@desy.de, valentina.forini@hu-berlin.de, vgmp@hi.is, luca.griguolo@unipr.it, seminara@fi.infn.it}

\preprint{DESY-25-193, HU-EP-25/41}

\abstract{
We study the generalized cusp anomalous dimension, or  quark-antiquark potential on the three-sphere,  in the presence of a large $R$-charge $L$ and at  strong coupling. Considering the insertion of a local scalar operator of charge $L$ on a cusped Wilson loop, we investigate the general solution of the dual string configuration for any value of cusp angles, paying particular attention to the different ranges of the physical parameters and obtaining a unified description that simultaneously captures the near BPS, the antiparallel lines and the large-$L$ regimes. 
We observe, in the limit of small separation between the lines, a non-analytic change in the behavior of the cusp anomaly at a critical value of $L$: one crosses from a region dominated by the familiar Coulomb-like singularity to a ``deconfined" situation describing the leading L\"uscher corrections at large $L$ of the generalized cusp anomalous dimension. We derive the equations for small fluctuations around the classical string solution, generalizing previous  analysis of the uncharged case, and we study the behavior of the spectrum near the
transition region. Finally, we speculate on the field theory origin of the transition and on its implications for the fusion of Wilson lines.

}

\begin{document}
\maketitle
\flushbottom

\section{Introduction}
\label{sec:intro}

Wilson loops are gauge-invariant observables that serve as fundamental probes of non-Abelian 
gauge theories. In conformal field theories like $\mathcal{N}=4$ supersymmetric Yang-Mills (SYM), their vacuum expectation values encode a large set of physical data: a particularly rich class emerges when the contour develops a cusp, i.e. a point of non-differentiability along the contour, with a finite opening angle $\phi$. Such a behavior induces a specific ultraviolet divergence, whose strength is governed by the cusp anomalous dimension $\Gamma_{\text{cusp}}(\lambda,\phi, \theta)$~\cite{Korchemsky:1991zp,Korchemsky:1987wg,DF,Correa:2012nk}, a function of the 't Hooft coupling $\lambda$ and of both the geometric angle~$\phi$ and an internal angle~$\theta$ associated with a sudden change in the coupling to the scalar fields. Far from being a regulator artifact, this quantity lies at the center of several core aspects of the theory. It is intimately related to the Bremsstrahlung function of an accelerated quark~\cite{Correa:2012at,Fiol:2012sg}, controls the infrared structure of planar scattering amplitudes through the Wilson loop/amplitude duality~\cite{Alday:2007hr, Drummond:2007aua,Brandhuber:2007yx,Henn:2010bk}, and can be interpreted as a heavy quark-antiquark potential on the three-sphere within a specific quantization scheme~\cite{DF}. In the planar limit, the exact functional form of $\Gamma_{\text{cusp}}(\lambda,\phi,\theta)$ is accessible through integrability techniques, providing a closed description for any value of the 't Hooft coupling~\cite{Correa:2012hh,Drukker:2012de,Gromov:2015dfa}. Furthermore, the AdS/CFT correspondence offers a dual geometric description in terms of minimal surfaces in $\text{AdS}_5 \times S^5$~\cite{Maldacena:1998im,Drukker:1999zq}, while supersymmetric localization yields exact results for specific protected BPS cusps~\cite{Pestun:2007rz,Correa:2012at,Drukker:2007qr}.

A profound feature of the cusp anomalous dimension, $\Gamma_{\text{cusp}}(\lambda,\phi,\theta)$, is its ability to connect distinct physical regimes in $\mathcal{N}=4$ SYM through the simple variation of its opening angle~$\phi$. In one extreme, the straight line limit ($\phi \to 0$), it 
corresponds to the scaling dimension of a local operator 
in the Defect Conformal Field Theory (DCFT) defined on a Wilson line~\cite{Drukker:2006xg,Cooke:2017qgm,Giombi, Kim:2017sju,Liendo:2018ukf}. This framework provides powerful insights into the DCFT's operator spectrum~\cite{Grabner:2020nis,Cavaglia:2021bnz,Cavaglia:2022yvv}. In the other extreme, the antiparallel limit ($\phi=\pi$), it directly governs the static quark-antiquark potential~\cite{Maldacena:1998im,Rey:1998ik,Gromov:2016rrp}. Thus, a single function elegantly bridges the abstract world of defect conformal data with the concrete dynamics of a fundamental force law.

A natural generalization of this system involves inserting a non-trivial local operator at the cusp. 
Actually, the original integrability-based description of the cusp anomalous dimension~\cite{Correa:2012hh,Drukker:2012de} relied on inserting a specific type of operator: a chiral primary with large $R$-charge, chosen to be orthogonal to the scalars already coupled to the Wilson loop. This large $R$-charge is crucial, as it maps the problem onto an integrable open spin chain with reflecting boundaries. 
The Thermodynamic Bethe Ansatz (TBA) equations for this spin chain then yield the cusp anomalous dimension exactly for any value of the coupling. The standard cusp anomalous dimension is recovered in the limit where this additional $R$-charge vanishes. From a physical standpoint, inserting such an operator in the context of the quark-antiquark potential corresponds to exciting the system into a state different from the standard vacuum. In the infinite $R$-charge limit, this state coincides with the BMN vacuum~\cite{Berenstein:2002jq}, see discussion in~\cite{Drukker:2006xg}.

One might expect that insertions carrying a small $R$-charge (small compared to the~'t Hooft coupling) would only perturb the quark-antiquark flux tube slightly, leading to a minor correction of the effective Coulomb charge that governs the interaction potential at both strong and weak coupling. However, integrability-based computations of the cusp anomalous dimension at large $R$-charge~\cite{Correa:2012hh,GS} show no evidence of a singularity in the antiparallel limit $\phi \to \pi$, which is a distinctive feature of the standard potential. This suggests a non-trivial interplay between the $R$-charge and the physics of the flux tube. So far, detailed studies of the strong coupling regime, employing the AdS/CFT correspondence~\cite{GS}, supersymmetric localization \cite{Giombi:2021zfb, Giombi:2022anm}, and the quantum spectral curve technique~\cite{Gromov:2015dfa}, have largely been confined to near-straight line or near-BPS configurations.  
This motivates an extensive investigation of the cusp anomalous dimension with $R$-charge for general values of the parameters. In this work, we focus on insertions constructed from scalars orthogonal to the Wilson loop. We perform a comprehensive analysis of this system using its dual string description in AdS/CFT, which includes studying the one-loop fluctuations around the classical string solutions.

\subsection{Setup: Cusped Wilson Loop with Scalar Insertions}
\label{subsec:setup}

In this work, we study a generalized cusped Wilson loop operator in $\mathcal{N}=4$ supersymmetric Yang-Mills (SYM) theory. The operator consists of two rays meeting at a cusp with opening angle $\phi$, where we insert $L$ copies of a specific complex scalar field $Z = \Phi_1 + i\Phi_2$ at the cusp point. The scalar coupling along each ray is specified by different directions in the internal $SO(6)_R$ space: the first ray couples to scalars projected along a direction $\vec{n}_q$, while the second ray couples to scalars projected along  $\vec{n}_{\bar q}$, with an internal angle $\theta$ between them~ ($\cos\theta=\vec{n}_{q}\cdot\vec{n}_{\bar q}$).
The operator is explicitly written as~\cite{Correa:2012hh,Drukker:2012de}
\begin{equation}\label{WL}
W_L =\text{Tr}\!\!\left( \!\!\mathcal{P}\exp \left\{\int_{-\infty}^{0} \!\!\!\!\!\!\!dt
\left[ iA \cdot \dot{x}_q + \vec{\Phi} \cdot \vec{n}_{q} \, |\dot{x}_q| \right]\!\!\right\}
\, Z^L(0)
\, \mathcal{P}\exp\left\{ \int_{0}^{\infty} \!\!\!\!\!dt
\left[ i A \cdot \dot{x}_{\bar{q}} + \vec{\Phi} \cdot \vec{n}_{\bar q} \, |\dot{x}_{\bar{q}}| \right]\right\}\!\!\right)\!\!,
\end{equation}
where $\mathcal{P}$ denotes path ordering, $x_q(t)$ and $x_{\bar{q}}(t)$ parameterize the two rays  and the complex scalar $Z = \Phi_1 + i\Phi_2$ is chosen to be orthogonal to both $\vec{\Phi} \cdot \vec{n}_{q}$ and $\vec{\Phi} \cdot \vec{n}_{\bar q}$.

\paragraph{The bare cusp and conformal mapping.}
For $L=0$, the expectation value of the cusped Wilson loop exhibits a logarithmic divergence,
\begin{equation}\label{logdiv}
\langle W_0 \rangle \simeq e^{-\Gamma_{\text{cusp}}(\lambda,\phi,\theta) \log \frac{\Lambda}{\epsilon}},
\end{equation}
where $\Gamma_{\text{cusp}}(\lambda,\phi,\theta)$ is the generalized cusp anomalous dimension, and $\Lambda$ and $\epsilon$ denote IR and UV cutoffs, respectively.

This system can be mapped via a conformal transformation to $\mathbb{R} \times S^3$, where the cusped Wilson loop becomes a pair of antiparallel lines separated by an angle $\pi - \phi$ along a great circle of $S^3$, with opposite time orientation. In this picture, the expectation value computes the potential $V(\lambda,\phi,\theta)$ between two heavy static sources propagating over a large time $T$:
\begin{equation}
\langle W_{\text{lines}} \rangle \simeq e^{-T V(\lambda,\phi,\theta)}.
\end{equation}
Conformal invariance relates the two descriptions via
\begin{equation}
\Gamma_{\text{cusp}}(\lambda, \phi, \theta) = V(\lambda, \phi, \theta),
\end{equation}
which allows one to recover the standard quark--antiquark potential in flat space as $\phi \to \pi$ (see, e.g.,~\cite{DF} for details).

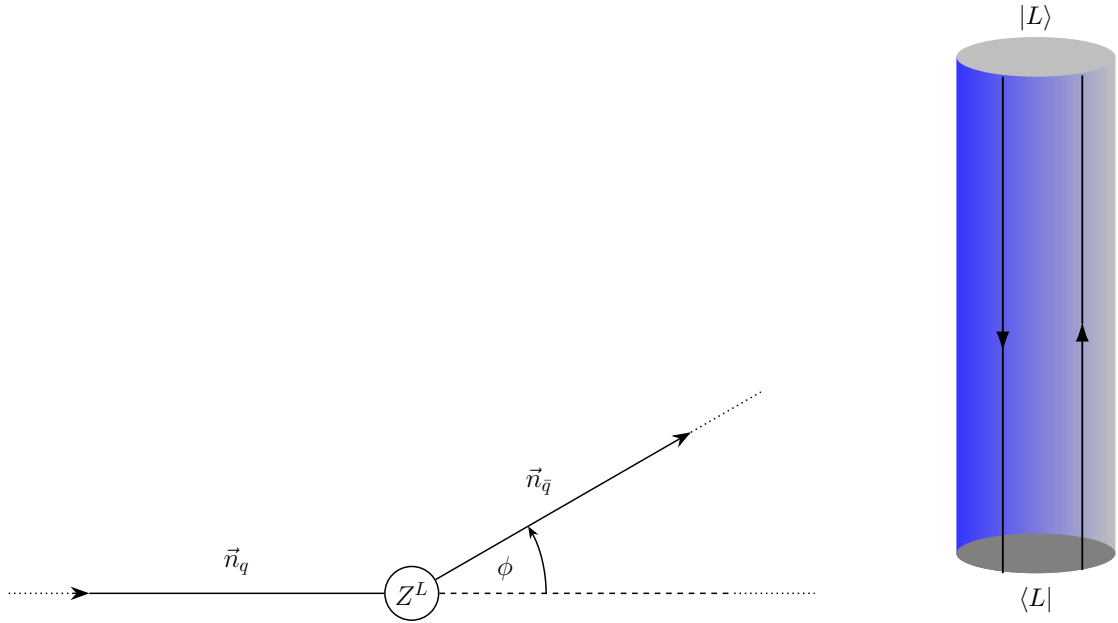
\begin{figure}
\resizebox{0.7\textwidth}{!}{
\begin{tikzpicture}[
    >=Stealth,
    thick,
    font=\Large
]

    \coordinate (Origin) at (0,0);
    \coordinate (LeftStart) at (-6,0);
    \coordinate (LeftFar) at (-7.5,0);
    \coordinate (RightFar) at (6,0);
    \coordinate (RightFarDotted) at (7.5,0);
    \coordinate (AngleEnd) at (30:6);
    \coordinate (AngleFar) at (30:7.5);

    \node[circle, draw, thick, fill=white, inner sep=2pt, minimum size=1cm] (Z) at (Origin) {$Z^L$};

    \draw[dotted] (LeftFar) -- (-6.3,0);
    \draw[-{Stealth[length=3.5mm, width=2.5mm]}] (-6.3,0) -- (LeftStart);
    \draw (LeftStart) -- (Z) node[midway, above=2mm] {$\vec{n}_q$};

    \draw[dashed] (Z) -- (RightFar);
    \draw[dotted] (RightFar) -- (RightFarDotted);

    \draw[-{Stealth[length=3.5mm, width=2.5mm]}] (Z) -- (AngleEnd) node[midway, above left=1mm] {$\vec{n}_{\bar {q}}$};
    \draw[dotted] (AngleEnd) -- (AngleFar);

    \draw[->] (2.5,0) arc (0:30:2.5);
    \node at (15:1.8) {$\phi$};

\end{tikzpicture}
}
\qquad\qquad
\resizebox{0.15\textwidth}{!}{
\begin{tikzpicture}

    \def\R{1.2}    
    \def\H{7.5}     
    \def\ry{0.3}     
    \def\xinner{-0.5}
    
    \node at (0,\H+0.6) {$|L\rangle$};
    \node at (0,0-0.7) {$\langle L|$};
    
    \pgfmathsetmacro{\yoffset}{-\ry * sqrt(1 - (\xinner/\R)^2)}

    \shade[left color=blue!80!white, right color=lightgray]
        (-\R, \H) -- (-\R, 0) 
        arc (180:360:{\R} and {\ry}) 
        -- (\R, \H) -- cycle;

    \fill[lightgray] (0, \H) ellipse ({\R} and {\ry});
    \fill[gray] (0, 0) ellipse ({\R} and {\ry});

    \draw[thick, -{Latex[length=3.0mm, width=2.0mm]}] 
      (\xinner, \H-0.3) -- (\xinner, \H/2-0.7);
    \draw[thick] 
       (\xinner, 0-0.3) -- (\xinner, \H/2-0.4);

    \draw[thick, -{Latex[length=3.0mm, width=2.0mm]}] 
        (\R-0.5, 0-0.25) -- (\R-0.5, \H/2-0.25);
    \draw[thick] 
        (\R-0.5, \H/2-0.28) -- (\R-0.5, \H-0.28);
        
\end{tikzpicture}
}

\caption{\small Cusped Wilson line with opening angle $\phi$ and a $Z^L$ insertion, see equation~\eqref{WL}. The two line segments in the figure couple differently to the scalar fields, corresponding to an angle $\theta$ in an internal space. Using a conformal transformation, this configuration can be mapped to a pair of antiparallel Wilson lines on the cylinder $\mathbb{R} \times S^3$, as in the figure on the right.}
\label{fig:setup}
\end{figure}

\paragraph{Insertions as excitations of the flux tube.}
When $L>0$ scalar insertions are present, the $L$-dependent generalized cusp anomalous dimension   $\Gamma_L(\lambda,\phi,\theta)$ defined by 
\begin{equation}\label{cuspdef}
    \langle W_{L}  \rangle \simeq e^{-\Gamma_{L}(\lambda,\phi,\theta) \log \frac{\Lambda}{\epsilon}}
    \end{equation}
has been shown to satisfy a Thermodynamic Bethe Ansatz (TBA) system in which $L$ plays the role of the volume~\cite{Correa:2012hh,Drukker:2012de}. The $L=0$ case is recovered in the limit of vanishing volume. In the $\mathbb{R} \times {S}^3$ picture, the insertions $Z^L$ impose nontrivial boundary conditions at past and future infinity, effectively placing the quark--antiquark system in an excited state with non-zero $R$-charge (see Fig.~\ref{fig:setup}). Consequently, $\Gamma_L(\lambda,\phi,\theta)$ can be interpreted as the interaction energy of two heavy particles immersed in a constant $R$-charge background.

\paragraph{Strong coupling dual.}

The AdS/CFT correspondence provides a dual geometric description in terms of strings in $\text{AdS}_5 \times S^5$. 
For large $R$-charge $L$, the system admits a semiclassical description in which the string carries angular momentum on $S^5$~\cite{Drukker:2006xg,Correa:2012at,GS}.
More precisely, the expectation value $\langle W_L \rangle$ can be obtained from the large-$L$ limit of the defect two-point function $\langle \bar{Z}^L(\infty) Z^L(0) \rangle_{W_0}$, where the scalar fields are inserted in the Wilson line~\eqref{logdiv}. In this limit, the cusp anomalous dimension $\Gamma_L(\lambda,\phi,\theta)$ is given by the Legendre transform of the classical string area with respect to the conserved $R$-charge $L$~\cite{Drukker:1999zq,Janik:2010gc,Bak:2011yy} (see also discussion in~\cite{Giombi:2021zfb,Giombi:2022anm})~\footnote{
In the $L=0$ case, the cusp anomalous dimension is proportional, at strong coupling, to the on-shell worldsheet action~\cite{DF}.   
When $L$ insertions are present, the relevant semiclassical saddle must be projected onto a state of fixed angular momentum $L$. As discussed in~\cite{Janik:2010gc}, this projection replaces the classical action by the Hamiltonian governing evolution in global AdS time. Equivalently, fixing $L$ via a Legendre transform leads to a Routhian~\cite{Bak:2011yy} that coincides exactly with the physical energy of the string configuration. The exponential behavior of the cusped Wilson loop is therefore controlled by this energy, yielding $\Gamma_L(\lambda,\phi,\theta) = E$ at strong coupling.}.

\subsection{Summary and Results}
\label{subsec:summary}

We have conducted a thorough analysis of the classical string solution dual to the cusped Wilson loop operator $W_L$ and of its quadratic fluctuations, extending previous investigations~\cite{Klebanov:2006jj,GS}~\footnote{In~\cite{Klebanov:2006jj} there is no presence of $R$-charge insertions; the solution of~\cite{GS} has been studied only in the near BPS limit.}, while clarifying several physical aspects that were previously overlooked. Our study focuses on the interplay between the geometrical parameters of the cusp and the effects of the $R$-charge insertion $L$.

\paragraph{Semiclassical regime and parametrization.}
In the semiclassical  regime, in which~$L \sim g=\frac{\sqrt{\lambda}}{4\pi} $, it is convenient to define the rescaled $R$-charge parameter
\begin{equation}
\mathcal{L} \equiv \frac{L}{4g},
\end{equation}
which, together with the geometric opening angle $\phi$ and the internal scalar angle $\theta$, controls the cusp anomalous dimension. Solving the string equations of motion subject to appropriate boundary conditions leads to a solution parameterized by four auxiliary parameters: $n$, $m$ (associated with the AdS$_5$ part) and $\nL$, $\mL$ (associated with the $S^5$ part). These are related to the physical parameters through the following equations
\begin{align}
\phi &= \pi - 2\frac{\sqrt{1-n}\sqrt{n-m}}{\sqrt{n}}\,
        \bigl[\,\Pi(n|m) - K(m)\,\bigr], \label{eq:phi} \\
\theta &= 2\sqrt{\frac{(1-\nL)(\nL-\mL)}{\nL}}\,
         \Pi(\nL|\mL), \label{eq:theta} \\
\mathcal{L} &= \sqrt{\frac{\nL}{\mL}}\,
              \bigl[\,K(\mL) - E(\mL)\,\bigr], \label{eq:Ldef}
\end{align}
in terms of elliptic functions defined in Appendix~\ref{app:elliptic}.

\paragraph{Effective $R$-symmetry angle and consistency.}
The apparent mismatch between the number of parameters in the classical string solution and in the dual gauge theory operator is resolved by introducing an \emph{effective} $R$-symmetry angle $\Theta_{\text{eff}}(\theta,\mathcal{L})$. This quantity emerges from the consistency condition between the AdS$_5$ and $S^5$ parts of the worldsheet evolution:
\begin{equation}
2\sqrt{1-n-m}\,K(m) = \Theta_{\text{eff}}(\theta,\mathcal{L})\,, \label{eq:thetaeff}
\end{equation}
see section~\ref{sec:angle-s5}. 

The energy of the classical solution, which gives the leading strong-coupling contribution to the cusp anomalous dimension $\Gamma_L(\lambda,\phi,\theta)$, depends \emph{only} on $\phi$ and $\Theta_{\text{eff}}$, and not separately on $\theta$ and $\mathcal{L}$:
\begin{equation}
\mathcal{E} \equiv \frac{E}{4g}
      = \frac{1}{\sqrt{n-m}}\,
        \bigl[\,(1-m)K(m) - E(m)\,\bigr]. \label{eq:E}
\end{equation}

\begin{figure}[h]  
    \centering
    \includegraphics[width=0.5\textwidth]{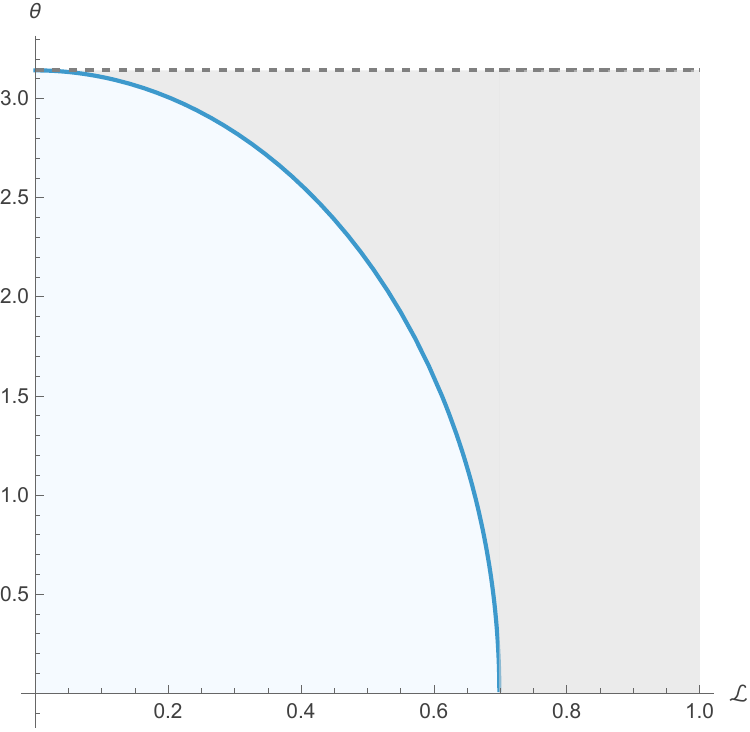} %
    \caption{\small Critical curve, defined implicitly by $\Theff(\theta, \mathcal{L})=\pi$, for the cusp internal angle~$\theta$ as a function of the rescaled $R$-charge $\L$. The shaded region in light blue corresponds to~Region~{\bf $A_L$}, the light gray one to~Region {\bf $B_L$}.}
    \label{fig:Transitioncurve}
\end{figure}

\paragraph{Two regimes of parameter space.}
The interpretation of $\Theta_{\text{eff}}$ as an angular variable leads to a natural division of the parameter space into two distinct regimes:
\begin{itemize}
    \item \textbf{Region A$_L$:} $\Theta_{\text{eff}}(\theta,\mathcal{L}) < \pi$. Here, the energy functional coincides with that of the generalized cusp solution without $R$-charge insertion~\cite{DF}, provided one substitutes $\theta \to \Theta_{\text{eff}}$. The insertion effectively dresses the internal scalar angle.
    \item \textbf{Region B$_L$:} $\Theta_{\text{eff}}(\theta,\mathcal{L}) > \pi$. In this region,  
    the effect of the $R$-charge cannot be absorbed into a simple redefinition of $\theta$; the solution changes character qualitatively and connects to the large charge regime studied in~\cite{Drukker:2006xg,Correa:2012hh,Drukker:2012de}.  
\end{itemize}
The implicit curve in the plane $(\L, \theta)$ defined by the value $\Theta_{\text{eff}}(\theta,\mathcal{L})=\pi$ is represented in Fig.~\ref{fig:Transitioncurve}. 
We have determined that there exists a critical value $\mathcal{L}_c^\text{\tiny{max}} \simeq 0.6986882(5)$ such that for $\mathcal{L} > \mathcal{L}^\text{\tiny{max}}_c$ one is always in region B$_L$, regardless of $\theta$.

\paragraph{The antiparallel limit and flux-tube interpretation.}
The limit $\phi \to \pi$ corresponds to antiparallel lines in flat space and reveals a striking difference between the two regimes:
\begin{itemize}
    \item In \textbf{region A$_L$}, the energy exhibits a Coulomb-like divergence:
    \begin{equation}
        \mathcal{E}^{(A)} = \frac{\mathcal{E}_{\text{Cas}}^{(A)}}{\pi-\phi}
                           + b^{(A)}\,(\pi-\phi) + O\bigl((\pi-\phi)^3\bigr), \label{eq:E_A}
    \end{equation}
    where $\mathcal{E}_{\text{Cas}}^{(A)}$ is a function of $\Theta_{\text{eff}}$ that monotonically decreases from a positive value at $\Theta_{\text{eff}}=0$ to zero at $\Theta_{\text{eff}}=\pi$ (the latter point formally coincides with the BPS configuration~\cite{DF}). From the viewpoint of the inserted $R$-charge, this implies that for any fixed $\theta$ there exists a critical $\mathcal{L}_c(\theta)$ at which the Coulomb potential \emph{dynamically} vanishes -- a phenomenon interpreted in earlier flat-space studies as the ``ionization'' of a quarkonium bound-state~\cite{Hong:2004gz}. The explicit form of the energy in this region is in~\eqref{cL}. 
    \item In \textbf{region B$_L$}, the energy remains finite as $\phi \to \pi$:
    \begin{equation}
        \mathcal{E}^{(B)} = \Delta^{(B)} + \beta\,(\pi-\phi)^2
                           + O\bigl((\pi-\phi)^4\bigr), \label{eq:E_B}
    \end{equation}
    with $\Delta^{(B)} = 0$ precisely at the transition line $\Theta_{\text{eff}}=\pi$, ensuring continuity of the energy across the two regions even in the antiparallel limit. Equation~\eqref{EnergyB} displays the explicit expression for the energy in this region. 
\end{itemize}
It is also instructive to contrast this behavior with expectations at weak coupling.  
For the ordinary cusp without insertions ($L=0$), perturbation theory yields the familiar Coulomb-like divergence in the antiparallel limit $\phi\to\pi$, corresponding to the standard quark–antiquark potential~\cite{DF}. When scalar insertions with fixed R-charge $L>0$ are present, however, the leading perturbative contribution to the generalized cusp anomalous dimension~\cite{Correa:2012hh} appears only at order~$\lambda^{L+1}$~\footnote{In this case the angular dependence involves Bernoulli polynomials, see Section 2.4.1, eq. (45), in~\cite{Correa:2012hh}. They vanish linearly as $\phi\to\pi$, cancelling  an explicit $1/\sin\phi$ factor and yielding a finite result in the antiparallel limit.}. This indicates that, for any fixed $L>0$, the weak-coupling behavior is already qualitatively similar to our Region B$_L$, where the Coulomb pole is absent and the energy remains finite as $\phi\to\pi$. All this suggests that the transition between Regions A$_L$ and B$_L$ observed at strong coupling might persist at weak coupling, and may reflect a genuine large-charge phenomenon emerging when $L\sim\sqrt{\lambda}$ rather than at fixed $L$.

\paragraph{Critical scaling.}
The behavior near the critical line $\mathcal{L} = \mathcal{L}_c(\theta)$ is non-analytic. Expanding on either side we find
\begin{align}
\mathcal{E}_{\text{Cas}}^{(A)} &= C_A\,(\mathcal{L}_c^\theta - \mathcal{L})^{3/2}
                                 + O\bigl((\mathcal{L}_c^\theta - \mathcal{L})^{5/2}\bigr), \label{eq:E_A_crit} \\
\Delta^{(B)} &= C_B\,(\mathcal{L} - \mathcal{L}_c^\theta)^{1/2}
               + O\bigl((\mathcal{L} - \mathcal{L}_c^\theta)^{3/2}\bigr)\,, \label{eq:E_B_crit}
\end{align}
where for compactness of notation we have defined $\L_c^\theta\equiv\L_c(\theta)$.
These scaling forms highlight the different nature of the solutions in the two regimes and signal a genuine transition in the classical string configuration as the effective angle $\Theta_{\text{eff}}$ crosses $\pi$.

\paragraph{Expansions in different regimes.}
We have verified that our expressions reproduce known results in several limits. In region B$_L$, our solution matches the large-charge expansion of~\cite{Correa:2012hh,Drukker:2012de} and correctly yields the leading Lüscher correction to the BPS energy. Furthermore, we have successfully compared our small-angle expansions with those presented in earlier works~\cite{GS,Gromov:2013qga}, finding complete agreement. 
We have also extended the analysis of the large-charge limit by computing explicitly the next two subleading exponential corrections to the energy, see~\eqref{ElargeL}.

\paragraph{Quantum fluctuations and normal modes.}

The next step of our investigation concerns the spectrum of the quadratic quantum fluctuations around the classical string solution, dual to the cusped Wilson loop with $R$-charge insertions. We focus on its semiclassical (WKB) approximation and its behavior in several physically interesting limits.

The eight physical transverse bosonic fluctuations $\xi^i$ ($i=1,\dots,8$) around the classical string background satisfy a system of coupled differential equations that can be schematically written as
\begin{equation}
\mathcal{K}_{ij} \,\xi^j = 0, \qquad i,j = 1,\dots,8,
\label{eq:fluct-eq}
\end{equation}
where the operator matrix $\mathcal{K}_{ij}$ is obtained following the general procedure outlined in~\cite{Drukker:2000ep,Forini:2015mca}.

The system naturally decomposes into two sectors:
\begin{itemize}
    \item \textbf{Uncoupled modes} ($i=1,\dots,4$): These four modes are diagonal and can be quantized exactly by mapping the problem to Schrödinger equations with Lamé potentials.
    \item \textbf{Coupled modes} ($i=5,\dots,8$): The exact quantization of these modes is considerably more challenging due to the presence of both a non-diagonal mass matrix and a non-trivial $SO(4)$ connection on the normal bundle. A possible approach toward their systematic WKB treatment is discussed in Appendix~\ref{app:WKB}.
\end{itemize}
Given these difficulties, we have employed the WKB approximation to study all bosonic frequencies, while obtaining more precise results for the uncoupled sector. For large quantum numbers $\ell \gg 1$, we find a universal asymptotic expression for the physical frequencies. At leading order, it reads
\begin{equation}
\omega_{\text{phys},\ell} = \frac{\pi\ell}{2\sqrt{n-m}\,K(m)} \,, \qquad \ell \in \mathbb{N}.
\label{eq:omega-asympt}
\end{equation}
This formula serves as the starting point for our analysis of various regimes.

\paragraph{The antiparallel limit.}

As with the classical energy, the limit $\phi \to \pi$ (corresponding to antiparallel lines in flat space) exhibits markedly different behaviors in regions A and B.

\textbf{Region A$_L$:}
From the exact quantization condition for the uncoupled modes, we obtain at leading order
\begin{equation}
\omega^{(A)}_{\text{phys},\ell} =
\frac{\pi\ell\bigl[E(\mbar) - (1-\mbar)K(\mbar)\bigr]}{\sqrt{\mbar(1-\mbar)}\,K(\mbar)\,(\pi-\phi)},
\label{eq:omega-regA}
\end{equation}
where $\mbar \in (0,\tfrac{1}{2})$   is a solution of the consistency condition 
in the  antiparallel ($n\to m$) limit, equation~\eqref{consistencybarA}, 
and parameterizes the position along the transition line between the two regions (see Fig.~\ref{fig:Transitioncurve}). The singular factor $1/(\pi-\phi)$ corresponds, in the decompactification limit, to the expected $1/L_\text{sep}$ dependence on the distance $L_\text{sep}$ between the two lines.

Setting $\mbar = 1/2$ recovers the result of \cite{Klebanov:2006jj} for the standard quark--antiquark flux tube:
\begin{equation}
\omega^{(A)}_{\text{phys},\ell} \big|_{\mbar=1/2} =
\frac{8\pi^3\ell}{\Gamma(1/4)^4\,(\pi-\phi)},
\label{eq:omega-KMT}
\end{equation}
upon the identification of $\pi-\phi$ with the distance between the heavy static particles.

\textbf{Region B$_L$:}
In this regime, at leading order, the frequencies remain finite as $\phi \to \pi$:
\begin{equation}
\omega^{(B)}_{\text{phys},\ell} =
\frac{\pi\ell}{2\sqrt{-\widehat{m}}\,K(\widehat{m})},
\label{eq:omega-regB}
\end{equation}
where $\widehat{m}<0$ is the solution of the consistency condition in this region, equation~\eqref{consistencyhatB}, and  parameterizes the transition point from the B side. This finiteness contrasts sharply with the divergent behavior in region A$_L$. The same behavior is found for the coupled modes in the WKB approximation, since the antiparallel limit is ultimately controlled by the constant velocity $\kappa = \sqrt{(n-m)/(1-n-m)}$ that governs the evolution of the decompactified global time, see~\eqref{ansatz2}.

\paragraph{Critical scaling near $\mathcal{L}_c^\theta$.}

A further interesting aspect is the behavior of the bosonic fluctuations near the critical line $\mathcal{L} \sim \mathcal{L}_c^\theta$. 
There, the normal mode frequencies exhibit a characteristic scaling behavior that mirrors that of the classical energy.

\textbf{Region A$_L$:} Here we are approaching the blue line in Fig.~\ref{fig:Transitioncurve} from below. 
The coefficient of the pole at $\phi = \pi$ vanishes as $\mathcal{L} \to \mathcal{L}_c^{\theta}$:
\begin{equation}
\omega^{(A)}_{\text{phys},\ell} \sim
\frac{\ell}{\pi-\phi}
\sqrt{\frac{2\pi}{3}}
\sqrt[4]{\frac{m_L^c}{m_L^c(n_L^c-1) + n_L^c}}
\sqrt{\mathcal{L}_c^\theta - \mathcal{L}}\, ,
\label{eq:omega-crit-A}
\end{equation}
where $n_L^c$ and $m_L^c$ denote the values of $n_L$, $m_L$ at the critical point.

\textbf{Region B$_L$:} 
Slightly above the blue transition curve, the finite frequencies diverge as $(\mathcal{L} - \mathcal{L}_c^\theta)^{-1/2}$:
\begin{equation}
\omega^{(B)}_{\text{phys},\ell} \sim
\frac{\ell\sqrt{\pi}\,\sqrt[4]{m_L^c(n_L^c-1) + n_L^c}}
{2\sqrt{2}\,\sqrt[4]{m_L^c}\,\sqrt{\mathcal{L} - \mathcal{L}_c^\theta}}.
\label{eq:omega-crit-B}
\end{equation}
This divergence originates from the factor $\kappa$ mentioned above, which itself becomes singular at the critical point.

\paragraph{Large $\mathcal{L}$ expansion and connection to BMN.}

To connect with earlier literature and to clarify the quantum fluctuations of the $R$-charged cusp, we have also studied the bosonic frequencies in the large-$\mathcal{L}$ limit. We expected, and indeed found, a relation to the BMN picture that underlies the integrability framework of~\cite{Correa:2012hh, Drukker:2012de}. Expanding the one-loop frequencies of the uncoupled transverse fluctuations for $\mathcal{L} \gg 1$ and $\ell\gg1$, keeping fixed the ratio $\eta = \frac{\sqrt{\lambda}\,\ell}{L}$, we obtain
\begin{equation}
\omega_\ell = \sqrt{1 + \frac{\pi^2\eta^2}{4}} + O\!\left(\frac{1}{\mathcal{L}}\right)\,.
\label{eq:omega-ppwave}
\end{equation}
This matches precisely the spectrum of an open string in a pp-wave background~\cite{Drukker:2006xg}. For the coupled modes (for which we can only perform a WKB analysis,  corresponding effectively to take $\eta\gg 1$) we reproduce as expected the leading order of the expansion in $\eta$ of~\eqref{eq:omega-ppwave}.

\paragraph*{Frequency densities and decoupling from the cusp.}
To further elucidate the relation between exact frequencies $\omega_\ell$ and the quantum number $\ell$ at large $\mathcal{L}$, we have introduced ``frequency densities'' ${\rho}_\omega$ (for AdS$_5$ modes) and $\tilde\rho_\omega$ (for $S^5$ modes) following \cite{Giombi:2022anm}:
\begin{equation}
\ell = \int_{\omega_0}^{\omega_\ell} \rho_\omega(\omega) \,\dif\omega, \qquad \ell \in \mathbb{N},
\label{eq:freq-density}
\end{equation}
and similarly for $\tilde{\rho}_\omega$.  
We find that the physical angles $\phi$ and $\theta$ contribute to both ${\rho}_\omega$ and $\tilde{\rho}_\omega$ only via exponentially suppressed terms at large $\mathcal L$. This indicates that the string fluctuations effectively decouple from the cusped Wilson loop geometry in the large $\mathcal{L}$ limit -- a fact anticipated from earlier studies \cite{Correa:2012hh, Drukker:2012de}. In other words, the leading terms in the frequency densities agree with those for a straight Wilson line \cite{Drukker:2006xg,Giombi:2022anm}, and the effective quantization condition becomes independent of the cusp angles up to exponentially small corrections.


\subsection{Discussion and Outlook}
\label{subsec:discussion}

It is instructive to compare our results with earlier related investigations. In the $L=0$  case  studied in~\cite{Klebanov:2006jj}, the frequencies \eqref{eq:omega-KMT} were interpreted as excitations of quark--antiquark bound states near the dissociation threshold $E=0$ at strong coupling ($\sqrt{\lambda} \gg 1$). The authors argued that the flux-tube dynamics at strong coupling differs dramatically from its weak-coupling counterpart: the string in AdS$_5$ possesses a rich spectrum of discrete energy levels between the ground state (determined by the Coulomb potential) and $E=0$, reflecting the many internal degrees of freedom of a fluctuating flux tube. Finding no evidence for such bound states at weak coupling, they concluded that a critical coupling ${\lambda_c}$ must exist, above which these excitations emerge. Remarkably, our transition line $\Theta_{\text{eff}}=\pi$ corresponds precisely to configurations with zero classical string energy, consistently with their picture.

Another setup we consider for comparison is the following. In the $D3$--$D7$ brane system of~\cite{Hong:2004gz}\footnote{The analysis of~\cite{Hong:2004gz} has been largely inspired by~\cite{Tseytlin:2002tr}. Despite the similarities, there is a crucial difference between \cite{Tseytlin:2002tr} and \cite{Hong:2004gz} (and thus between us) on the boundary conditions imposed on the rotational angle in $S^5$.}
 -- which describes dynamical quarks interacting with scalar bosons -- a similar interaction potential was derived. In that context, the fluctuation modes of the string should be viewed as excitations of a generalized quarkonium system near dissociation. However, \cite{Hong:2004gz} did not analyze the behavior of quantum fluctuations, nor the region corresponding to our region B$_L$, where the bound state disappears and the interpretation of \eqref{eq:omega-regB} becomes less straightforward.

On the other hand, both \cite{Klebanov:2006jj} and \cite{Hong:2004gz} effectively overlooked region B$_{L}$: in the former it reduces to a single point, while in the latter the dynamical quark mass begins to dominate. In contrast, we have shown that the qualitative distinction between regions A$_{L}$ and B$_{L}$ persists at the quantum level. This indicates that the classical transition is not washed out by one-loop effects, even though a complete analysis would require exact closed-form expressions for all fluctuation frequencies.

Coming back to the fluctuation spectra discussed in this work, they are naturally defined in global AdS$_5$, whose boundary is $\mathbb{R}\times S^3$. As a result, all frequencies are initially measured in units of the inverse curvature radius  $a$ of the three-sphere. The flat-space quark-antiquark potential is recovered only after taking the decompactification limit $a\to\infty$ while keeping the physical separation $L_\text{sep}=a(\pi-\phi)$ fixed. In this limit, the distinction between the two regimes becomes sharp. In Region A$_L$, the {\it physical} fluctuation frequencies scale as $\omega_{\rm phys}\sim 1/(\pi-\phi)$, so that the factor of $a$ cancels and finite flux-tube excitations survive in flat space, in agreement with the standard picture of a string stretched between two static sources. In Region B$_L$, by contrast, the frequencies approach a constant as $\phi\to\pi$ and therefore scale as $\omega_{\rm phys}\sim  1/a$, vanishing in the decompactification limit. This behavior signals that Region B$_L$ is dominated by large $R$-charge dynamics: the classical string effectively localizes in the AdS directions while remaining extended on the internal $S^5$, and its excitations are governed by the global curvature scale $a$ rather than by the quark-antiquark separation $L_\text{sep}$. Such modes are analogous to BMN excitations of semiclassical strings carrying large angular momentum~\cite{Berenstein:2002jq, Metsaev:2002re, Frolov:2002av}, and to local excitations of defect states in large-charge sectors of defect CFTs~\cite{Giombi:2022anm}. 

Operator insertions have been used to describe scattering amplitudes on the Coulomb branch of 
$\mathcal{N}=4$ SYM via cusped Wilson lines~\cite{Henn:2010bk,Bruser:2018jnc}. This approach reveals deep connections between the cusp anomalous dimension, Regge-limit behavior, and the spectrum of flux-tube excitations. In particular, subleading Regge trajectories are encoded by specific operator insertions at the cusp, involving scalars whose couplings are either parallel or orthogonal to the Wilson loop~\cite{Flieger:2025ekn,Alday:2025pmg}.
Using quantum spectral techniques, in \cite{Alday:2025pmg} the authors show that the coefficient of the Coulomb potential in the presence of parallel insertions vanishes below a certain critical $\hat\lambda_c$, becomes finite above it, and asymptotes to the value without insertions as $\lambda \to \infty$. Our analysis is consistent with this picture: our large-$\mathcal{L}$ regime (region B$_{L}$) corresponds parametrically to weak coupling (where $R$-charge is a finite integer and $\lambda$ is small), and indeed we find no Coulomb potential there, with quantum fluctuations being of BMN type. Conversely, for small $\mathcal{L}$ (or large $\sqrt{\lambda}$ at fixed integer $L$), a flux tube emerges and the Coulomb coefficient approaches its standard $L=0$ value. Although the nature and strength of the operator insertions differ between the two setups, we believe the underlying transition shares a common origin.

Cusped Wilson loops also provide a tool for studying the fusion of conformal defect lines~\cite{Kravchuk:2024qoh,Cuomo:2024psk}. The antiparallel limit $\phi \to \pi$ relates to the effective field theory of defect fusion, yielding explicit predictions for the leading terms in the expansion. Inserting a local operator promotes this fusion from occurring in the vacuum to taking place in an excited state \cite{Cuomo:2024psk}. We briefly discuss this scenario at strong coupling in Section~\ref{sec:field}, where we also propose a prediction, see equation~\eqref{three-point-pred}, for the asymptotic spectrum of defect three-point functions involving heavy bulk operators in the presence of a straight Wilson line with $Z^L$ insertions.

Several direct extensions of our work present themselves as natural next steps. The most concrete follow-up would be to study the system using the Quantum Spectral Curve technique. This would determine whether the transition we observed at strong coupling persists at intermediate coupling and for finite values of the \(R\)-charge. Such an analysis could reveal whether the classical transition corresponds to a genuine effect in the full quantum theory, or is merely an artifact of the semiclassical approximation. Understanding the quantum field theory origin of the two distinct regimes remains an open challenge. One promising approach would be to identify specific classes of Feynman diagrams that dominate the generalized quark--antiquark potential in a suitable limit\footnote{In \cite{Gromov:2021ahm}, a fishnet-like class of diagrams was used to study the all-coupling behavior of a related system.}. Such a diagrammatic analysis might elucidate how the transition emerges from the microscopic dynamics of \(\mathcal{N}=4\) SYM.

From the string theory perspective, it would be interesting to study correlation functions involving operator insertions on the cusped Wilson loop. These correlators could probe aspects of the transition not captured by the vacuum expectation value alone, potentially revealing signatures in higher-point functions or in the operator product expansion. 

More ambitiously, one could explore whether large orthogonal insertions play a role in the description of scattering amplitudes on the Coulomb branch of \(\mathcal{N}=4\) SYM. The connection between cusped Wilson loops and Regge trajectories~\cite{Flieger:2025ekn,Alday:2025pmg} suggests that our findings might have implications for the high-energy limit of amplitudes in the presence of vacuum expectation values.

Recently, Thermodynamic Bethe Ansatz (TBA) equations for the cusp anomalous dimension have been derived in ABJM theory~\cite{Correa:2023lsm}. In that context, operator insertions at the cusp have only been studied at leading order in weak coupling. It would be valuable to explore their effects on the quark--antiquark potential in ABJM, both to uncover universal features and to identify theory-specific aspects of the transition.

\vskip 0.7cm

The paper proceeds as follows.
In Section~\ref{sec-classical} we review the classical string solution dual to the generalized cusped Wilson loop and fix our conventions and parameters. In Section~\ref{sec:potentia-and-BPS} we study the main kinematical regimes of the classical solution, including the antiparallel lines limit, the near-BPS limit, and the large $\L$ expansion. Section~\ref{sec:quantum} contains the quantum analysis: we set up the quadratic fluctuation problem, derive the corresponding frequency/quantization conditions, and analyze them in the relevant limits, with particular emphasis on the antiparallel lines regime and on the large $\L$ (BMN-type) expansion. In Section~\ref{sec:field} we discuss the field-theory interpretation, focusing on the small-angle limit in terms of integrated correlators and on the antiparallel lines limit in terms of defect fusion. Several technical details are collected in the appendices: 
Appendix~\ref{app-classical} summarizes the classical setup and the relation to parameterizations used in the literature; Appendix~\ref{app:B} collects explicit expansions of the classical energy in the various regimes; Appendix~\ref{app-quantum} presents the full bosonic and fermionic fluctuation operators derived from the Green–Schwarz action; Appendix~\ref{app:D}  is devoted to the analysis of fluctuation frequencies; Appendix~\ref{app:elliptic} collects conventions and useful formulae for elliptic functions employed throughout the paper.

\section{The classical string solution }
\label{sec-classical}

In this section we review the classical string solution  dual to the Wilson loop $W_L$ in~\eqref{WL}~\cite{Drukker:2006xg,Correa:2012at,GS}, referring to  Appendix~\ref{app-classical} for some details.
In terms of embedding coordinates, the ansatz for the classical solution reads
\begin{equation}
\begin{aligned} 
\text{AdS}_5:&~ X_1+i X_2=  \cosh\!{\rho(\sigma)}\,e^{i \kappa\tau},~~ &X_3+i X_4=\sinh\!\rho(\sigma)\,  e^{i \varphi(\sigma) }, ~~ X_5= X_6= 0\,,\\
S^5:&~Y_1+i Y_2= \cos\psi(\sigma)\,e^{i\gamma\tau}, &\!\!\!Y_3+i Y_4= \sin\psi(\sigma) \,e^{i \vartheta(\sigma)}, ~~ Y_5=Y_6=0\,,
\end{aligned}
\label{ansatz}
\end{equation}
where we have set the AdS (and $S$) length to 1.  Above, $Y_1$ and $Y_2$ are turned on to account for the insertion of $Z^L=(\Phi_1+i\Phi_2)^L$ at the cusp vertex, while  $Y_3$ and $Y_4$ represent the two scalars, $\Phi_3$ and $\Phi_4$, which in our case are chosen to appear  in the generalized connection of the Maldacena-Wilson loop~\eqref{WL}. The ansatz \eqref{ansatz} probes an $\text{AdS}_3\times S^3$ subspace~\footnote{In the ansatz~\eqref{ansatz}, with the choice $X_5= X_6= 0$ and $Y_5=Y_6=0$ one exploits the invariance under rotation in the two  spheres complementary to the $\text{AdS}_3\times S^3$ subspace.}.  
Equivalently, given the $\text{AdS}_5\times S^5$ parametrization~\eqref{parametrization1}, 
it can be written as
\begin{equation}
\begin{aligned}
&t=\kappa\, \tau, \qquad \rho=\rho(\sigma),\qquad~ \varphi=\varphi(\sigma),\qquad x_1=1,\qquad y_1=0\,,\\
&\nu=\gamma\,\tau,\qquad \psi=\psi(\sigma)\,,\qquad \vartheta=\vartheta(\sigma),\qquad x_2=1,\qquad y_2=0\,.
\end{aligned}
\label{ansatz2}
\end{equation}
Worldsheet coordinates are $\tau\in\mathbb{R}$ and $\sigma\in[-s/2,s/2]$. The range $s$ of~$\sigma$ is determined dynamically, as we explain in details in the following subsections. 
The decompactified global time ($-\infty<t<\infty$) evolves with constant velocity $\kappa$ and the $\text{AdS}$ boundary, where the Wilson loop is defined,  is reached at infinite radial direction $\rho\to\infty$. 
The geometrical and internal angles, $\phi$ and  $\theta$ respectively,  are described respectively by the $\varphi$-rotation in the $(X_3, X_4)$-plane and  the $\vartheta$-rotation in the $(Y_3, Y_4)$-plane, with $\varphi\in[0,2\pi)$ and $\vartheta\in[0,2\pi)$.   The $R$-charge $L$ of the inserted operator $Z^L$ is realized as conserved angular momentum associated to a uniform $\nu$-rotation in the $(Y_1, Y_2)$-plane  characterized by constant $\gamma$, with $\nu\in[0,2\pi)$ and $\psi\in[0,\frac{\pi}{2})$. Notice that, within $S^5$, the rotations in $\theta$ and in $\nu$ take place in mutually orthogonal planes. 
 The boundary conditions of the problem are such that, at the AdS boundary $\rho(\pm s/2)\to\infty$,  the cusp width in $\text{AdS}_5$ and in $S^5$  reach their limiting values:  $\varphi(s/2)=\pi-\phi/2$, $\varphi(-s/2)=\phi/2$, $\psi(\pm s/2)=\pi/2$ and  $\vartheta(\pm s/2)=\pm \theta/2$.%
 \footnote{In the case of $\psi(\pm s/2)$, we have a shift of $\pi/2$ with respect to the notation of~\cite{Drukker:2006xg}.}
 The parameter $\kappa$ in \eqref{ansatz2} can be restricted to positive values, since a change of sign can be absorbed into the redefinition $\tau \to -\tau$. Likewise, we choose $\gamma > 0$ in \eqref{ansatz2}, as it controls the sign of $L$ (see eq.~\eqref{Lconserved2}), which in our setup is taken to be positive.

To make contact with the relevant literature and obtain compact expressions for the  Noether charges and the integrals of motion of the problem~\cite{GS}, it is useful to proceed with the parametrization and metric in~\eqref{parametrization2}-\eqref{metric2}, corresponding to the identification
\be\label{transf-rhotilde}
\sinh\rho=r\,,\qquad\qquad \sin\psi=\tilde\rho\,.
\ee

The bosonic part of the GS string action is 
\be
S_{b}=- {1\over 4\pi \ell_s^2} \int \dif \s\dif \tau \sqrt{h} h^{\alpha\beta} \pa_\alpha X^\mu \pa_\beta X^\nu G_{\mu\nu}\,,
\ee
where the $h_{\alpha\beta}$ is the worldsheet metric (with curved indices $\alpha=0,1$), $G_{\mu\nu}$ the target-space metric (with curved indices $\mu, \nu=0, \dots, 9$), and here $X^\mu$ represents collectively the embedding coordinates. 
In conformal gauge, using the metric \eqref{ansatz} and the ansatz \eqref{ansatz2}, then string Lagrangian reads
\be
\label{Polyakov}
\mathcal{L}_b=
\frac{g}{2}\,\left[\kappa^2 (1+r^2)+\frac{r^{\prime 2}}{1+r^2}+r^2\varphi^{\prime 2}-\gamma^2(1-\tilde\rho^2)+\frac{\tilde\rho^{\prime 2}}{1-\tilde\rho^2}+\tilde\rho^2 \vartheta^{\prime 2}\,\right]\equiv\mathcal{L}_{\mbox{\tiny{AdS$_3$}}}+\mathcal{L}_{\mbox{\tiny{S$^3$}}}\,,
\ee
where we have included the overall factor 
$$g :\, = {1\over 4\pi\ell_s^2}=\frac{\sqrt{\lambda}}{4\pi}$$ in $\mathcal L_b$ (the $\text{AdS}$ radius is set to 1), and primes denote derivatives with respect to $\sigma$. The Lagrangian $\mathcal{L}_b$   naturally splits into two separate pieces, $\mathcal{L}_{\mbox{\tiny{AdS$_3$}}}$ and  $\mathcal{L}_{\mbox{\tiny{S$^3$}}}$, whose corresponding Hamiltonians, $\mathcal{H}_{\mbox{\tiny{AdS$_3$}}}$ and $\mathcal{H}{\mbox{\tiny{S$^3$}}}$, are separately conserved. The dynamics is further constrained by the Virasoro condition
\be\label{Virasoro}
\mathcal{H}_{\mbox{\tiny{AdS$_3$}}}+\mathcal{H}_{\mbox{\tiny{S$^3$}}}=\frac{r^{\prime 2}}{1+r^2}+r^2\varphi^{\prime 2}-\kappa^2 (1+r^2)+ \frac{\tilde\rho^{\prime^2}}{1-\tilde\rho^2}+ \tilde\rho^2 \vartheta^{\prime 2}+\gamma^2(1- \tilde\rho^2)=0\,\,,
\ee
which makes the induced metric conformally flat
\be\label{inducedflat}
h_{\alpha\beta} =e^{2\Lambda(\s)} \eta_{\alpha\beta}\,, \qquad e^{2\Lambda(\s)}=\kappa^2(1+r^2)-\gamma^2(1-\tilde\rho^{\,2}) = \kappa ^2 \cosh ^2\rho -\gamma ^2 \cos ^2\psi\,.
\ee
Above, we  used~\eqref{transf-rhotilde} in the last step, to have the  Weyl factor $2\Lambda(\sigma)$  written in both the parametrizations used in this paper.  There are also two cyclic coordinates, $\varphi$ and $\vartheta$, which give rise to two conserved ($\sigma$-conserved) canonical momenta,
\be
\label{conservedmomenta}
r^2\varphi^\prime=\ell_\varphi \ \ \ \ \ \mathrm{and}\ \ \ \ \  \tilde\rho^2 \vartheta^\prime=-\ell_\vartheta.
\ee
Substituting these relations into the conserved Hamiltonians, we obtain two first-order differential equations for $r$ and $\tilde\rho$
\begin{eqnarray}\label{req}
&&\mathcal{H}_{\mbox{\tiny{AdS$_3$}}}=\frac{r^{\prime 2}}{r^2+1}+\frac{\ell _{\varphi }^2}{r^2}-\kappa^2 (1+r^2)=-1\,,\\\label{rhoeq}
&&\mathcal{H}_{\mbox{\tiny{S$^3$}}}=\frac{\tilde\rho^{\prime 2}}{1-\tilde\rho^2}+\frac{\ell _{\vartheta }^2}{\tilde\rho^2}+\gamma^2(1-\tilde\rho^2)=1.
\end{eqnarray}
Here we have used the invariance of our ansatz under a rescaling of $\sigma$ and $\tau$ by a common positive constant $c_0$, together with a rescaling of the parameters $\ell_\varphi$ and $\ell_\vartheta$ by $1/c_0$, to fix the constant values of the Hamiltonians to $\mp 1$ \cite{Correa:2012hh}.\footnote{The vanishing of the Virasoro constraint ensures that the two integration constants are opposite~\cite{Drukker:2005cu}. Moreover,  the right-hand side of the $\mathcal{H}{\mbox{\tiny{S$^3$}}}$ equation must be positive, since its left-hand side is positive definite for $\tilde \rho \leqslant 1$.}

\subsection{$\text{AdS}_5$ sector}
\label{sec:classical-ads}

We now trade the parameters $\kappa,\,\ell_\varphi$ of the solution in $\text{AdS}_5$ for two modular parameters $n$ and $m$ that govern the elliptic integrals appearing in the solution~\footnote{{The vanishing of $\Delta$, i.e.\ when $\kappa^2 = 1$ and $\ell_\phi = 0$, corresponds to a singular point in the above parametrization. While $m$ always diverges to $-\infty$, the behavior of $n$ depends on the relative scaling between $(\kappa^2 - 1)$ and~$\ell_\phi$. For instance, if $\kappa^2\to 1^+$ and the ratio $(\kappa^2 - 1)^2 / \ell_\phi$ remains finite, the system approaches the straight line configuration.}}, 
\be\label{nm}
n\equiv\frac12+\frac{\kappa^{2}-1}{2\sqrt{\Delta}}\,,\qquad\qquad m\equiv\frac12-\frac{\kappa^{2}+1}{2\sqrt{\Delta}}\,,\qquad\qquad
 \Delta \equiv (\kappa^{2}-1)^{2}+4\kappa^{2}\ell_{\varphi}^{\,2}\,.
\ee
Given that $\Delta\geq0$
 and $\sqrt{\Delta}\geq|\kappa^2-1|$, %
 the two auxiliary parameters  in~\eqref{nm} have the range
\be\label{rangenm}
0\leq n\leq1\,,\qquad\qquad m\leq\frac{1}{2}\,,
\ee
and the following inequalities hold
\begin{equation}\label{par-relations}
n-m=\frac{\kappa^{2}}{\sqrt{\Delta}}\ge 0, \qquad\qquad
1-n-m=\frac{1}{\sqrt{\Delta}}\ge 0\,.
\end{equation}
Both of these parameters vanish as $\kappa$ approaches zero; however, their ratio $n/m$ remains finite and, in this limit, it is equal to $\frac{\ell_\varphi^2}{\ell_\varphi^2 - 1}$.
It is useful to write also the inverse relations among the two sets of parameters, that is
\be\label{formula-kappaellephi-of-nm}
\kappa^2 = {n-m \over 1-n-m}\,, \qquad \ell_\varphi^2= {n(1-n)\over (n-m)(1-m-n)}\,.
\ee

Rather than $\sigma$, one can use  $\varphi$ to parameterize the spatial direction of the worldsheet~\cite{DF}. 
The equation~\eqref{req} for the $\text{AdS}$ radial coordinate $r$ takes then the form
\be\label{reqnew}
(\partial_\varphi r)^2= \frac{r^2 (r^2 + 1)}{\,(1-n)} \,\Big(r^2 \,\Big(1 -\frac{m}{n}\Big)  + 1\, \Big) \,\big( r^2\,(n -m) -1+n\big)\,.
\ee
Because of~\eqref{rangenm} and~\eqref{par-relations}, only the last factor above can vanish.   Then 
\be\label{rmin}
r_\text{min}=\textstyle{\frac{\sqrt{1 - n}}{\sqrt{n -m}}}
\ee
 is the minimal value, or deepest penetration of the minimal surface in the $\text{AdS}$ bulk. Notice that this is reached at $\s=0$.

Equation~\eqref{reqnew} can be inverted and solved  for the cusp angle \( \varphi \). We choose \( \varphi \) to span the interval \( \frac{\phi}{2} \leqslant \varphi \leqslant \pi - \frac{\phi}{2} \), so that the geometric opening of the cusp is \( \pi - \phi\). The bulk turning point $r_\text{min}$ is obtained when \( \varphi = \frac{\pi}{2} \), the midpoint of the interval. For the branch of the surface where $\varphi$ runs from $\frac{\phi}{2}$ to $\frac{\pi}{2}$, one can then write
 \be\label{varphi2}
\varphi(r)=\frac{\phi}{2}+ \int^{\infty}_{r^2} \frac{ dy\,\sqrt{n(1-n)}}{2 y \, \sqrt{(y+1) \,\big(y \,(n-m)+n\big)\,\big (y \,(n-m)-1+n\big) }}\,,
\ee
where the integral has been put in a standard elliptic form. We perform the integral
assuming the following ordering of the roots in the denominator:  ${\textstyle r^2>\frac{1-n}{n-m}>-\frac{n}{n-m}>-1}$, which is  
equivalent, due to~\eqref{rangenm} and~\eqref{par-relations},  to assume $m>0$. 
One then obtains
\be\label{varphi-a}
 \varphi(r)=\frac{\phi}{2}+
\frac{\sqrt{1-n} \sqrt{n-m}}{\sqrt{n}} \left[\Pi \left(n;\xi\left|m\right.\right)-F\left(\xi\left|m\right.\right)\right]\,,~~\,\,\,\, \xi=\arcsin\left(\textstyle\sqrt{\frac{1}{r^2 (n - m)+n}}\right)\,.
\ee
The other possible ordering (${\textstyle r^2>\frac{1-n}{n-m}>-1>-\frac{n}{n-m}}$) corresponds to negative values of $m$  in the elliptic functions appearing above.  
At the bulk turning point $r_\text{min}=\textstyle{\frac{\sqrt{1 - n}}{\sqrt{n - m}}}$, it is $\xi=\frac{\pi}{2}$ and $\varphi(r_\text{min})=\frac{\pi}{2}$. This boundary condition determines then $\phi$ in terms of $n$ and $m$ as
\be
\label{phi01}
\phi=\pi -2 \frac{\sqrt{1-n} \sqrt{n-m}}{\sqrt{n}} \big(\,\Pi \big( n | m ) - K(m)\,\big)\,.
\ee
The invariance under global time translations in \text{AdS}$_5$ ($K_t = \partial_t$) fixes the conserved energy density and the corresponding conserved charge.
The resulting integral is power-like divergent and  only the finite part (f.p.) should be retained, 
\begin{equation}
\label{Energy}
E =\left[ \int_{-s/2}^{s/2} d\sigma  \,\Pi_t^0 \right]_{\rm f. p.}\!\!\!\!\!\!\!\!= 2 g \kappa\left[ \int_{-s/2}^{s/2} d\sigma (1 + r^2(\sigma))\right]_{\rm f. p.}
\!\!\!\!\!\!\!\!=2 g \left[{ \int_{r_\text{min}}^{\infty} \frac{d x\sqrt{x+1}}{ \sqrt{\big(x+\frac{n}{n-m}\big)\,\big(x-\frac{1-n}{n-m}\big)}}}\right]_{\mathrm{f.p.}}
\end{equation}
Introducing a cut-off $x=1/\epsilon^2$, expanding for $\epsilon\to 0$ and dropping the purely divergent part~\cite{DF}, one obtains
\be\label{E}
\mathcal{E}\equiv\frac{E}{4g}=\frac{1}{\sqrt{n-m}} \big(\,(1-m) K (m) - E(m)\,\big)\,.
\ee 
Taking into account the relations~\eqref{nmtobp}, these are formally the same expressions of Ref.~\cite{DF}. Since all classical conserved charges are proportional to $g\sim\sqrt{\lambda}$,  in what follows it is more natural to work with the rescaled variable $\mathcal{E}$ introduced above~\footnote{Similarly, rather than with the angular momentum $L$,  we will work below with a rescaled momentum $\L=L/(4g)$, see~\eqref{L}.}.   

Finally, the range of the  wordsheet  coordinate $\sigma$ is fixed by integrating  $d\sigma=dr/r'$ with $r^\prime$ given by the solution of \eqref{req} (i.e. \eqref{rprime}). We obtain
\be
\label{sAdS5}\!\!\!
s_\text{\tiny $\text{AdS}_5$}\!\!=2\int_{r_\text{min}}^\infty   \frac{r \, dr\,\sqrt{n-m}\,\sqrt{1-n-m}}{
\sqrt{(r^2+1) \,\big(r^2(n-m)+n\big)\big(r^2(n-m)-1+n\big)}}= \,2 \, \sqrt{1 - n - m} \,K(m).
\ee

\subsection{$S^5$ sector}
\label{sec:classical-s}

Two convenient auxiliary moduli, functions of the parameters $\gamma$ and $\ell_\vartheta$ of the solution in $S^5$, are in this case defined as\footnote{The apparent singularity in the expression for $n_{\text{\tiny L}}$ as $\gamma \to 0$ is removable: although the denominator vanishes, the numerator scales with the same power of $\gamma$, ensuring that the limit remains finite.}
\be\label{nLmL}
n_{\text{\tiny L}}\equiv\frac{1}{2}+\frac{1-\sqrt{\Delta_{\text{\tiny L}}}}{2 \gamma ^2}\,,\qquad 
m_{\text{\tiny L}}\equiv1-\frac{2 \sqrt{\Delta_{\text{\tiny L}}}}{\gamma ^2 +1+\sqrt{\Delta_{\text{\tiny L}}}}\,,\qquad 
 \Delta_{\text{\tiny L}} \equiv (\gamma^{2}-1)^{2}+4\gamma^{2}\ell_{\vartheta}^{\,2}\,.
\ee
They obey the following inequalities
\begin{equation}\label{par-relations-L}
  \frac{ n_{\text{\tiny L}}}{m_{\text{\tiny L}}} =\frac{\gamma ^2+1+\sqrt{\Delta_{\text{\tiny L}}}}{2
   \gamma ^2}\geqslant 1, \qquad\qquad
1 + m_{\text{\tiny L}} \Big(1 - \frac{1}{n_{\text{\tiny L}}}\Big)
=\frac{2}{\gamma^2+1+\sqrt{\Delta_{\text{\tiny L}}}} \geqslant 0\,.
\end{equation}
Since the three terms in equation~\eqref{rhoeq} are non–negative for
$\tilde\rho^{2} \leqslant 1$~\footnote{Recall that $\tilde\rho^{2} \leqslant 1$ by construction (see eq.~\eqref{transf-rhotilde}).} and sum to $1$, one has in particular
$\ell_\vartheta^{2}/\tilde{\rho}^{2} \leqslant 1$. Hence
$\ell_\vartheta^{2} \leqslant 1$. Using the inverse relations
\be\label{formula-gammaelletheta-of-nmL}
 \gamma^2 = \frac{m_{\text{\tiny L}}}{n_{\text{\tiny L}}-m_{\text{\tiny L}}(1-n_{\text{\tiny L}})}\,, \qquad 
 \ell_\vartheta^2= \frac{(1-n_{\text{\tiny L}})(n_{\text{\tiny L}}-m_{\text{\tiny L}})}{n_{\text{\tiny L}}-m_{\text{\tiny L}}(1-n_{\text{\tiny L}})}\,,
\ee 
combined with the inequalities~\eqref{par-relations-L}, the requirements $\gamma^2\ge 0$ and $0\leqslant \ell_{\vartheta}^2\leqslant 1$ translate into the following ordering for the moduli $m_{\text{\tiny $L$}}$ and $n_{\text{\tiny $L$}}$
\be\label{rangenLkL}
 0 \leqslant m_{\text{\tiny $L$}}\leqslant \nL \leqslant1\,.
\ee
Using the second relation in~\eqref{conservedmomenta}, we rewrite equation~\eqref{rhoeq} by replacing the derivative with respect to $\sigma$ by the derivative with respect to $\vartheta$. In terms of the parameters~\eqref{nLmL}, we then obtain
\be\label{rhoeqnew}
 (\partial_{\vartheta}\tilde\rho)^{2}=\frac{\tilde\rho^2\left(1-\tilde\rho^2\right)\,\big(\,\tilde\rho ^2-(1-n_{\text{\tiny $L$}})\,\big) \left(\tilde\rho ^2
   m_{\text{\tiny $L$}}-m_{\text{\tiny $L$}}+n_{\text{\tiny $L$}}\right)}{\left(1-n_{\text{\tiny $L$}}\right) \left(n_{\text{\tiny $L$}}-m_{\text{\tiny $L$}}\right)}\,.
   \ee
Due to eq. \eqref{transf-rhotilde}, we look for solutions where  $\tilde\rho^2 \leqslant 1$. Moreover, the right-hand side of eq. \eqref{rhoeqnew} must be non-negative in order for real solutions to exist. This requirement is satisfied when 
$\tilde\rho_\text{min}\leqslant  \tilde\rho^2 \leqslant 1$, with 
\be\label{def-rhotilde-min}
\tilde\rho_\text{min}\equiv\sqrt{1-n_{\text{\tiny $L$}}}\,.
\ee
For the branch of the solution in which $\tilde\rho$ decreases monotonically from its maximal value $1$ to its minimal value $\sqrt{1 - n_{\text{\tiny $L$}}}$, symmetry allows us to take $\vartheta$ to run from $-\theta/2$ to $0$. Integrating equation~\eqref{rhoeqnew} with the boundary condition $\vartheta|_{\tilde\rho=1} = -\theta/2$, we find
\be
\label{cirotto}
\vartheta(\tilde\rho)=-\frac{\theta}{2}+\frac{1}{2 }\int^1_{\tilde\rho^2} \frac{dw\,\sqrt{(1 - n_{\text{\tiny $L$}}) (n_{\text{\tiny $L$}} - m_{\text{\tiny $L$}})}}{w\sqrt{\left(1-w\right)(w - 1 +n_{\text{\tiny $L$}}) (w \,m_{\text{\tiny $L$}} + n_{\text{\tiny $L$}}- 
  m_{\text{\tiny $L$}})}}.
\ee
 Performing the integral we obtain
\be
\vartheta(\tilde\rho)=\textstyle
-\frac{\theta}{2}+\frac{\sqrt{1-n_{\text{\tiny $L$}}} \sqrt{n_{\text{\tiny $L$}}-m_{\text{\tiny $L$}}}}{\sqrt{n_{\text{\tiny $L$}}}}
\Pi \left(n_{\text{\tiny $L$}};\chi|m_{\text{\tiny $L$}}\right)\,,\qquad\qquad   \chi=\arcsin\left(\sqrt{\frac{1 - \tilde\rho^2}{n_{\text{\tiny $L$}}}}\right)
   \ee
   Setting $\tilde\rho=\tilde\rho_\text{min}\equiv\sqrt{1-n_{\text{\tiny $L$}}}$, the requirement that $\vartheta |_{\tilde\rho=\tilde\rho_\text{min}}=0$ determines the value of $\theta$ in terms of the parameters $n_{\text{\tiny $L$}}$ and $m_{\text{\tiny $L$}}$
   \be
   \label{theta}
\theta=
2\textstyle\sqrt{\frac{(1-n_{\text{\tiny $L$}})(n_{\text{\tiny $L$}}-m_{\text{\tiny $L$}})}{n_{\text{\tiny $L$}}}}\,\Pi \left(n_{\text{\tiny $L$}}\,|m_{\text{\tiny $L$}}\right).
   \ee
We then evaluate the conserved angular conserved momentum $L$, corresponding to rotations along the $\nu$ direction in $S^5$, see~\eqref{Lconserved2}, as 
\be
\label{L00}
L= - \int_{-s/2}^{s/2} d\sigma \, \Pi_\nu^0 =2 g\, \gamma \int_{-s/2}^{s/2} d\sigma (1 - \tilde\rho^2(\sigma))=4 g\,
  \int_{\tilde\rho_\text{min}}^1 d\tilde\rho \frac{\tilde\rho \sqrt{1-\tilde\rho^2} \sqrt{m_{\text{\tiny $L$}}}}{\sqrt{n_{\text{\tiny $L$}}+\tilde\rho^2-1}
   \sqrt{\tilde\rho^2m_{\text{\tiny $L$}}+n_{\text{\tiny $L$}}-m_{\text{\tiny $L$}}}}\,,
\ee
where we used~\eqref{rhotildeprime}. Performing the integral leads to 
      \be\label{L}
 \mathcal{L}\equiv \frac{L}{4 g}= \, \sqrt{\frac{n_{\text{\tiny $L$}}}{m_{\text{\tiny $L$}}}} \big(\,K (m_{\text{\tiny $L$}})-E(m_{\text{\tiny $L$}})\,\big)\,.
\ee

When \( \L \) vanishes the results in~\cite{DF} must be recovered.  From Eq.~\eqref{L}, this can be achieved taking \( \nL \to 0 \), which in terms of the original parameters corresponds to \( \ell_\vartheta^2 \to 1 \). 
When \( \ell_\vartheta^2 \) approaches 1,   \( \mL \) also vanishes, causing the ratio \( \nL / \mL \) to remain finite  
\be\label{xi}
\left.\frac{\nL}{\mL}\right|_{\L\to 0} =\frac{1}{1-\frac{\theta^2}{\pi^2}}. 
\ee
Notice that \( \L \) still goes to zero because the expression in brackets in Eq.~\eqref{L} vanishes as \( \mL \to 0 \).  
Alternatively, one might attempt to reach this limit by directly taking \( \mL \to 0 \), since $K (m_{\text{\tiny $L$}})-E(m_{\text{\tiny $L$}})$ vanishes faster than \( \sqrt{\mL} \). The limit \( \mL \to 0 \) is however again equivalent to \( \ell_\vartheta^2 \to 1 \) and consequently \( \nL \to 0 \). Thus, we return to the previous case.

\subsection{An effective cusp opening in $S^5$}
\label{sec:angle-s5}

The range  $s$ of the spatial worldsheet coordinate  $\sigma$ can also be evaluated using~\eqref{rhotildeprime}, obtaining
\be
\label{sS5}
\!\!\!\!\!\!
s_{\text{\tiny $S^5$}}=
2\int_{\tilde\rho_\text{min}}^1
\frac{d\tilde\rho \,\tilde\rho \, \sqrt{m_{\text{\tiny $L$}} n_{\text{\tiny $L$}}-m_{\text{\tiny $L$}}+n_{\text{\tiny $L$}}}}{\sqrt{1-\tilde\rho ^2} \sqrt{n_{\text{\tiny $L$}}+\tilde\rho ^2-1} \sqrt{\tilde\rho^2
   m_{\text{\tiny $L$}}-m_{\text{\tiny $L$}}+n_{\text{\tiny $L$}}}}   =2 \textstyle \sqrt{1 + m_{\text{\tiny $L$}} (1 - \frac{1}{n_{\text{\tiny $L$}}})}\,K(m_{\text{\tiny $L$}})\,.
\ee
Clearly, the two different ways of computing $s$, equations ~\eqref{sAdS5} and~\eqref{sS5},  must produce the same result. This gives the constraint 
\be\label{s=s}
s_{\text{\tiny $\text{AdS}_5$}}=s_{\text{\tiny $S^5$}}\qquad\text{or}\qquad 2 \, \sqrt{1 - n - m} \,K(m)=2 \textstyle \sqrt{1 + m_{\text{\tiny $L$}} (1 - \frac{1}{n_{\text{\tiny $L$}}})}\,K(m_{\text{\tiny $L$}})\,.
\ee
This equation  obviously  solves the apparent mismatch between the number of  parameters in the classical string solution and in the dual gauge theory operator. There are four parameters in the string description, two free constants from the ansatz for the minimal surface ($\kappa$ and~$\gamma$) and two integration constants from conserved quantities ($\ell_\vartheta$ and $\ell_\varphi$), which we traded for convenience with $(n,m)$ and $(\nL,\mL)$ in~\eqref{nm} and~\eqref{nLmL}.  The constraint~\eqref{s=s} makes only three of them independent and related to the  three physical quantities defining the dual gauge theory operator,   the cusp angle $\phi$ in $\text{AdS}_5$,  the angular width $\theta$ in $S^5$, and  the number~$L$ of~$Z$ insertions at the cusp.

This consistency condition is the key distinction with respect to the setup in~\cite{DF} and defines its generalization. Indeed,  it is easy to see that in the limit  $L\to0$ the expression for~$s_{\text{\tiny $\text{AdS}_5$}}$ stays unvaried, while the expressions for $s_{\text{\tiny $S^5$}}$ in~\eqref{sS5} and  for $\theta$ in~\eqref{theta} coincide~\footnote{Namely, in the limit \( \nL \to 0 \), \( \mL \to 0 \), with \( \nL / \mL =(\gamma^2 + 1)/\gamma^2 \),  $s_{\text{\tiny $S^5$}}$ in~\eqref{sS5} and  $\theta$ in~\eqref{theta}  are both equal to $\textstyle\pi/\sqrt{1 + \gamma^2}$.}. 
For $L=0$ equation~\eqref{s=s}  reduces to
 \be\label{L0}
 \theta=2 \, \sqrt{1 - n - m} \,K(m)\,,
\ee
thus reproducing the result of~\cite{DF}~\footnote{See formula B.14 there, together with~\eqref{nmtobp} and~\eqref{pq}.}. It is then natural, in the case of general $L$ (or general~$\mathcal{L}$) of interest here, to interpret the right-hand-side of~\eqref{s=s} as an~\emph{effective} $R$-symmetry  opening in $S^5$, encoding the effect of $\theta$ and $\L$ on the $\text{AdS}_3$ sector of the minimal surface. We then define
\be\label{theffdef}
\Theff(\theta,\L)\equiv 2 \textstyle \sqrt{1 + m_{\text{\tiny $L$}} (1 - \frac{1}{n_{\text{\tiny $L$}}})}\,K(m_{\text{\tiny $L$}})\,,
\ee
and, below,  use the consistency condition~\eqref{s=s} written in the form
\be\label{thetaeff}
2 \, \sqrt{1 - n - m} \,K(m)=\Theff(\theta,\L)\,\,.
\ee
While $\theta$ is bounded to take values in the interval~$[0,\pi]$,~$\Theff(\theta,\L)$ can span the entire positive line.  
For a fixed value of \(\mathcal{L}\), the function~$\Theff(\theta,\L)$ is monotonic with respect to~\(\theta\)~\footnote{This behavior has been verified numerically for various values of \(\mathcal{L}\).}. 
It has a minimum value when \(\theta = 0\),  
which implies \(\nL= \mL\), so that $\Theff^\text{\tiny min}(\L)=2\,\sqrt{\mL} K(\mL) $. 
Its maximum value is realized when \(\theta = \pi\),  which corresponds to 
\(\nL= 1\), so that $\Theff^\text{\tiny max}(\L)= 2 K(\mL)$.
%
\begin{figure}[h]  
    \centering
    \includegraphics[width=0.8\textwidth]{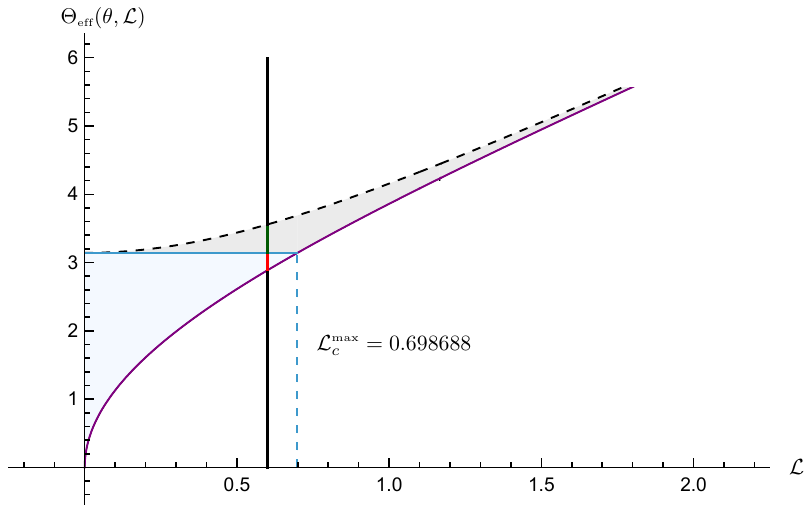} %
\caption{\small
The light-blue and light-gray shaded regions are the images, in the $(\mathcal{L},\Theff)$ plane, of Regions $A_L$ and $B_L$ of Fig.~\ref{fig:Transitioncurve}. 
For a fixed value of~$\mathcal{L}$, the effective angle $\Theff(\theta,\mathcal{L})$ spans the interval between the lower envelope
$\Theff^{\text{\tiny min}}(\mathcal{L}) = 2\sqrt{\mL}\,K(\mL)$ (purple)
and the upper envelope
$\Theff^{\text{\tiny max}}(\mathcal{L}) = 2\,K(\mL)$ (black dashed).
The curve $\Theff^{\text{\tiny min}}(\mathcal{L})$ is the image of the $\theta$-axis in Fig.~\ref{fig:Transitioncurve}, whereas
$\Theff^{\text{\tiny max}}(\mathcal{L})$ is the image of the dashed boundary of the light-gray region $(\theta=\pi)$ in Fig.~\ref{fig:Transitioncurve}.
The blue line segment separating the two shaded regions is the image of the transition curve in Fig.~\ref{fig:Transitioncurve}.
The plot clearly highlights the two regimes as $\mathcal{L}$ is varied. For $\mathcal{L}>\mathcal{L}_c^{\text{\tiny max}}$, one has $\Theff(\theta,\mathcal{L})>\pi$ for all $\theta$.
For $\mathcal{L}<\mathcal{L}_c^{\text{\tiny max}}$, varying $\theta$ at fixed~$\mathcal{L}$ yields two distinct branches: the red segment corresponds to Region $A_L$ in the main text, while the green segment corresponds to Region $B_L$ (for the chosen~$\mathcal{L}$).
}
    \label{rangethetaeff}
\end{figure}
In Fig.~\ref{rangethetaeff}, the shaded region between the curves represents the range of allowed values for \(\Theff(\theta, \mathcal{L})\) as \(\mathcal{L}\) varies. One can observe the existence of the critical value $\mathcal{L}^\text{\tiny{max}}_c = 0.6986882(5)$,
above which \(\Theff(\theta, \mathcal{L}) > \pi\) for all values of \(\theta\). 

Below the critical value $\mathcal{L}_c^\text{\tiny{max}}$,  for each value of $\mathcal{L}$ two distinct regions can be identified:
\vskip 0.5 cm
\paragraph{\noindent {Region A$_L$}:
  $\Theff(\theta, \mathcal{L}) \leqslant \pi$.} 
  This region exists only for $\L<\L_c^\text{\tiny{max}}$, and it corresponds to the red vertical segment in Fig.~\ref{rangethetaeff}.  

\vspace{-3mm}
\paragraph{\noindent {Region B$_L$}:
   $\pi < \Theff(\theta, \mathcal{L})$.} 
  This region is present for $\L > 0$. If $\L<\L_c^{\text{\tiny{max}}}$, it corresponds to the green segment in Fig.~\ref{rangethetaeff}. For 
  $\L>\L_c^{\text{\tiny{max}}}$ any point betwen the blue and brown curve belongs to this region.

 \vskip 0.5 cm
 Given $0 \leqslant \L \leqslant \L_c^{\text{\tiny max}}$, the transition between the two regions occurs when one reaches the value of $\theta$  for which  $\Theff(\theta, \mathcal{L})=\pi$, 
an equation which implicitly defines the curve in Fig.~\ref{fig:Transitioncurve}.  
A plot analogous to the one in Fig.~\ref{fig:Transitioncurve} is in Appendix B of~\cite{Hong:2004gz}~\footnote{
\label{footnote:strassler}
In~\cite{Hong:2004gz} the authors study the potential for a ``quarkonium'' state with the insertion of $J$ particles in the adjoint representation. In our notation, their problem is mapped to finding the potential in the $\theta=0$  case. They also notice that the potential vanishes when they approach a critical value of $\L$ which coincides with our value at $\theta=0$, and interpret this fact with the breaking of the fluxtube binding the quarkonium. 
In Appendix B they very briefly described that it is possible to generalize this analysis to the case $\theta\neq0$, recovering as a main result the same transition curve. However,  they do not realize that the solution continues to exist also above the critical line.}. 
The Region B$_L$ allows us to investigate the large $\L$ limit of our solution, used in~\cite{Correa:2012at} to compute L\"uscher corrections to the BMN vacuum from the string point of view.


\section{Antiparallel and near BPS limit} 
\label{sec:potentia-and-BPS}

We are now ready to discuss  two interesting  behaviors, the limits of antiparallel lines ($\phi\to\pi$, or ``potential'' limit) and the near BPS  ($\phi^2\to\theta^2$), of the  energy for the $R$-charged flux tube  in the leading strong coupling regime. Below, we first discuss the distinct parameter limits to which they correspond at the level of the classical string solution, and then evaluate the energy.


\subsection{Antiparallel lines limit}
\label{subsec-antiparallel}

The limit of antiparallel lines can be studied looking at the condition $\phi\to\pi$ in~\eqref{phi01},
\begin{equation}\label{zerocond}
  \frac{\sqrt{1-n}\,\sqrt{\,n-m}}{\sqrt{n}}
    \big(\,\Pi(n\,|\,m)-K(m)\,\big)\to 0\,.
\end{equation}
This can be achieved in two different ways: 
\paragraph{\noindent{\textbf{1. \(n\to m\):}}}      When the two moduli coincide,
      \(
      \Pi(n|n)=E(n)/(1-n)
      \)
      and the factor \(\sqrt{\,n-m}\) in~\eqref{zerocond} forces the product to vanish.  Because of~\eqref{rangenm}, in this case the range of $m$ is restricted to the interval~$0\leqslant m\leqslant \frac{1}{2}$. 

\paragraph{\noindent{\textbf{2. \(n\to0\):}}} The expansion
    \be\label{nto0}
      \Pi(n\,|\,m)=K(m)            +\frac{K(m)  -E(m)}{m}\,n
            +\mathcal O(n^{2}),
               \ee
      shows that in this limit the bracket in~\eqref{zerocond} decays faster than \(\sqrt{n}\). Recalling the first of the inequalities~\eqref{par-relations}, the value \(\phi=\pi\) is obtained when \(n\to0\) for any~\(m< 0\).\\  
      
Notice that $n\to1$ does not satisfies the condition~\eqref{zerocond}. This is because, though the prefactor \(\sqrt{1-n}\) vanishes,  \(\Pi(n|k^{2})\) diverges like \(1/\sqrt{1-n}\).  In fact,  the limit $n\to1$ in~\eqref{phi01} implies $\phi\to0$, thus defining the limit of straight line.
 \newcommand{\E}{\mathcal{E}}
\bigskip
\begin{figure}[h]  
    \centering
    \includegraphics[width=1\textwidth]{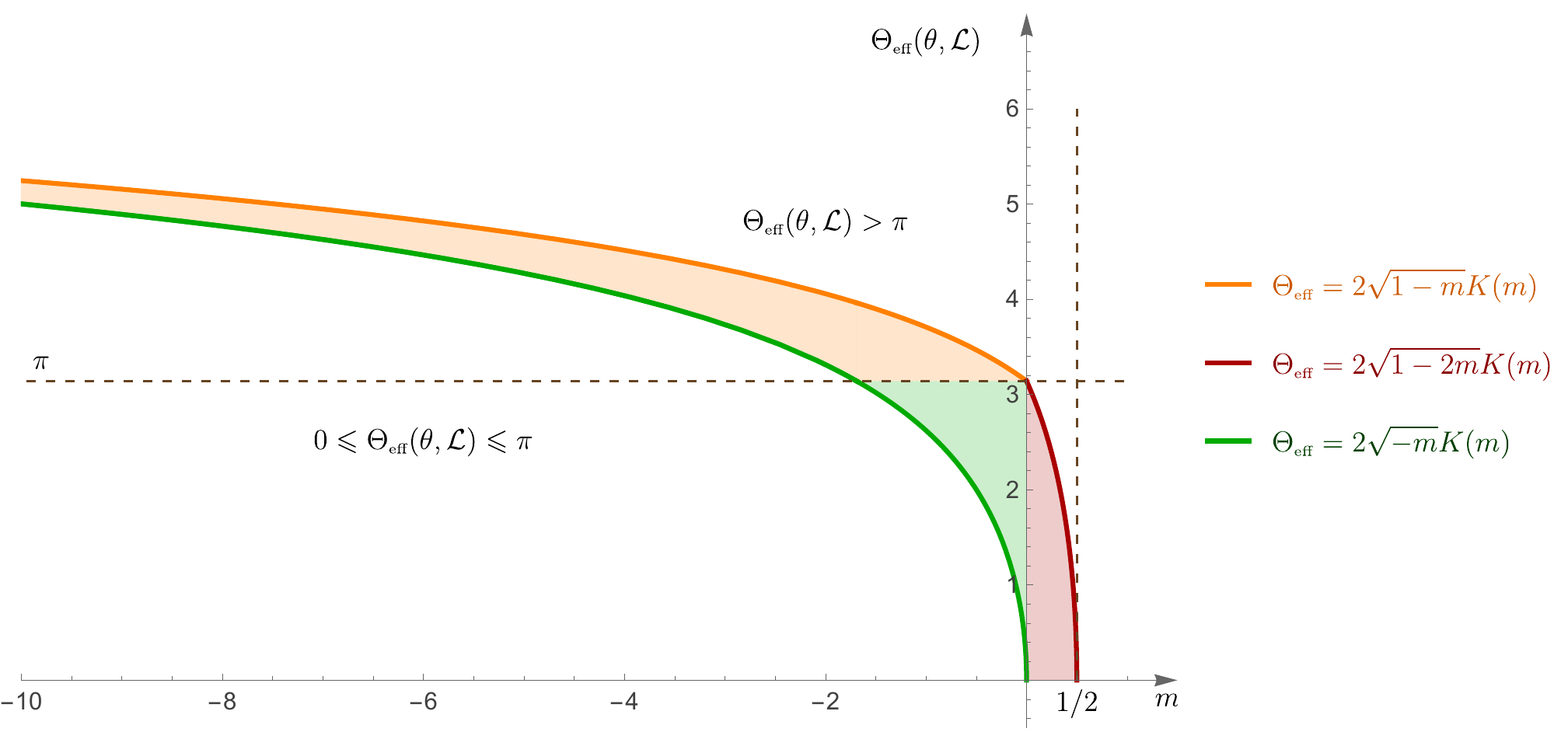} %
     \caption{\small Range of the possible values of $m$ in the plane $(m,\Theta_{\mathrm{eff}})$, as defined by the consistency condition~\eqref{thetaeff}.}.
  \label{fig:transition-mvsth}
\end{figure}

\bigskip

To understand the behavior of the  energy $\mathcal{E}$~\eqref{E} in the antiparallel line limit, it is instructive to look at the accessible values of~ 
$m$ when $\Theta_{\mathrm{eff}}$ is varied, see Fig.~\ref{fig:transition-mvsth}.  The combined red and green regions form Region A$_L$, while the orange region corresponds to Region B$_L$. The red portion specifically identifies the part of Region A$_L$ that is relevant in the antiparallel line limit.
The lower green curve represents the consistency condition~\eqref{thetaeff} in the limit $\phi\to0$ (i.e.~$n\to1$).
The red curve reproduces~\eqref{thetaeff} at $\phi=\pi$ in the Region  A$_L$ (i.e. $n\to m$), whereas the orange curve does the same in the Region  B$_L$ (i.e. $n\to 0$).  The red and orange curves meet at the transition point $\Theta_{\mathrm{eff}}=\pi$, but with different slopes. The fact that the condition $\phi=\pi$ is realized by two distinct curves in Region A$_L$ and Region B$_L$ anticipates a qualitatively different energy profile in the two regions.

\bigskip
Let us now analyze the energy  in the two different regimes  producing the antiparallel line configuration.   Despite $\E$ being formally identical to the $\L=0$ case considered in~\cite{DF}, its evaluation gets modified by the new conserved charge $\L$~\eqref{L}  because of the consistency condition~\eqref{thetaeff}. 
The presence of $\L$ opens the possibility  to explore the Region~{\bf$B_L$}, which otherwise collapses to a point when $\L$ is set to $0$.

We start by  considering the consistency condition~\eqref{thetaeff} in the case {\textbf 1}.
When $0\leq n\leq1$ and $0\leq m\leq\frac{1}{2}$, the left-hand-side of~\eqref{thetaeff}  takes  values between $0$ and $\pi$. 
Therefore~$\theta$ and~$\L$  must belong to a subregion of the Region {\bf A$_L$} defined in the previous section. 

The behavior of  $\mathcal{E}$ as a function of $\phi$, $\theta$ and $\L$ is obtained following the  steps in~\eqref{piminusphi}-\eqref{maroundmbar}. First we obtain the energy as an expansion in odd powers of $\pi - \phi$ 
\begin{eqnarray}
\label{Einterm}\nn
\mathcal{E}^{(A)}\!&= &
-\frac{2 \left(\left(\mbar-1\right)
\overline{  \mathds{K}}+\overline{  \mathds{E}}\right){}^2}{\sqrt{\mbar \left(1-\mbar\right)}
   \left(\pi -\phi\right)}+
   \frac{\left(\pi -\phi\right) \left(\left(\mbar \left(4-3
   \mbar\right)-1\right)\overline{\mathds{K}}+\left(1-2 \mbar\right)
   \overline{  \mathds{E}}\right)}{12 \sqrt{\mbar \left(1-\mbar\right)}
   \left(\left(\mbar-1\right)
\overline{  \mathds{K}}+\overline{  \mathds{E}}\right)}+O \big(\pi \!-\!\phi)^3 \\
&\equiv &\frac{\mathcal{E}^{(A)}_\text{\tiny Cas}}{\pi-\phi}+b^{(A)}\,(\pi-\phi)+\!O \big(\pi -\phi)^3\,,
\end{eqnarray}
where $\overline{\mathds{E}}=E(\mbar),~\overline{\mathds{K}}=K(\mbar)$, and $\mbar$ is a solution of the consistency condition~\eqref{thetaeff} in the same antiparallel ($n\to m$) limit
\be\label{consistencybarA}
2 \, \sqrt{1 - 2 {\mbar}} \, K(\mbar)=\Theff(\theta,\L)\,.
\ee
The expression~\eqref{Einterm} is formally identical to the results of~\cite{Cuomo:2024psk}, in particular  one observes the familiar pole singularity -- controlled by the ``Casimir energy"~$\mathcal{E}^{(A)}_\text{\tiny Cas}$ --  and the absence of a constant term.
The difference lies in the modular parameters: While here \( \mbar \) solves~\eqref{consistencybarA}, the modulus in the corresponding formula of~\cite{Cuomo:2024psk} - which we denote $\mbar_0$ here - satisfies the $\L\to0$ limit of \eqref{consistencybarA}
\be\label{consistency0}
2 \, \sqrt{1 - 2 \mbar_0} \,K(\mbar_0)=\theta\,.
\ee
The explicit corrections in $\mathcal{L}$ to the results of~\cite{Cuomo:2024psk} are then obtained expanding the consistency condition around small~$\L$, see~\eqref{nLmLexp}-\eqref{mbararoundm0} for details.  The result for the coefficients in~\eqref{Einterm} of both the pole  and  the linear term reads then, in this region, 
\begin{eqnarray}\nonumber
&&
\mathcal{E}^{(A)}_\text{\tiny Cas}=-\frac{\,\left(\left(\mbar_0-1\right)
    \mathds{K}_0+ \mathds{E}_0\right)^2}{\sqrt{\mbar_0
   \left(1-\mbar_0\right)}}\Biggl[{2 }{}-\frac{2 \sqrt{1-2 \mbar_0} \sqrt{\pi ^2-\theta^2}\mathcal{L}}{
  \theta\left(\left(\mbar_0-1\right)
    \mathds{K}_0+ \mathds{E}_0\right)}\!+\!O(\mathcal{L}^2)\Biggr]\,,
\\ \label{cL}
&&
b^{(A)}=\!\frac{1}{\sqrt{\mbar_0
   \left(1-\mbar_0\right)}}\Biggl[\frac{\left(\mbar_0 \left(4-3 \mbar_0\right)-1\right)  \mathds{K}_0+\left(1-2
   \mbar_0\right)  \mathds{E}_0}{12 
   \left(\left(\mbar_0-1\right)
    \mathds{K}_0+ \mathds{E}_0\right)}\\
  &&+\frac{\sqrt{1-2 \mbar_0} 
   \left(\left(\left(\mbar_0-1\right) \mathds{K}_0+\mathds{E}_0\right){}^2+\mbar_0(1-\mbar_0)\mathds{E}_0(\mathds{K}_0-\mathds{E}_0)\right)\sqrt{\pi ^2-\theta^2}\mathcal{L}}{12  
   \left(\left(\mbar_0-1\right)  \mathds{K}_0+ \mathds{E}_0\right){}^2
   \left((1-\mbar_0) \mathds{K}_0+\left(2
   \mbar_0-1\right)  \mathds{E}_0\right)\,\theta}+O(\mathcal{L}^2)\Biggr]\,,\notag
   \end{eqnarray} 
where $\mathds{E}_0\equiv  E(\mbar_0)$ and $\mathds{K}_0\equiv K(\mbar_0)$. 

\bigskip

In  the case {\textbf 2.} ($n\to 0$ and negative $m$) the left-hand-side of the consistency condition~\eqref{thetaeff} is always greater than $\pi$, so that~\eqref{thetaeff} can only be solved in the Region {\bf B$_L$}. 
To obtain the energy, one can  similarly proceed with the steps summarized in~\eqref{nexp}-\eqref{maroundmhat}. The novel, anticipated feature is the~\emph{absence} of the Coulomb-like potential term, and the appearance of a constant one
\begin{subequations}
\begin{eqnarray}\label{EnergyB}
\mathcal{E}^{(B)}
&=&\frac{(1-\widehat m)
   \widehat{\mathds{K}}- \widehat{\mathds{E}}}{\sqrt{-\widehat m}}+\frac{1
  }{8 \sqrt{-\widehat m}
   ( \widehat{\mathds{E}}- \widehat{\mathds{K}})}(\pi - \phi)^2+O\left(\pi - \phi\right)^4\,\\\label{DeltaB}
  & \equiv& \Delta^{(B)}+\beta\,(\pi - \phi)^2+O\left(\pi - \phi\right)^4\,.
\end{eqnarray}
\end{subequations}
Above, $ \widehat{\mathds{K}}\equiv K(\widehat{m})$, $\widehat{\mathds{E}}=E(\widehat{m})$ and $\widehat m$ is a solution of the consistency condition~\eqref{thetaeff} in this limit,
\be\label{consistencyhatB}
2 \, \sqrt{1 -  {\widehat m}} \,  \widehat{\mathds{K}}=\Theff(\theta,\L)\,.
\ee
The expansion in Region B$_L$ contains  only even powers in $\pi-\phi$, in contrast with what occurs in Region A$_L$. 
Also, while in Region A$_L$ the energy~\eqref{E} is negative, leading to a negative dominant contribution in the parallel line limit~\eqref{Einterm}, in Region B$_L$ the energy is always positive.

\begin{figure}[h] 
    \centering
    \includegraphics[width=0.8\textwidth]{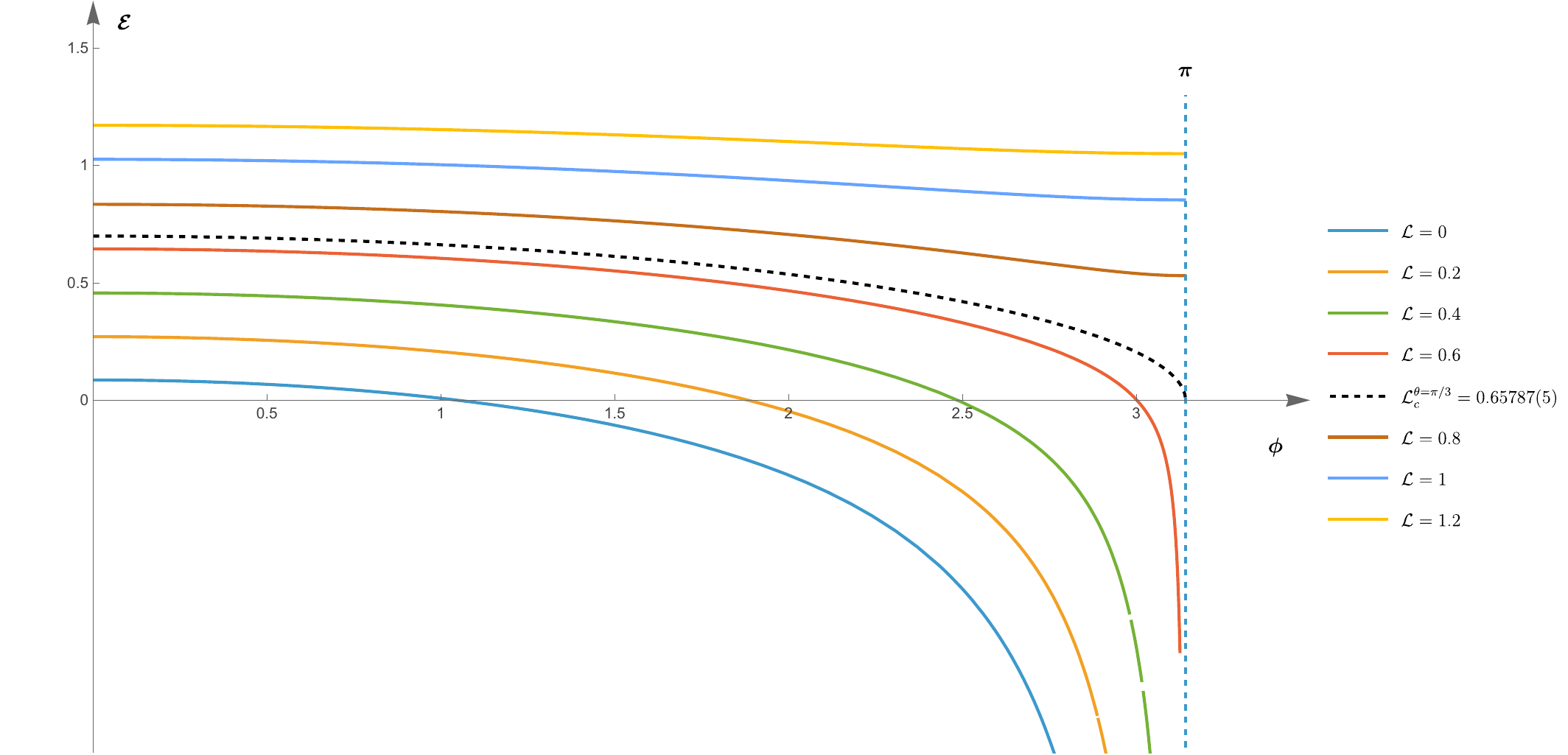} %
     \caption{\small Energy $\E$ in~\eqref{E} as a function of the geometric angle $\phi$ at fixed internal angle $\theta = \pi/3$, shown for several values of the $R$-charge $\L$. The first four curves correspond to $\L$ values below the transition, where the system exhibits the familiar Coulomb-like divergence: as $\phi \to \pi$ (antiparallel lines limit), the energy grows without bound. In contrast, the curves corresponding to larger $\L$ lie above the transition and remain finite as $\phi \to \pi$. Their limiting value increases with $\L$, asymptotically approaching $\E \simeq \L$ for large $R$-charge, consistent with the expectation~\eqref{ElargeL}. The dashed line marks the critical value $\mathcal{L}=\mathcal{L}_c^{\theta=\pi/3}$, for which the zero of the energy lies exactly at $\phi=\pi$, leaving no interval where a divergence can arise.
}
 \label{fig:enversphi}
\end{figure}

%
%
Fig.~\ref{fig:enversphi} illustrates the qualitative change of the energy in the antiparallel line behavior across the critical $R$-charge~$\L$.  When $\L$ approaches $\L^\theta_c$ from below, the point at which the energy vanishes is pushed arbitrarily close to $\phi=\pi$,  and  the fall-off of the curve becomes increasingly steep.
At $\L=\L_c^\theta$, the zero of the energy sits exactly at $\phi=\pi$, leaving no interval in which a divergence can develop.
For larger values of $\L$, the curves lie above the transition and remain finite in the limit $\phi \to \pi$. The limiting value increases with $\L$ and, at large $R$-charge, asymptotically tends to $\mathcal{E} \simeq \L$, consistently with the behavior predicted in~\eqref{ElargeL}. 

The vanishing of the Coulomb-like term in the expansion of the energy can be interpreted, in the language of~\cite{Klebanov:2006jj}, as the breaking of the flux tube binding the generalized quarkonium state due to an excessive rotation of the string in $S^5$. The exact value at which the breaking occurs for $\theta=0$ was firstly found in~\cite{Hong:2004gz}. 

\begin{figure}[t]
\centering
\begin{minipage}{0.48\textwidth}
\centering
\includegraphics[width=\linewidth]{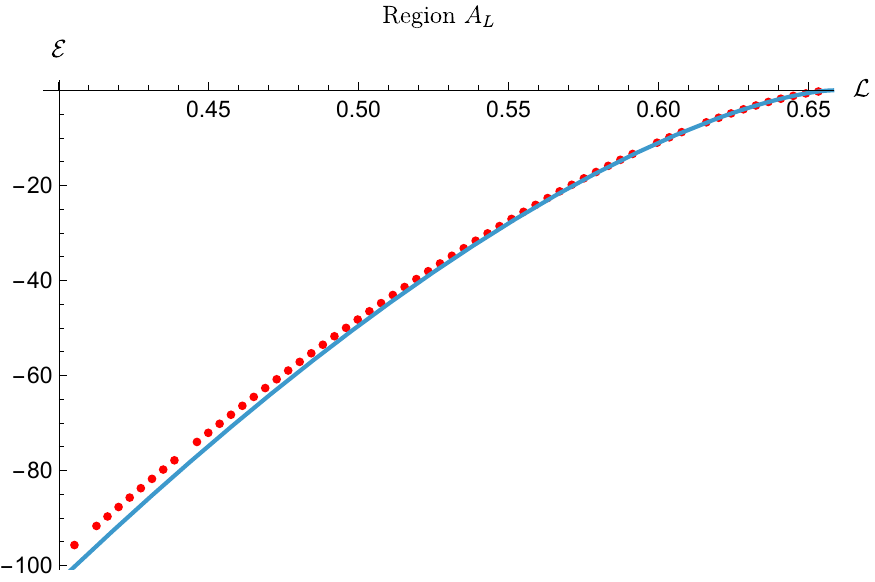}
\end{minipage}
\hfill
\begin{minipage}{0.48\textwidth}
\centering
\vspace{0.3cm}
\includegraphics[width=\linewidth]{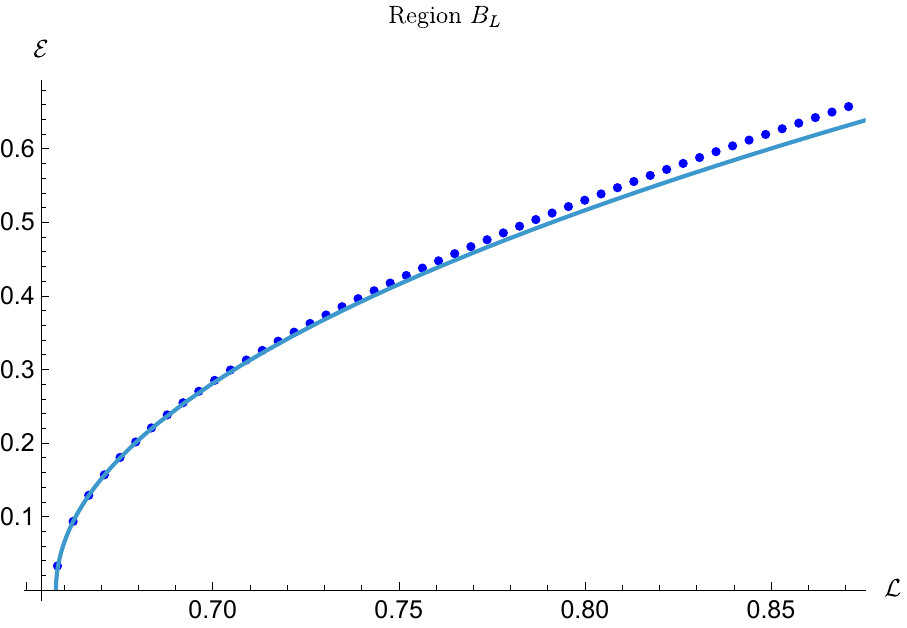}
\end{minipage}
\caption{\small $\E$ versus $\L$ for  $\theta=\pi/3$ and $\phi$ very close to $\pi$, i.e. $\phi=3.14$, both in the Region A$_L$ and in the Region B$_L$. The numerics, in red and blue dots,  start following the analytical blue curves, equations \eqref{EAcritical} and~\eqref{EBcritical}, as $\L_c^\theta$ is approached.}
\label{fig:EvsL_A-B}
\end{figure}

\bigskip

Clearly,  a small-$\L$ expansion of~\eqref{EnergyB} for general $\theta$ is not possible. When $\L=0$, $\Theff$ coincides with~$\theta$, whose range never exceeds the value of $\pi$ and thus does not have access to Region B$_L$. This is clear when looking at the curve in Fig.~\ref{fig:Transitioncurve}. There, the Region B$_L$ cannot explore arbitrary small values of~$\L$, since $L=0$ is possible only when $\theta$ is exactly equal to $\pi$. In other worlds, $\L_c^{\theta=\pi}=\L=0$, and the Region B$_L$ actually shrinks to a point $(\L=0, \theta=\pi)$ discussed in \cite{DF}.
In this region,  meaningful expansions  are the ones around the minimal and maximal values of $\mathcal{L}$ here, respectively $\mathcal{L}=\mathcal L_c^\theta$ and $\L=\infty$.

\bigskip
The transition from Region A$_L$ to  Region B$_L$  occurs when the moduli have critical values $\nL^c$ and $\mL^c$  which solve, see~\eqref{theffdef} and~\eqref{theta},   
   \be\label{mncritical}
    2 K\left(\mL^c\right) \sqrt{\mL^c\left(1-\frac{1}{\nL^c}\right) + 1} =\pi\,,\qquad
  2 \sqrt{\frac{\left(1-\nL^c\right) \left(\nL^c-\mL^c\right)}{\nL^c}}\Pi \left(\left. \nL^c\right|\mL^c\right)=\theta\,,
   \ee
and which then fix a critical value $\mathcal{L}_c^\theta$ for the angular momentum via
\be
\label{Lcritical}
   \mathcal{L}_c^\theta=\frac{\sqrt{\nL^c}}{\sqrt{\mL^c}} [K(\mL^c)-E(\mL^c)].
   \ee
As a function of~$\theta$, the transition is captured by the curve displayed in Figure~\ref{fig:Transitioncurve}.
The behavior of the potential at fixed $\theta$ and just below the transition can be obtained following the steps summarized in~\eqref{aroundcr}-\eqref{mcriticalA}.

We can study the behavior of the energy $\mathcal E$ as a function of $\L$ close to the transition point at a given value of the internal angle $\theta$. 
Next to $\mathcal{L}_c$, one finds that the coefficient  $\mathcal{E}^{(A)}_\text{Cas}(\theta,\mathcal{L})$ of the pole in $\pi-\phi$ scales as 
 \be\label{EAcritical}
\mathcal{E}^{(A)}_\text{\tiny Cas}=
 -\sqrt{\frac{8 \pi }{27}}\frac{ {\mL^c}^{3/4}}{ \left(-\mL^c+\mL^c \nL^c+\nL^c\right){}^{3/4}} \left(\mathcal{L}_c-\mathcal{L}\right)^{3/2} +O\left(\left(\mathcal{L}_c-\mathcal{L}\right){}^{5/2}\right)\,.
 \ee
 \bigskip
Similarly, one can obtain the behavior close to $\mathcal{L}_c$  in the region B$_L$ 
\be\label{EBcritical}
\mathcal{E}^{\,B)}=
\frac{\sqrt{\frac{\pi }{2}} \sqrt[4]{\mL^c} }{\sqrt[4]{-\mL^c+\mL^c \nL^c+\nL^c}}\left(\mathcal{L}-\mathcal{L}_c\right)^{1/2}+O\left((\mathcal{L}-\mathcal{L}_c)^{3/2}\right)\,.
\ee
A numerical analysis confirms this leading behavior close enough to the critical region, see~Fig.~\ref{fig:EvsL_A-B}.

 \begin{figure}[h]  
    \centering
    \includegraphics[width=0.7\textwidth]{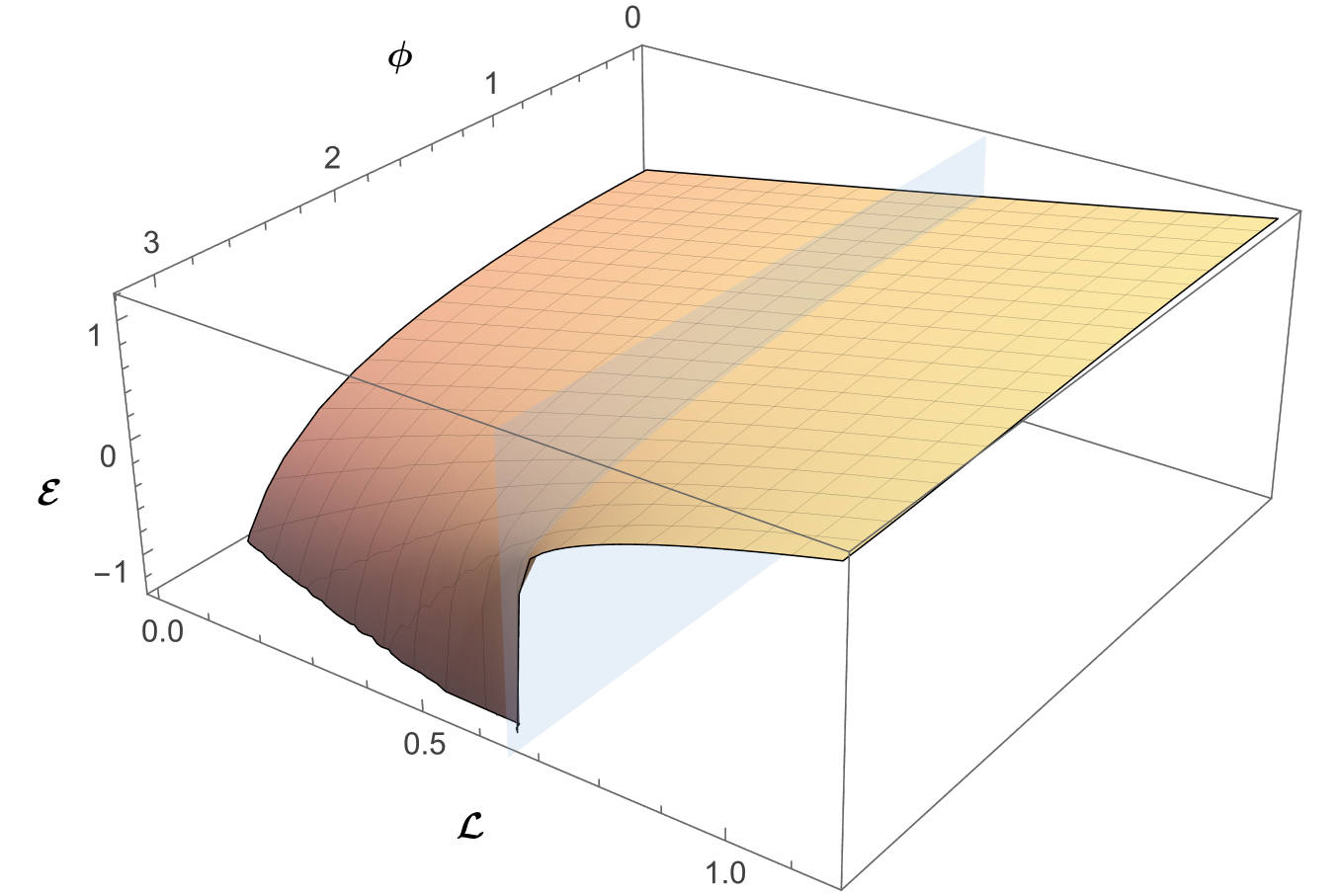} %
     \caption{\small Plot of the energy $\E$ versus $\phi$ and $\L$ at fixed $\theta=\pi/3$.  This three-dimensional figure  manifestly shows the change of behavior across the vertical light-blue plane  indicating the location of $\L_c^\theta$. On the left the energy is unbounded and becomes arbitrarily negative when $\phi$ approaches $\pi$, on the right it is smooth and positive for all values of $\phi$. The transition in the behavior of the classical solution arises when $\phi$ approaches $\pi$, i.e. when the configuration probes the boundary of the parameter space shown in  the red and orange curves of Fig.~\ref{fig:transition-mvsth}. As the figure illustrates, for values of $\phi$ away from $\pi$ the energy displays a smooth dependence on the parameters.}   
    \label{fig:3d}
\end{figure}


\subsection{Near BPS limit}
\label{sec:nearBPS}

In the case $\phi^2=\theta^2$ the Wilson loop is BPS~\cite{Zarembo:2002an}. 
The goal of this subsection is to expand the energy around this BPS configuration. Specifically, we fix $\theta$ and $\L$ to finite values and expand~$\E$ for~$\phi\sim\theta$. Given $\nL$ and $\mL$ producing those values of $\theta$ and $\L$,  the first step is to determine $n$ and $m$ such that $\phi=\theta$.  Starting from~$\theta$ in~\eqref{theta} and using the identities~\eqref{Pimodul} and~\eqref{Pitransf} one obtains the equivalent expression
  \be
  \label{thetamore}
  \begin{split}
\theta
=&\pi - 
2 \sqrt{\frac{\left(n_{\mbox{\tiny $L$}} -m_{\mbox{\tiny $L$}}\right)}{n_{\mbox{\tiny $L$}} \left(1-n_{\mbox{\tiny $L$}} \right) \left(1-m_{\mbox{\tiny $L$}} \right) }}
\left[\Pi \left(\left. \frac{m_{\mbox{\tiny $L$}}(n_{\mbox{\tiny $L$}}-1)}{n_{\mbox{\tiny $L$}}(m_{\mbox{\tiny $L$}}-1)}
\right| \frac{m_{\mbox{\tiny $L$}}}{m_{\mbox{\tiny $L$}}-1}\right)-K\left(\frac{m_{\mbox{\tiny $L$}}}{m_{\mbox{\tiny $L$}}-1}\right)\right]\,.
  \end{split} 
  \ee
In terms of the new parameters
  \be
    \label{newparameters}
  \widetilde{m}\equiv\frac{m_{\mbox{\tiny $L$}}}{m_{\mbox{\tiny $L$}}-1}\,,\quad \ \    \widetilde{n}\equiv \frac{m_{\mbox{\tiny $L$}}(n_{\mbox{\tiny $L$}}-1)}{n_{\mbox{\tiny $L$}}(m_{\mbox{\tiny $L$}}-1)}\,,\quad \text{with inverse}\quad m_{\mbox{\tiny $L$}}=\frac{ \widetilde{m}}{ \widetilde{m}-1}\,,\ \quad \ n_{\mbox{\tiny $L$}}= \frac{ \widetilde{m}}{ \widetilde{m}- \widetilde{n}}\,,
  \ee
equation~\eqref{thetamore} reads
 \be\label{thetaphi}
  \begin{split}
\theta
=&\pi - 
2 \sqrt{\frac{(1- \widetilde{n})( \widetilde{n}- \widetilde{m})}{ \widetilde{n}}}
\left[\Pi \left(\left.  \widetilde{n}
\right|  \widetilde{m} \right)-K\left( \widetilde{m}\right)\right]\,,
  \end{split} 
  \ee
which is identical in form to the equation~\eqref{phi01} that computes $\phi$. Therefore, choosing $(n,m)=(\widetilde{n},\widetilde{m})$ realizes the BPS point. In terms of the original parameters entering the classical string solution, the BPS point is obtained for $\kappa=\gamma$ and $\ell_\varphi^2=\ell_\vartheta^2$. 

We can check that the energy at the BPS point coincides with $\L$. 
Setting the modular parameters $(m,n)$ in~\eqref{E} equal to $(\widetilde{m},\widetilde{n})$  in~\eqref{newparameters}, one obtains
\begin{equation}
\!\!\!
\mathcal{E} 
\!=\! 
\smash{\sqrt{{\frac{n_{\mbox{\tiny $L$}}(\!1-\!m_{\mbox{\tiny $L$}})}{m_{\mbox{\tiny $L$}}}}}}\bigg( \!\frac{1}{1-m_{\mbox{\tiny $L$}}}K\Big(\!\frac{m_{\mbox{\tiny $L$}}}{m_{\mbox{\tiny $L$}}-1}\!\Big) - E\Big(\!\frac{m_{\mbox{\tiny $L$}}}{m_{\mbox{\tiny $L$}}-1}\!\Big)\!\bigg)
\!=\!\sqrt{\frac{n_{\mbox{\tiny $L$}}}{m_{\mbox{\tiny $L$}}}} 
\left( K\left(m_{\mbox{\tiny $L$}}\right) - E\left(m_{\mbox{\tiny $L$}}\right)\right)\!\equiv\mathcal{L}\,,
\end{equation}
where in the last equality  the modular transformation~\eqref{modtransfKE} was used.

The expansion of the energy close to the BPS points and at small $\mathcal{L}$ reads
\begin{eqnarray}\label{EBPS}
\mathcal{E}&&\!\!\!\!=
 \mathcal{L} +\frac{\epsilon~\theta}{\sin\theta \sqrt{\pi^2 -\theta^2}}  \biggl(\frac{1  }{2}+\frac{3   \mathcal{L}}{2 \sqrt{\theta
   ^2-\pi ^2}}-\frac{3   \left(\theta ^2-3 \pi ^2\right) \mathcal{L}^2}{8 \pi ^2 (\pi^2 -\theta^2 )}+\frac{ \left(5 \pi ^4-3 \theta
   ^4+6 \pi ^2 \theta ^2\right)\mathcal{L}^3}{8 \pi ^4 \left(\pi ^2-\theta ^2\right)^{3/2}}+\biggr.\nonumber \\
 &&+\biggl.\frac{15  \left(-3 \theta ^6+3 \pi ^2 \theta ^4+11 \pi ^4 \theta ^2-3 \pi
   ^6\right) \mathcal{L}^4}{128 \pi ^6 (\pi^2 -\theta^2)^{2}}+O(\mathcal{L}^5)\biggr)+O(\epsilon^2)\,,
\end{eqnarray}
where $\epsilon=\cos\phi-\cos\theta$, and we used~\eqref{mBPS}-\eqref{nBPS}. The $O(\epsilon^2)$ term in the expansion is algebraically cumbersome and is displayed in~\eqref{EBPSlong}.     
    
Starting from~\eqref{EBPS}, one can compute  $L$-generalized Bremsstrahlung~$B_L(g,\phi)$ and Curvature~$C_L(g,\phi)$ functions defined as follows 
\be\label{BCdef}
\frac{\Gamma_L(\lambda,\phi,\theta)}{4g}\equiv\mathcal{E}=
\frac{1}{4g} \Big[\frac{ \cos \phi-\cos \theta}{\sin \phi}\frac{2 \phi}{1-\frac{\phi^2}{\pi ^2}}B_L(g,\phi)+ \left(\frac{\cos (\phi)-\cos (\theta)}{\sin (\phi)}\right)^2\phi^2 C_L(g,\phi)+\ldots\Big]\,.
\ee
At the leading order in the strong coupling expansion,  we can compute $B_L(g,\phi)$ as a series in small $\mathcal{L}$: 
\begin{equation}
\begin{split}\label{Brem}
B_L(g,\phi) =&\,g\,\Big[\frac{\sqrt{\pi ^2-\phi ^2}}{\pi ^2}-\frac{3
   \mathcal{L}}{\pi ^2}+\frac{3 \left(3 \pi ^2-\phi
   ^2\right) \mathcal{L}^2}{4 \pi ^4 \sqrt{\pi ^2-\phi
   ^2}}+\frac{\left(-3 \phi ^4+6 \pi ^2 \phi ^2+5 \pi
   ^4\right) \mathcal{L}^3}{4 \pi ^6 \left(\pi ^2-\phi^2\right)}\\
  & +
   \frac{15 \left(-3 \phi ^6+3 \pi ^2 \phi
   ^4+11 \pi ^4 \phi ^2-3 \pi ^6\right) \mathcal{L}^4}{64
   \pi ^8 \left(\pi ^2-\phi
   ^2\right)^{3/2}}+O\left(\mathcal{L}^5\right)\,\Big]\,. 
  \end{split}
  \ee
In~\cite{Gromov:2013qga}, a strong-coupling expansion up to $O(g^{-4})$ and arbitrary $L$ was presented. With the  identification $L \to 4 g \mathcal{L}$ and expanding to leading order in $g$, we find exact agreement with~\eqref{Brem}, if we take into account an overall multiplicative factor which arises from a slightly different parametrization of the linear coefficient in~$(\phi-\theta)$. Our expression~\eqref{EBPSlong} allows us to also evaluate the curvature function in~\eqref{BCdef} at the leading order in $g$ as a series in $\L$, that is
\allowdisplaybreaks
\begin{align}
&
C_L(g,\phi)=g\Big[\frac{-3 \phi ^2+2 (\phi -\pi ) (\phi +\pi ) \phi  \cot (\phi
   )+2 \pi ^2}{2 \phi ^2 \left(\pi ^2-\phi
   ^2\right)^{3/2}}+\frac{3 \left(\phi ^2+4 \pi ^2
   \phi  \cot (\phi )-4 \pi ^2\right)}{4 \pi ^2 \phi ^2
   \left(\pi ^2-\phi ^2\right)}\,\mathcal{L} \nonumber\\\nonumber
   &
   +\frac{ \big(-9
   \phi ^6+45 \pi ^2 \phi ^4-48 \pi ^4 \phi ^2-6 \pi ^2
   (\phi ^4-4 \pi ^2 \phi ^2+3 \pi ^4) \phi  \cot
   (\phi )+18 \pi ^6\big)}{8 \pi ^4 \phi ^2 (\pi
   ^2-\phi ^2)^{5/2}}\,\mathcal{L}^2\\\nonumber
   &
   +\frac{\big(\!-\!45 \phi
   ^8\!+\!231 \pi ^2 \phi ^6\!-\!303 \pi ^4 \phi ^4\!+\!141 \pi ^6 \phi
   ^2\!-\!8 \pi ^2  (3 \phi ^6\!-\!9 \pi ^2 \phi ^4\!+\!\pi ^4 \phi
   ^2\!+\!5 \pi ^6 ) \phi  \cot (\phi )\!+\!40 \pi ^8\big)}{32
   \pi ^6 \phi ^2  (\pi ^2-\phi
   ^2 )^3}\,\mathcal{L}^3 \\\nonumber
   &
   +\frac{1}{128 \pi ^{10} (\pi^2 -\phi^2 )^{7/2} \phi
   ^2 (\phi +\pi )}\Big(\big(30 \pi ^4 (\pi -\phi
   ) \phi  (\phi +\pi )^2 \big(3 \phi ^6-3 \pi ^2 \phi ^4-11
   \pi ^4 \phi ^2+3 \pi ^6\big) \cot (\phi )\\\nonumber
   &
   -45 \big(4851
   \phi ^{13}+4851 \pi  \phi ^{12}-24790 \pi ^2 \phi
   ^{11}-24790 \pi ^3 \phi ^{10}+51419 \pi ^4 \phi ^9+51419
   \pi ^5 \phi ^8\\\nonumber
&
   -55279 \pi ^6 \phi ^7-55279 \pi ^7 \phi
   ^6+33680 \pi ^8 \phi ^5+33680 \pi ^9 \phi ^4\\\label{curvature}
 &  -19747 \pi
   ^{10} \phi ^3-35 \pi ^{11} \phi ^2+2 \pi ^{12} \phi +2 \pi
   ^{13}\big)\big)\,\L^4+O(\mathcal{L}^5)\Big)\Big]\,.
 \end{align}
At $\phi \rightarrow0$, the generalized Bremsstrahlung and Curvature functions reduce to
\begin{eqnarray}
\begin{split}\label{BCsmall}
\!\!\!  \mathbf{B}_L(g)&\!\!=\!g\Big[\frac{1}{\pi }\!-\!\frac{3 \mathcal{L}}{\pi ^2}+\frac{9
   \mathcal{L}^2}{4 \pi ^3}\!+\!\frac{5 \mathcal{L}^3}{4 \pi
   ^4}\!-\!\frac{45 \mathcal{L}^4}{64 \pi
   ^5}\!+\!O(\mathcal{L}^5)\Big]\,, \\
\!\!\! \mathbf{C}_L(g)&\!\!=\! g \Big[\frac{2 \pi ^2-3}{6 \pi ^3}\!+\!\frac{\left(3-4 \pi ^2\right)
   \mathcal{L}}{4 \pi ^4}\!+\!\frac{3 \left(\pi ^2-4\right)
   \mathcal{L}^2}{4 \pi ^5}\!+\!\frac{\left(399+40 \pi ^2\right)
   \mathcal{L}^3}{96 \pi ^6}\!+\!\frac{\left(3465-90 \pi ^2\right)
   \mathcal{L}^4}{384 \pi ^7}\!+\!O(\mathcal{L}^5)\Big]\,. 
   \end{split}
\end{eqnarray}
The leading order term of $\mathbf{B}_L(g)$ reproduces the result of \cite{Correa:2012at}. The subleading terms in $\mathbf{B}_L(g)$, together with the expansion of $ \mathbf{C}_L(g)$, are in perfect agreement with the strong coupling results of~\cite{Gromov:2013qga}, upon identifying $L \to 4 g \mathcal{L}$.

\bigskip

It is instructing to look  at the behavior of the energy when $\phi=\Theff$, which reduces to the actual BPS limit for $L=0$. 
Formally, the calculation is identical to the  $L=0$ case (identifying $\theta$ with $\Theff$), obtaining
\begin{align}\label{EfakeBPS}
\mathcal{E}&=
(\cos\phi-\cos\Theff)   \frac{\Theff  }{2 \sin\Theff\sqrt{\pi^2 -\Theff^2}} +\\\nn
   &+(\cos\phi-\cos\Theff)^2  \frac{(-3 \Theff ^2+2 (\pi^2 -\Theff^2 )
  \Theff \cot \Theff   -2 \pi ^2) }{8 \sin^2 \Theff(\pi^2 -\Theff^2 )^{3/2}} + O\big((\cos\phi- \cos\Theff)^3\big)\,.
   \end{align}
This expression is just a rewriting of~\eqref{EBPS}, where all the dependence on $L$ is absorbed  in~$\Theff$, as can be checked inserting~\eqref{theffsmallL} and expanding for small~$\L$.

\subsection{Large-$\L$ expansion}
\label{subsec:largeL_summary}

The large-$\mathcal L$ regime is controlled by the endpoint behavior of the elliptic moduli entering
\eqref{L}.  The ordering $0\le m_{\mbox{\tiny $L$}}\le n_{\mbox{\tiny $L$}}\le1$ implies the bounds
\eqref{rangeinequality}.  Together with the inequality \eqref{ellipticbound} for $K$ and $E$ on
$[0,1]$, these bounds show that $m_{\mbox{\tiny $L$}}\to1$ as $\mathcal L\to\infty$ with exponentially
small corrections, and consequently $n_{\mbox{\tiny $L$}}\to1$ at the same rate.  This motivates an
ansatz of the form \eqref{mLnLlargeL} in terms of a small parameter $\eL\sim e^{-2\mathcal L}$.
The leading term of \eqref{L} fixes the normalization of $\eL$,
\begin{equation}\label{epsilonL}
\mathcal{L}=\frac{1}{2}\left(\log\frac{16}{\eL}-2\right)
\qquad\Longrightarrow\qquad
\eL=\frac{16}{e^2}\,e^{-2\mathcal{L}}\,,
\end{equation}
while higher coefficients are obtained by solving \eqref{theta} and \eqref{L} order by order using
the expansion of the complete elliptic integral of the third kind $\Pi$ given in
Appendix~\ref{app:expansionPi}.

In the $\text{AdS}_5$ problem, the consistency condition places the solution in region~$B$ where $m<0$ and
$m\to-\infty$ as $\mathcal L\to\infty$, with the scaling $m=-\eL^{-1}+O(1)$, while $n$ remains
finite.  To keep the elliptic-modulus expansions within $[0,1]$, a modular transformation is
performed,
\be\label{mprimenprime}
m'=\frac{m}{m-1},
\qquad
n'=\frac{n-m}{1-m},
\ee
under which the defining equation for $\phi$ retains its form (with $(m,n)\to(m',n')$) and
the consistency condition becomes \eqref{cons1}.  Since $m',n'\to1$ for large $\mathcal L$,
one expands
\eqref{mhatnhatargeL} and solves \eqref{phi01} and \eqref{cons1} perturbatively in $\eL$, using
Appendix~\ref{app:expansionPi} together with the standard $m\to1$ asymptotics of $K$ and $E$.
The explicit expansions for $m'$ and $n'$ up to $O(\eL^4)$ are collected in
Appendix~\ref{app:largeL}; here we quote the final result for the energy. In terms of the transformed parameters, the energy reads
\be
\mathcal E=\frac{1}{\sqrt{n'}}\bigl(K(m')-E(m')\bigr),
\ee
and its large-$\mathcal L$ expansion takes the compact form
\begin{align}\label{ElargeL}
\mathcal{E}=&\,
\mathcal L+\frac{1}{4}\,(\cos\phi-\cos\theta)\Bigg[
\eL
+\frac{\eL^2}{8}\Big((4\mathcal L+1)\cos\theta+(4\mathcal L+5)\cos\phi\Big)
\notag\\
&\hspace{2.9cm}
-\frac{\eL^3}{128}\Big(
2\big(32\mathcal L(\mathcal L+1)+1\big)\cos\theta\cos\phi
+\big(16\mathcal L(2\mathcal L+1)+5\big)\cos2\theta
\notag\\
&\hspace{4.2cm}
+\big(16\mathcal L(2\mathcal L+5)+49\big)\cos2\phi
+4\mathcal L(4\mathcal L-1)
\Big)
+O(\eL^4)
\Bigg].
\end{align}
The leading term agrees with the result first obtained in \cite{Correa:2012hh}.
Additional details of the derivation, including the intermediate expansions of the moduli and of
$\Theta_{\mathrm{eff}}$, are presented in Appendix~\ref{app:expansionPi}.
The expansion of the energy also allows one to extract straightforwardly the functions
$B_L(\phi)$ and $C_L(\phi)$ defined in \eqref{BCdef}:
\begin{align}\label{Blarge1}
 B_L(\phi)=&4g
\left(1-\frac{\phi ^2}{\pi ^2}\right) \frac{\sin \phi }{\phi }\Bigg[\frac{\eL }{8}-\frac{\eL ^2 }{32} (4 \mathcal{L}+3) \cos
   \phi +\frac{\eL ^3}{1024} ((32 \mathcal{L} (3 \mathcal{L}+4)+55)
   \cos 2 \phi +\notag\\
   &+4 \mathcal{L} (12
   \mathcal{L}+7)+1)+O(\eL ^4)\Bigg]\,,
\end{align}
and
\begin{align}\label{Clarge1}
 C_L(\phi)=&4g\frac{\sin^2\phi}{\phi^2}
\left[\frac{1}{64} g (32 \mathcal{L}+8) \eL ^2-\frac{1}{64} g (32
   \mathcal{L} (3 \mathcal{L}+2)+11) \eL ^3 \cos \phi
   +O\left(\eL ^4\right)\right]\,.
\end{align}

\section{String normal modes}
\label{sec:quantum}

In this section we analyze quadratic quantum fluctuations around the classical configuration defined by the ansatz \eqref{ansatz2}. The classical string geometry is specified by the two functions $\rho(\sigma)=\mathrm{arcsinh}(r(\sigma))$ and $\psi(\sigma)=\arcsin(\tilde \rho(\sigma))$, whose explicit forms are given in \eqref{rsol} and \eqref{rhotildesol}. 
In particular, we derive the quadratic Lagrangian for the bosonic fluctuations in this background (see section~\ref{sec:fluctuations}), and use it to determine the corresponding normal mode frequencies (section~\ref{sec:freq}) in the semiclassical (WKB) regime. 
The quadratic Lagrangian for the fermionic fluctuations is computed in app. \ref{app:fermions}. 
We then examine the behavior of these frequencies in the antiparallel lines limit, in the vicinity of the transition curve (the blue line in fig. \ref{fig:Transitioncurve}), and in the large-$\mathcal{L}$ expansion.

\subsection{Fluctuations}
\label{sec:fluctuations}

In this section we compute the quadratic string action for the bosonic fluctuations. In what follows, we use $\alpha, \beta, \dots=0,1, $ to denote curved worldsheet indices, and $a, b, \dots=0,1,$ to denote flat worldsheet indices. Target spaces indices are denoted by $\mu, \nu, \dots=0, \dots, 9$ for the curved case and $A, B, \dots=0, 1, \dots 9$ for the flat ones, respectively. The indices on the normal bundle are: $i, j=1, \dots, 8$. 
In order to simplify the exposition we define the following positive functions
\bea\label{def-XandY}
 \mathcal X(\s) &:\, =& \sqrt{ \kappa^2 \cosh^2\rho(\s)-1}\,=  \sqrt{  {n-m \over 1-n-m} \cosh^2\rho(\s)-1}\,, 
 \\ \nn
 \mathcal Y(\s) &:\, =& \sqrt{ 1-\gamma^2 \cos^2\psi(\s)}\,= \sqrt{ 1-\frac{\mL}{\nL-\mL(1-\nL)} \cos^2\psi(\s)}\,. 
\eea
where we have used~\eqref{formula-kappaellephi-of-nm},~\eqref{formula-gammaelletheta-of-nmL} in the second equality. 
From the discussion in section \ref{sec-classical}, it follows that both functions are well defined and real for all the range of $\s$. The Virasoro constraints \eqref{req}-\eqref{rhoeq} in terms of $\mathcal X, \mathcal Y$ become
\be
\begin{split}
\mathcal X \mathcal X' 
&= \sqrt{1+\mathcal X^2} \sqrt{\Big(\mathcal X^2 -\frac{n}{1-m-n}\Big)\Big(\mathcal X^2+ \frac{1-n}{1-m-n}\Big)}\,,
\\
\mathcal Y \mathcal Y' 
&=  \sqrt{1-\mathcal Y^2} \sqrt{\Big(\mathcal Y^2 -\frac{\nL-\mL}{\nL-\mL(1-\nL)}\Big)\Big(\mathcal Y^2+\frac{\mL(1-\nL)}{\nL-\mL(1-\nL)}\Big)}\,.
\end{split}
\ee
Notice that the conformal factor of the induced worldsheet metric \eqref{inducedflat} can be rewritten as~
$e^{2\Lambda(\s)} = \mathcal X^2+\mathcal Y^2$.

We expand the bosonic string action around the classical solution \eqref{ansatz2} (where $\rho, \psi$ satisfy the eom  \eqref{req}-\eqref{rhoeq}) following the general procedure illustrated in \cite{Forini:2015mca, Singh:2023olv}. Choosing a convenient basis in the normal bundle, see Appendix \ref{app:bosons-oneloop}, the Lagrangian for the physical eight normal bosonic modes reads
\be
\label{L-bos-oneloop}
\mathcal L_{ (2)\rm b} = \sqrt{h} \left(h^{\alpha\beta} \mathcal D_\alpha \xi^i \mathcal D_\beta \xi_i- \mathcal M_{ij} \xi^i \xi^j\right)\,,
\ee
where $h_{\alpha \beta}$ is the on-shell induced worldsheet metric \eqref{inducedflat}, the covariant derivative acting on a section of the normal bundle is given by
\be
\mathcal D_\alpha \xi^i :\, = \partial_\alpha \xi^i-A^i_{j\, \alpha} \xi^j\,, 
\ee
with $A^i_{j\, \alpha}$ a connection on the normal bundle. The ``mass'' matrix $\mathcal M_{ij}$ contains contributions from the (projected) Riemann tensor and the extrinsic curvature. We refer the reader to  \cite{Forini:2015mca, Singh:2023olv} for more details. 
Given the conformal gauge choice of~\eqref{inducedflat}, we can write the bosonic operators acting on the fluctuations $\xi^i$ as 
\be
\label{eom-bosons-1loop-all}
\xi^i \mathcal K_{ij} \xi^j = \xi^i \left(- \eta^{\alpha \beta} \mathcal D_\alpha \mathcal D_\beta -\sqrt{g} \mathcal M\right)_{ij}\xi^j\,,
\ee
where $i,j$ are SO(8) indices. 
With our choice of the basis in the normal bundle, the operators acting on the first four bosonic modes turn out to be rather simple, since the covariant derivative is actually a partial derivative, and the mass matrix is diagonal. Explicitly we have 
\begin{subequations}
\be\label{K11}
\mathcal{K}_{ii}  = -\eta^{\a\b} \pa_\a \pa_\b- 1+2 \kappa^2 \cosh^2\rho =-\eta^{\a\b} \pa_\a \pa_\b+1+2 \mathcal X^2 \,,  \qquad i=1,2 \,, 
\ee
\be\label{K33}
\mathcal{K}_{ii}  = -\eta^{\a\b} \pa_\a \pa_\b- 1+2 \g^2 \cos^2\psi\, = -\eta^{\a\b} \pa_\a \pa_\b +1-2 \mathcal Y^2\,,  \qquad i=3,4 \,,
\ee
\end{subequations}
where we remind the reader that $\rho$, $\psi$ (and so $\mathcal X, \mathcal Y$) are functions of $\s$. 
The remaining four transverse directions are all coupled, via normal bundle connections as well as the extrinsic (squared) curvature which makes the mass matrix $\mathcal M$ non-diagonal. 
The gauge connections are
\begin{subequations}
\begin{eqnarray}
\label{Atau}
&& A_\tau^{\, 5\,6} =  \gamma \ell_\vartheta\, e^{-\Lambda}\,  \frac{ \sqrt{1+\XX^2}}{\YY}\,,
\qquad
A_\tau^{\, 5\,7} =  \kappa  \ell_\varphi \,  e^{-\Lambda}\, \frac{\sqrt{1-\YY^2}}{\XX} \,, 
\\ \nn
&& A_\tau^{\, 5\,8} = -\frac 12 e^{-2\Lambda}\,\sqrt{1-\YY^2}\sqrt{1+\XX^2} \left( \frac{\YY}{\XX} \pa_\s \log(1+\XX^2)-\frac{\XX}{\YY} \pa_\s \log(1-\YY^2)\right)\,,
\\
\label{Asigma}
&&A_\s^{\,6\,8} =-  \gamma  \ell_\vartheta \, e^{-\Lambda} \, \frac{ \XX \sqrt{1-\YY^2}}{\YY^2}\,,
\qquad
A_\s^{\,7\,8}= \kappa  \ell_\varphi \, e^{-\Lambda}\,  \frac{ \YY\sqrt{\XX^2+1} }{\XX^2}\,.
 \end{eqnarray}
\end{subequations}
The elements of the mass matrix are given by
\begin{subequations}
\label{massmatrix-last4}
\bea
-\sqrt{g} \mathcal M_{55} &=&-1+\frac 12 \, e^{-4\Lambda}\,\left(\mathcal X^2+1\right) \left(1-\YY^2\right) \left(\pa_\s\log{1+\mathcal X^2\over 1-\YY^2}\right)^2\,,
\\
-\sqrt g \mathcal M_{66} &=& 1-2 \YY^2 +2 \gamma ^2 \ell_\vartheta^2 \, e^{-2\Lambda}\,  \frac{ 1-\YY^2}{\YY^2}\,,
\\
-\sqrt{g} \mathcal M_{67} &=&- 2 \gamma  \kappa  \ell_\vartheta \ell_\varphi \, e^{-2\Lambda}\, \frac{\sqrt{\XX^2+1} \sqrt{1-\YY^2}}{\XX\, \YY}\,, 
\\
-\sqrt{g} \mathcal M_{68} &=&   2 \gamma\,  \ell_\vartheta \, e^{-3\Lambda}\,  \XX \sqrt{1-\YY^2}\,\pa_\s \log{\XX\over \YY}\,,
\\
-\sqrt{g} \mathcal M_{77} &=&1+2 \XX^2 -  2 \kappa ^2 \ell_\varphi^2   \, e^{-2\Lambda}\,  \frac{  \XX^2+1}{\XX^2}\,,
\\ 
-\sqrt{g} \mathcal M_{78} &=& 2 \kappa\, \ell_\varphi  \, e^{-3\Lambda}\, \YY \sqrt{1+\XX^2}\,\pa_\s \log{\XX\over \YY}\,,
\\
-\sqrt{g} \mathcal M_{88} &=&1- 2 \,e^{-4\Lambda}\,  \XX^2\, \YY^2 \, \left(\pa_\s \log{\XX\over \YY}\right)^2\,.
\eea
\end{subequations}
In the expression above the parameters $(\kappa, \ell_\varphi, \gamma, \ell_\vartheta)$ can be easily converted into $(n, m, \nL, \mL)$ using the formulae \eqref{formula-kappaellephi-of-nm} and \eqref{formula-gammaelletheta-of-nmL}. 
In Appendix~\ref{app:fermions} we also compute the quadratic fermionic Lagrangian, and we check that the trace relation among the bosonic and fermionic masses is satisfied, ensuring a consistent string worldsheet theory.

\subsection{Frequencies}
\label{sec:freq}

As explained in the Discussion (see Sec. \ref{sec:intro}), the classical string solution~\eqref{ansatz2}  describes two heavy W-bosons in a large $R$-charge background. Alternatively, following~\cite{Hong:2004gz} we can think about this system as a quarkonium with the insertion of  $L$ scalar particles in the adjoint representation.  
The fluctuations computed in section \ref{sec:fluctuations} represent small perturbations around this solution~\eqref{ansatz2}, which can then be interpreted as excitations of this bound system~\cite{Klebanov:2006jj, Drukker:2006xg, Correa:2012hh}. In this section we study the normal mode frequencies of these excitations. We focus on the frequencies of the eight scalar fields $\xi^i$ in view of their interpretation as excitations of ``mesonic'' states following~\cite{Callan:1999ki, Hong:2004gz, Klebanov:2006jj}.

The physical transverse bosonic fluctuations are governed by the Lagrangian computed in section \ref{sec:fluctuations}. 
From there we can compute the equations of motion, which can be schematically written as
\be\label{eom-schem}
\mathcal K_{ij} \xi^j=0\,, \qquad i, j=1, \dots, 8\,,
\ee
according to \eqref{eom-bosons-1loop-all}.
In this section we analyze the frequencies of these eight bosonic mode $\xi^i$ in a WKB approximation. 
The first four bosonic modes $\xi^i$ ($i=1, \dots, 4$) are uncoupled, see  equations \eqref{K11} and \eqref{K33}, and their computation can be performed exactly, as shown in Appendix~\ref{app:Lame}. However, computing exactly the frequencies for the remaining coupled modes ($i=5, \dots, 8$) turned out to be rather challenging, and for this reason we choose to present here the WKB approximation for all bosonic frequencies. 
In app. \ref{app:Lame} we show that the semiclassical expansion of the exact frequencies for the uncoupled scalar fields agree with the WKB analysis described in this section. 

We start by illustrating the general procedure for the transverse uncoupled AdS modes~\eqref{K11}. By Fourier transforming the modes $\pa_\tau \to e^{i \omega\tau}$, one obtains a one-dimensional Schroedinger equation for the radial component, which we rename $\hat\xi(\s)$:
\be\label{1d-K11}
-\pa_\s^2 \hat\xi^i(\s)-\omega^2 \hat\xi^i(\s)+(1+2\XX^2(\s)) \hat\xi^i(\s)=0\,, \qquad i=1, 2\,.
\ee
The potential $V(\s)=1+2\XX^2(\s)$ behaves as a well potential, where at the center of the worldsheet ($\s\to 0$) approaches a constant, while it blows up at the boundaries $\s\to \pm s/2$. 
We impose Dirichlet conditions at the boundaries, that is $\hat\xi(\pm s/2)=0$. 
We want to compute the frequencies $\omega$ in a semiclassical approximation, which implies a large $\omega$ expansion. Then, we expand the modes $\hat\xi^i(\s)$ in a WKB approximation where the parameter of the expansion is indeed $\omega$:
\be\label{xihat-exp}
\hat\xi(\s) \sim e^{i \int_{-s/2}^\s  P(\s') \dif \s' }\,, \qquad \text{with}\quad P(\s)=\sum_{\ell=0}^M \omega^{1-\ell}  { P_\ell(\s)}\,, 
\ee
where $M$ determines the order of the approximation, and the overall amplitude does not matter for our discussion. The second-order equation \eqref{1d-K11} becomes a first-order Riccati equation for the function $P$, which can be solved order by order in $\omega$. 
The odd functions $P_{2\ell+1}$ (with $\ell\ge 0$) contribute only to the overall amplitude of the wave function $\hat\xi$, while the first even functions are given by
\be
P_{0, \pm}(\s) =\pm 1\,, \qquad P_{2,\pm}(\s) = \mp (\tfrac12-\kappa^2 \cosh^2\rho(\s)) =\mp (\tfrac12+\XX^2(\s))\,, \qquad 
\ee
and higher terms can be easily determined recursively. The Dirichlet boundary conditions for  $\hat\xi$ result in a sine wave function, which has to vanish at $\s=s/2$:
$$
\hat \xi^i(s/2) \sim \sin\Big(\omega \int_{-s/2}^{s/2} \dif \s' \Big(P_{0,+}(\s')+{1\over \omega^2} P_{2, +}(\s')+\dots \Big) \Big) =0\,, \qquad i=1,2\,.
$$
This is equivalent to impose the following quantization condition 
\be
\omega_\ell \int_{-s/2}^{s/2} \dif \s' \Big(P_{0,+}(\s')+{1\over \omega_\ell^2} P_{2, +}(\s')+\dots \Big) = \ell \pi, \qquad \ell \in \mathbb{N}\,.
\ee
The integrals of the functions $P_{0,+}, P_{2,+}$ have been already computed in section \ref{sec:classical-ads}, and we have 
\be
 \int_{-s/2}^{s/2} \dif \s' P_{0,+}(\s')=s\,, \qquad  \int_{-s/2}^{s/2} \dif \s' P_{2,+}(\s') =  \frac12 \left(s -{\kappa\over g} E\right)\,,
\ee
if we use  equation \eqref{Energy}. The integral of $P_{2, +}$ is divergent, and we choose to regularizing it as in  \eqref{Energy}, such that the quantization condition reads now
\be
\omega_\ell \left(s+\frac{1}{2\omega_\ell^2} \left(s - 4 \kappa  \mathcal E\right)+\dots\right)  = \ell\pi\,, \qquad \ell=1, 2\,, \dots\,, \qquad \text{i.e.}~~ \ell \in\mathbb N\,,
\ee
in terms of the rescaled regularized energy $\mathcal E$ \eqref{E}. 
The higher order corrections $P_{2\ell}$ cannot be written in such a neat form in terms of $s$ and the conserved charge $\mathcal E$.
 However, it should be kept in mind that our semiclassical analysis is only valid at very large $\ell$, hence solving the above condition perturbatively for large quantum numbers $\ell\gg 1$, we obtain
 \be\label{omega12-semicl}
 \omega_\ell ={\pi \ell \over s}-{1\over 2\pi \ell} \left(s -4\kappa\,\mathcal E\right)+\dots\,, \qquad \ell\in \mathbb N\,, ~\ell\gg 1\,. 
 \ee
 
 The analysis for the uncoupled transverse directions in the five-sphere is completely analogous. Given the operator \eqref{K33}, the Schrodinger equation reads now
  \be\label{1d-K33}
-\pa_\s^2 \hat\xi^i(\s)-\omega^2 \hat\xi^i(\s)+(1-2\YY^2(\s)) \hat\xi^i(\s)=0\,, \qquad i=3, 4\,,
\ee
and the semiclassical expansion \eqref{xihat-exp} of the wave function gives 
\be
P_{0, \pm}(\s) =\pm 1\,, \qquad P_{2,\pm}(\s) = \pm (\tfrac12-\gamma^2 \cos^2\psi(\s)) =\mp (\tfrac12-\YY^2(\s))\,. 
\ee
Repeating the same steps as before, the quantization condition is now expressed in terms of the conserved $R$-charge $\mathcal L$ \eqref{L} (see equation \eqref{L00}):
\be
\omega_\ell \left(s +\frac{1}{2\omega_\ell^2} \left(s-4\gamma \mathcal L\right)+\dots\right) = \ell\pi, \qquad \ell\in \mathbb N\,,
\ee
which for large quantum number $\ell$ is
\be\label{omega34-semicl}
\omega_\ell  = {\pi \ell \over s}-{1\over 2\pi \ell} \left(s - 4 \gamma \mathcal L\right)+\dots\,, \qquad \ell\in \mathbb N\,, ~~\ell\gg 1\,. 
\ee

In Appendix \ref{app:Lame}, we obtain the exact frequencies corresponding to the modes $i=1, 2, 3,4$ and we check that their semiclassical expansions reproduce \eqref{omega12-semicl} and \eqref{omega34-semicl} at the leading order.

For the remaining coupled bosonic modes ($i=5, \dots, 8$) we can proceed in a similar way, keeping in mind that we are now dealing with a $4\times 4$-matrix second-order coupled differential equation. The method used is summarized in Appendix~\ref{app:WKB}, here we report only the leading order result for the frequencies $\omega$ in the large quantum number $\ell$, that is
\be\label{omega58-semicl-main}
\omega_\ell={\pi \ell \over s}+{\theta_I \over 2 s}+\dots\,, \qquad \ell\in \mathbb N\,,~~\ell\gg 1\,,  \quad I=1,  \dots, 4\,. 
\ee
Here $e^{ i \theta_I}$ denotes the eigenvalues of the $\grSO(4)$ holonomy matrices constructed from the background gauge connections $(A_\tau, A_\sigma)$ of the normal bundle. For their appearance in the semiclassical solution of the coupled differential equations for the modes $i = 5, \dots, 8$, we refer the reader to appendix~\ref{app:WKB}.

In summary, a WKB analysis of the equations of motion of the physical bosonic fluctuations has lead to the semiclassical computation of the related frequencies, that is equations \eqref{omega12-semicl}, \eqref{omega34-semicl}, and \eqref{omega58-semicl-main}. 
This naive implementation of the WKB analysis for the uncoupled modes misses the corrections of order $\ell^{\,0}$~\footnote{Recovering this correction would require a more refined analysis of the boundary conditions. Within the
EBK/WKB approximation, Dirichlet boundary conditions at both endpoints behave as two hard walls, leading to a unit shift of one of the integers $\ell$ in the leading quantization condition. This is confirmed by the exact analysis in Appendix~\ref{app:energydensity}, see discussion around equation~\eqref{r0const}.}, as expected~\cite{Klebanov:2006jj}, see  \eqref{omega12-semicl}, \eqref{omega34-semicl}.
 For the coupled modes, our analysis captures the additional shift induced by the gauge fields, see \eqref{def-U1}.

We should keep in mind that the actual physical frequencies, that is the frequencies measured by an observer at time $t$, differ by $\omega$ due to the rescaling among the worldsheet and target space time, see  \eqref{ansatz2}. Hence, for all the bosonic frequencies, at large quantum numbers, we have 
\be \label{omega-phys}
\omega_{\rm phys, \ell} = {\omega_\ell\over \kappa}\, = {\pi \ell\over s \kappa}+\dots \,, \qquad \ell\in \mathbb N\,, \quad \ell \gg 1\,. 
\ee
In the next sections we discuss their behavior  in some relevant regimes. In all cases, we proceed by first examining the exact frequencies for the uncoupled modes, and then for the semiclassical frequencies, valid for all the bosonic string modes. 
To understand the relation between exact frequencies $\omega_\ell$ and quantum number $\ell$, as well as to facilitate the various expansions, we employ the ``frequency densities'' $\tilde\rho_\omega, \rho_\omega$, as in~\cite{Giombi:2022anm}:
\be\label{all-rho}
\ell =\int_{\omega_0}^{\omega_\ell} \rho_{\omega}(\omega) \dif \omega\,, \qquad \ell\in\mathbb N\,,
\ee
and similarly for $\tilde\rho_\omega$. The densities as well as their corresponding extremes of integration $\omega_0, \tilde{\omega}_0$ are worked out in Appendix~\ref{app:Lame}. 

\subsubsection{Antiparallel lines limit}
\label{sec:omega-antiparallel}

It is interesting to understand how the {\it physical} frequencies \eqref{omega-phys} behave in the limit of antiparallel lines discussed in section \ref{subsec-antiparallel}, close enough to the transition curve (see fig. \ref{fig:Transitioncurve}).

We begin our analysis by studying the exact frequencies for the transverse modes in AdS. 
In Region A$_L$, introduced in section~\ref{sec:angle-s5}, i.e. below the transition, the antiparallel lines limit is reached by taking 
$$n \to m\,, \qquad m \in (0,1/2)\,,$$ 
as discussed in case {\bf 1} in section~\ref{subsec-antiparallel}.
It is useful to understand the exact relation between the parameter $\kappa$ and the geometric opening angle of the cusp in this limit. 
We use the expression of $\kappa$ as a function of the parameters $(n, m)$ \eqref{formula-kappaellephi-of-nm}, and the expansions \eqref{nminusm} and \eqref{maroundmbar},  to find
\be\label{antiparallel-rega-omega}
{1\over \kappa}= 2 \sqrt{\frac{1-2\mbar}{\mbar(1-\mbar)}}{\left(E(\mbar)-(1-\mbar)K(\mbar)\right)\over \pi-\phi}+O((\pi-\phi)^2)\,, \qquad \mbar\in \left(0, \tfrac 12\right)\,,
\ee
where $\mbar$ is determined in terms of $\theta$ and $\L$ via the consistency condition~\eqref{consistencybarA}.
As we will see, this plays a crucial role for the frequencies close to the transition. 
Expanding  the density $\rho_\omega$ \eqref{density-ads} according to \eqref{nminusm} and \eqref{maroundmbar}, then integrating this from $\omega_0 = \sqrt{{\mbar\over 1-2 \mbar}}+\mathcal O((\pi-\phi)^2)$ \eqref{omega0ads-A} to large $\omega$, we obtain 
\be\label{cond-quant-regA}
\ell=\int^{\omega_\ell}_{\sqrt{{\mbar\over 1-2 \mbar}}} \rho_\omega(\omega) \dif \omega= {2\over \pi}\sqrt{1-2 \mbar} \, K(\mbar)\, \omega_\ell + O((\pi-\phi), \omega^2) \,, \qquad \ell\in \mathbb N\,.
\ee
Then, at the leading order in the large quantum number $\ell$, dividing the above expression by $\kappa$ \eqref{antiparallel-rega-omega}, we have
\be\label{omega-phys-rega}
\omega^{(A)}_{\rm phys, \ell} = \frac{\pi  \ell  \left(E\left( \mbar\right)-\left(1-\mbar\right) \, K\left(\mbar\right)\right)}{ \sqrt{\mbar(1-\mbar)}
\,    K\left(\mbar\right) \, (\pi -\phi)}+ O({(\pi-\phi), \ell^{-1}})\,, \qquad \mbar\in \left(0, \tfrac 12\right)\,, 
\ee
where the superscript $A$ underlines that the limit is valid in Region A$_L$. 
The  $1/(\pi-\phi)$ behavior in the decompactification limit  becomes the usual $1/L_{\text{sep}}$ behavior (with $L_{\text{sep}}$ distance between the two lines)  present in the plane~\cite{Maldacena:1998im,Rey:1998ik}.%
\footnote{This replacement should be really thought  as $a(\pi-\phi) \to L_{\text{sep}}$ where $a$ is the radius of the sphere, which approaches to $\infty$, such that $a(\pi-\phi) $ remains finite and can be interpreted as the distance among the quarks on the flat space.}
We stress that the coefficient of the pole in \eqref{omega-phys-rega} is not corrected by the next-to-leading order terms in $\pi-\phi$ and $\ell^{-1}$. 

In the expression above we set $\mbar\in(0, 1/2)$. However, there are two delicate and yet interesting  limits to consider.
When $\mbar =1/2$, we have 
\be\label{omega-phys-rega-KMT}
\omega^{(A)}_{\rm phys, \ell} = \frac{8 \pi ^3 \ell}{\Gamma \left(\frac{1}{4}\right)^4 (\pi-\phi)}+O({(\pi-\phi), \ell^{-1}})\,,
\ee
which agrees with the result of \cite{Klebanov:2006jj} once we replace $\pi-\phi$ with $L_{\text{sep}}$, where $L_{\text{sep}}$ is the distance between the two probe charges on the plane. In this limit, the condition of quantization \eqref{cond-quant-regA} seems rather problematic. Notice that also  $s_{\text{AdS}}$  \eqref{sAdS5} vanishes. This simply signals that our reparameterization is not suited for the limit. In fact, in this limit the length of the string interval stays finite in \cite{DF}, the vanishing of $s_{\text{AdS}}$ \eqref{sAdS5} is due to the vanishing of the pre-factor $\sqrt{1-n-m}$, which is scaled out in \cite{DF}. The same factor should be scaled out in the frequencies $\omega$, and thus in \eqref{cond-quant-regA}, leaving again a finite result.

Let us consider $\mbar\to 0$. The coefficient of the pole in \eqref{antiparallel-rega-omega}, and thus in \eqref{omega-phys-rega} vanishes, as well as $\mathcal E_{\rm Cas}^{(A)}$ in \eqref{Einterm}. 
In this limit, the effective angle $\Theff=\pi$, according to \eqref{consistencybarA}, and looking at figure~\ref{fig:transition-mvsth}, we expect to find pairs of $(\nL, \mL)$ parameters which satisfy $\Theff=\pi$ \eqref{theffdef}. In fact these are nothing but those describing the curve plotted in figure \ref{fig:Transitioncurve}. 
Such a singular expansion for $\mbar\to 0$  and the vanishing of the Casimir energy were already observed in \cite{Cuomo:2024psk} for the generalized cusp (that is $L=0$), and this is nothing but an effect of the BPS condition $\theta=\phi$. 
Here, we observe an analogous behavior once we replace $\theta$ by $\Theff$, in agreement with what discussed in section \ref{sec:nearBPS}.

Approaching the critical point from above means we are in the region B$_L$, that is in the limit of case {\bf 2} as discussed in section \ref{subsec-antiparallel}, where $n\to 0$ and $m<0$. According to the expansions \eqref{nexp} and \eqref{maroundmhat}, the parameter $\kappa$ scales as 
\be
\label{antiparallel-regb-omega}
{1\over \kappa}= \sqrt{1-\frac{1}{\widehat m}}+ O\left((\pi-\phi)^2\right) \,,
\ee
where $\widehat{m}<0$ is the solution of the consistency condition, equation~\eqref{consistencyhatB}. Expanding the density $\rho_{\omega}$ \eqref{density-ads} by means of \eqref{nexp} and \eqref{maroundmhat}, and in the large $\omega$ approximation, we have 
\be
\ell= \int^{\omega_\ell}_{\sqrt{\widehat m}/\sqrt{\widehat m-1} }\rho_\omega(\omega) \dif \omega= \frac{2 \sqrt{1-\widehat m}  K(\widehat m)}{\pi }\, \omega_\ell  +O((\pi-\phi)^2, \omega^2) \,, \qquad \ell\in \mathbb N\,.
\ee
Combining the above relations we obtain 
\be\label{omega-phys-regb}
\omega^{(B)}_{\rm phys, \ell} = \frac{\pi  \ell}{2 \sqrt{-\widehat m} \,K(\widehat m)}+ O((\pi-\phi)^2, \ell^{-1})\,, \qquad \widehat m<0\,,
\ee
where again the superscript $B$ stresses the validity of this formula in Region B$_L$,  showing how the physical frequencies \eqref{omega-phys} are finite when we approach the antiparallel lines from above the transition. 

\vskip 0.5 cm
All the other frequencies \eqref{omega34-semicl} and \eqref{omega58-semicl-main} are not affected by the antiparallel lines limit, only indirectly via the relation \eqref{consistencybarA}, or \eqref{consistencyhatB} and via the value of the parameter $m$, that is either $\mbar$ or $\widehat m$,  for region A$_L$ and B$_L$ respectively, thus remaining finite quantities. 
It is solely due to the overall factor $\kappa^{-1}$, and its different behavior above and below the transition, see equations \eqref{antiparallel-rega-omega} and \eqref{antiparallel-regb-omega} respectively, that the physical frequencies \eqref{omega-phys} will develop (or not) a pole when $\phi \to \pi$.

In summary, the semiclassical normal mode frequencies of the bosonic fluctuations preserve the features shown by the classical string energy close to the transition point. It is important to stress that this analysis is valid only sufficiently close to the transition point. 
In \cite{Klebanov:2006jj}, for $L = 0$, the frequencies \eqref{omega-phys-rega-KMT} were interpreted as those of excited quark–antiquark bound states near the unbinding threshold $E = 0$. Extending this to our setup and to \cite{Hong:2004gz}, they can be viewed as the excitation frequencies of a quark–antiquark–scalar bound system near the unbinding transition point. Above the transition, the bound system disappears, and the interpretation of \eqref{omega-phys-regb} is less clear.

\paragraph{Expansion around the critical point $\mathcal{L}_c$.}

As in section \ref{subsec-antiparallel}, we can investigate the behavior of normal mode frequencies when $\mathcal L \sim \mathcal L_c$.  
We recall that this means to be in the antiparallel lines limit, at fixed $\theta$, but $\mathcal L\to\mathcal L_c$. 
In terms of the modular parameters, we specify $\mbar$ in \eqref{omega-phys-rega} as in \eqref{mcriticalA}, and $\widehat m$ in \eqref{omega-phys-regb} as in \eqref{mcriticalB}, with the modular parameters of the 5-sphere $(\nL^c, \mL^c)$ satisfying the critical condition \eqref{mncritical}. 
For the region A$_L$, the coefficient of the pole at $\phi=\pi$ in \eqref{omega-phys-rega} approaches zero when $\mathcal L\to \mathcal L_c$ as
\bea
\omega_{\rm{phys}, \ell}^{(A)} \sim  {\ell\over \pi-\phi} \sqrt{\frac{2 \pi }{3}} \sqrt[4]{\frac{\mL^c}{\mL^c (\nL^c-1)+\nL^c}} \sqrt{\mathcal{L}_c-\mathcal{L}}+\dots\,,
\eea
while for the region B$_L$, the physical frequencies \eqref{omega-phys-regb} behave as
\bea
\omega_{\rm{phys}, \ell}^{(B)} \sim  \frac{\ell \sqrt{\pi } \sqrt[4]{\mL^c (\nL^c -1)+\nL^c}}{2 \sqrt{2} \sqrt[4]{\mL^c} \sqrt{\mathcal{L}-\mathcal{L}_c}}+\dots 
\eea
The $(\L-\L_c^\theta)^{-1/2}$ divergence of the physical frequencies above the transition is entirely driven by the $\kappa^{-1}$ dependence.

\subsubsection{Large $\mathcal L$ expansion and BMN}
\label{sec:largeL-frequencies}

In this section we investigate the large $\mathcal L$ regime of the normal mode frequencies obtained in the previous section. 
In this limit we need to expand the parameters $(\mL, \nL)$ according to~\eqref{mLlargeL}-\eqref{nLlargeL}, see  discussion in Appendix~\ref{app:largeL}. We start by looking at the decoupled modes $i=1, 2,3,4$. In this case, the frequencies are known exactly, see  Appendix~\ref{app:Lame}, and we can use the expressions for the related densities \eqref{density-ads}, \eqref{density-s}, obtained in Appendix~\ref{app:Lame}.

Considering the transverse modes in S$^5$, expanding $\tilde\rho_\omega$ \eqref{density-s} according  to \eqref{mLlargeL}-\eqref{nLlargeL}, we obtain
\bea
\label{tilderho-largeL}
\tilde\rho_\omega =\frac{2\, \omega  \mathcal{L}}{\pi  \sqrt{\omega ^2-1}}+\frac{2\sqrt{\omega ^2-1}}{\pi  \omega }
+ e^{-2 \mathcal{L}-2} \frac{8 \cos\theta}{\pi  \sqrt{\omega ^2-1} }\left[
\frac{2 \omega }{\omega ^2-1}+\frac{ \frac{1}{\omega }-\frac{2}{\omega ^3}-\omega}{\mathcal{L}}
\right]
+ O(e^{-4 \mathcal L})\,,\qquad
\eea
and the extreme of integration $\tilde{\omega}_0$ \eqref{omega0s} approaches $\tilde{\omega}_0=1+ 8 e^{-2 \mathcal{L}-2} \cos\theta+O(e^{-4 \mathcal L})$. 
In the large $\mathcal L$ regime, the effective angle $\Theff$ \eqref{theffdef} approaches to 
\be
\Theff \to \log{16\over \eL}\,,
\ee
as it can be seen in \eqref{thetaeff-largeL}, taking into account the definition of~$\eL$ in~\eqref{epsilonL}. Then, in order to satisfy the consistency condition \eqref{s=s},  $m$ has to approach $-\infty$ for fixed $n\in (0,1)$, as amply discussed in App. \ref{app:largeL}. That is, 
\bea
\label{m-n-largeL}
m &=&-\frac{1}{\eL }+ \frac{1}{2} (1+\cos\theta-(\mathcal{L}+1) \cos\phi)
+\frac{\eL }{4}  (1-\cos \theta+(\mathcal{L}+1) \cos\phi)^2+\dots\,,
   \\\nn
n &=& 1-\frac{1}{2} \sin ^2\left(\frac{\phi }{2}\right)+\frac{\eL }{4}\sin ^2\left(\frac{\phi }{2}\right) 
\Big (2 \mathcal{L}+2-\cos\theta+(3 \mathcal{L}+2) \cos\phi\Big)+\dots\,.
\eea
Using the above expansion for the parameters $(n, m)$, we can then expand the density $\rho_\omega$ for the transverse AdS modes: 
\bea
\label{rho-largeL}
\rho_\omega &=&\frac{2\, \omega  \mathcal{L}}{\pi  \sqrt{\omega ^2-1}}+\frac{2\sqrt{\omega ^2-1}}{\pi  \omega }
\\ \nn
&-& \frac{8 e^{-2 \mathcal{L}-2}}{\pi \sqrt{\omega ^2-1}}  \Bigg[
\frac{\omega ^4-\omega ^2+2}{2 \omega ^3}+\frac{\omega  \left(\omega ^2-3\right) \mathcal{L}}{2 \left(\omega ^2-1\right)}
+\omega\cos\theta
-\Bigg(\frac{\omega ^4+\omega ^2-2}{2 \omega ^3}+\frac{\left(\omega ^3+\omega \right) \mathcal{L}}{2 \left(\omega ^2-1\right)}\Bigg)\cos\phi
\Bigg]
\\ \nn
   &+&O(e^{-4 \mathcal L})\,.
\eea
The lowest extreme of integration $\hat\omega_0$ \eqref{omega0ads-B} approaches to $\hat\omega_0=1
+\mathcal O(e^{-2\mathcal L})$, while $\omega_0$ defined in \eqref{omega0ads-A} for the region A$_L$ is exponentially suppressed.

It is interesting to notice that, at the leading order, there is no contribution from the internal and physical angles $\theta, \phi$ in $\tilde\rho_\omega$ \eqref{tilderho-largeL} and $ \rho_\omega$ \eqref{rho-largeL}, that is, the string does not see anymore the cusped Wilson loop as expected \cite{Correa:2012hh, Drukker:2012de}. Said in  other words, the two leading terms in  \eqref{tilderho-largeL} and \eqref{rho-largeL} agree with the result in~\cite{Giombi:2022anm} for the straight line. 
The integral \eqref{def-rho-density} gives us the relation between the normal mode frequencies $\omega_\ell$ and the quantum number. Hence, neglecting exponentially suppressed terms, the integral of $\tilde\rho_\omega$ \eqref{tilderho-largeL}   and $ \rho_\omega$ \eqref{rho-largeL} between one and $\omega_\ell$ result in 
\be\label{eq531GKO}
\ell= \frac{2}{\pi} (\mathcal{L}+1) \sqrt{\omega_\ell ^2-1}-\frac{2}{\pi} \arctan\left(\sqrt{\omega_\ell ^2-1}\right)+ O(e^{- 2\mathcal{L}})\,,
\ee
which is in agreement with eq. (5.31) in~\cite{Giombi:2022anm} once we take into account that $\mathcal L\to \mathcal J/4$.%
\footnote{Notice that for us $\ell=1, 2, \dots$, while in~\cite{Giombi:2022anm} the quantum number is an integer. The difference is due to the fact that we are imposing homogenous Dirichlet boundary conditions~\cite{Klebanov:2006jj}, see  Appendix~ \ref{app:Lame}.}
We stress that for the transverse modes in AdS the correct extreme of integration is $\hat\omega_0$ \eqref{omega0ads-B} which is approaching one. Integrating from $\omega_0\sim e^{- \mathcal{L}-1}$ \eqref{omega0ads-A} would lead to an imaginary integral, signaling an instability. This is simply saying that in the large $\mathcal L$ limit the correct region of phase space is the region B$_L$ as expected, see also fig. \ref{fig:Transitioncurve}.

Given the relation \eqref{eq531GKO}, it is possible to compute the various interesting regimes such as $\ell \ll \mathcal L, \ell \gg  \mathcal L\,, \ell \sim \mathcal L$. Since at the leading order there is no difference with the straight line, we refer the reader to~\cite{Giombi:2022anm}. 
We only mention here the BMN limit, obtained when $\ell\sim \mathcal L$, that is rescaling the quantum number $\ell=\eta\, \mathcal L$~\cite{Giombi:2022anm}, we have 
\be
\omega_\ell= \sqrt{1+\frac{\pi ^2 \eta ^2}{4}} + O\left({1\over \mathcal L}\right)\,, 
\ee
where $\frac{\pi \eta}{2}=\frac{\sqrt\lambda \ell}{2 L}$ as obtained in \cite{Drukker:2006xg}. 

All the above analysis has be done for the uncoupled frequencies. However, the WKB result for the remaining normal modes \eqref{omega58-semicl-main} is equivalent to take $\eta =\frac{\ell}{\mathcal L} \gg 1$. Indeed, once $s$ is expanded according to \eqref{tilderho-largeL}, we have
\be
\omega_\ell= {\pi \eta \over 2}+O\left({1\over \mathcal L}\right)\,= \frac{\sqrt\lambda \ell}{2 L}+O\left({1\over \mathcal L}\right)\,,
\ee
agreeing with the above findings at the leading order. 

In summary, we expanded the one-loop frequencies of the uncoupled transverse worldsheet fluctuations in the large $\mathcal L$ limit (discussed in Appendix~\ref{app:largeL}). In particular, when the $R$-charge $\mathcal L$ and the quantum number $\ell$ are comparable, this expansion reproduces the excitations of an open string in a pp-wave background~\cite{Drukker:2006xg}.%
\footnote{This is true also for the  semiclassical frequencies of the coupled worldsheet scalar fields, once we assume $\eta :\,=\frac{\ell}{\mathcal L} \gg 1$.}
From the worldsheet perspective, in this regime, the deepest penetration of the string in AdS, i.e. $r_{\rm min}$  \eqref{rmin}, is approaching zero as $r_{\rm min}\sim 4 e^{-1-\mathcal L} \sqrt{1-n}$, receding at the center of AdS, while reaches the north pole on the sphere $\tilde\rho_{\rm min}\sim 4 e^{-1-\mathcal L} \cos\theta/2$, with $\tilde\rho_{\rm min}$ defined in \eqref{def-rhotilde-min}.
In this limit, the string is effectively describing a very long open chain made of scalars $\rm{Tr}\,Z^L$~\cite{Drukker:2006xg}, namely a BMN-like vacuum, where there is no hint of the initial cusped Wilson loop, where the scalars were inserted, at the leading order. Only the exponentially suppressed terms in $\mathcal L$ encode information on the angles $\phi, \theta$, see equations \eqref{tilderho-largeL} and \eqref{rho-largeL}.



\section{The Cusp with $\mathbf{Z^L}$
  insertion: field theory aspects }
\label{sec:field} 

We discuss here the field theory interpretation of the results presented above. As reviewed in Section~\ref{sec:intro},  the generalized cusp anomalous $\Gamma_L(\lambda,\phi,\theta)$ defined by~\eqref{cuspdef} measures the dimension associated with a cusped Wilson line containing an insertion of a chiral primary operator $Z^L$, chosen to be orthogonal to the scalars already coupled to the Wilson line.
This observable is dual to the energy~\cite{Drukker:1999zq,Janik:2010gc,Bak:2011yy} (see also~\cite{Giombi:2021zfb,Giombi:2022anm}) of the corresponding string configuration in the holographic description.

There are two physically interesting limits of this quantity: the small-angle limit, where either the geometrical or the internal opening angle is small, and the antiparallel line limit. 
The former case, with  $\phi\sim0$ and $\theta\sim0$, corresponds to a small deformation of the 1/2-BPS straight line~\cite{DF}, and 
 defines a one-dimensional superconformal defect. A small deformation of the Wilson line in the physical (or internal) space can be interpreted as the insertion of a displacement (or tilt) operator~\cite{Billo:2016cpy}, integrated over the defect worldline. The leading order in the small-angle expansion of $\Gamma_L(\lambda,\phi,\theta)$ is thus equivalent to the expectation value of a correlator of displacement or tilt operators inserted along the straight line defect, and averaged over the defect. Such integrated correlators have recently attracted considerable attention~\cite{Cavaglia:2022yvv, Drukker:2022pxk, Gabai:2025zcs, Girault:2025kzt, Kong:2025sbk, Drukker:2025dfm}. 
When expanded at small angles,  the BPS limit result in~\eqref{EBPS} and the large-$\L$ expansion~\eqref{ElargeL}  impose nontrivial constraints on these integrated defect correlators at strong coupling.

The second limit of interest is the antiparallel line limit, corresponding to $\phi \rightarrow \pi$. 
As discussed in Section~\ref{sec:intro}, 
in this regime $\Gamma_L(\lambda,\phi,\theta)$  can be interpreted as the energy of a generalized quark–antiquark pair in an excited background carrying $R$-charge $L$. The limit $\phi \rightarrow\pi$, where the two Wilson lines approach each other, admits  an effective description in terms of defect fusion~\cite{Kravchuk:2024qoh,Cuomo:2024psk, Diatlyk:2024qpr}. Our results~\eqref{Einterm}  and~\eqref{EnergyB}  encode information about the Casimir energy (or potential) associated with the defect fusion, as well as the scaling dimension of the resulting fused conformal defect, in various regimes of the parameter $\mathcal{L}$.

\subsection{The small-angle limit and integrated correlators}

At $\phi=\theta=0$ and $L=0$, the cusped $W_L$~\eqref{WL} reduces to the straight $1/2$-BPS Wilson line $\mathcal{W}$ defined as\cite{Maldacena:1998im}
\begin{equation}\label{wilsondef}
    \mathcal{W}= \text{Tr}\, \left(\mathcal{P}\,e^{\int_{-\infty}^\infty d\tau\,(i A_0+\Phi_6)}\right)\,,
\end{equation}
where one takes the contour to lie along $x=(\tau, 0,\ldots,0)$ and $\vec{n}_{q}=\vec{n}_{\bar q}=(0,0,\ldots,1)$.
This line defect preserves an $\grSL(2,\mathbb{R})$ subgroup of the bulk conformal group, and is therefore a conformal line defect. Correlation functions of operators inserted along the Wilson line are defined as
\begin{equation}
    \langle\!\langle \mathcal{O}_1 ... \mathcal{O}_n \rangle\!\rangle = \frac{\langle \text{Tr} \,\mathcal{P}\left\{ \mathcal{O}_1 (\tau_1) \mathcal{W}(\tau_1,\tau_2) \mathcal{O}_2(\tau_2)\mathcal{W}(\tau_2,\tau_3) ... \mathcal{O}_n (\tau_n) \mathcal{W}(\tau_n,\tau_1)\right\}\rangle}{\langle \mathcal{W} \rangle}\,,
\end{equation}
where $\mathcal{W}(\tau_i,\tau_j)$ denotes the Wilson line segment \eqref{wilsondef} between points $\tau_i$ and $\tau_j$.
The presence of the line defect breaks translational invariance in the directions orthogonal to the line, an effect captured by the displacement operator $\mathbb{D}^i$ defined through the Ward identity\cite{Billo:2016cpy}
\begin{equation}\label{warddisp}
    \partial^\mu T_{\mu i}(\tau,x_{\perp}) \sim \delta^3\left(x_{\perp}\right) \mathbb{D}_i(\tau) \,, \qquad i=2,...,4 \,,
\end{equation}
where $T_{\mu i}(\tau,x_{\perp})$ is the stress-energy tensor and $x_{\perp}$ are the coordinates orthogonal to the line. The displacement operator has protected dimension $\Delta_{\mathbb{D}}=2$ and generates infinitesimal deformations of the Wilson line in the orthogonal directions. Explicitly, a deformation of the contour $x^i(\tau) \rightarrow x^i(\tau)+\delta x^i(\tau)$ induces the variation
\begin{equation}\label{disp}
    \delta \mathcal{W}=\mathrm{Tr}\left( \mathcal{P} \int d \tau \, \delta x^i(\tau) \, \mathbb{D}_i(\tau) \mathcal{W} \right)+O(\delta x^2)\,. 
\end{equation}
One can also consider deformations of the line in the internal space. In that case, the deformation is generated by the tilt operator $t_I$, which plays a role analogous to the displacement operator but acts on the internal $S^5$ directions. The tilt operator is associated with the breaking of the global $\grSO(6)$ $R$-symmetry to $\grSO(5)$ caused by the defect \eqref{wilsondef} and satisfies the Ward identity
\begin{equation}
    \partial_\mu J^\mu_I (\tau, x_\perp) \sim  \delta^3\left(x_{\perp}\right) t_I (\tau) \,, \quad I=1,...,5\,,
\end{equation}
where $J^\mu_I(\tau, x_\perp)$ is the  $SO(6)$ $R$-symmetry current and $\Delta_{t_I}=1$. This equation implies the analogue of~\eqref{disp},
\begin{equation}\label{tilt}
    \delta \mathcal{W}= \mathcal{P} \int d \tau \, \delta n^I(\tau) \, t_I(\tau) \mathcal{W} +O(\delta n^2)\,, \qquad I=1,...,5\,,
\end{equation}
where $\delta n^I(\tau)$ parametrizes an infinitesimal change of the coupling to the scalar fields.
In the case of the Wilson line \eqref{wilsondef}, the tilt operator has a particularly simple expression in terms of the fundamental scalars of the theory inserted on the defect, 
   that is $ t_I = \Phi_I$ for $I=1,\ldots,5$.

We can use the tilt and displacement operators to express the cusped Wilson line~\eqref{WL} at small physical or internal angles as a deformation of the straight line~\eqref{wilsondef}. This perspective naturally leads to an expansion of the generalized cusp anomalous dimension in terms of integrated correlators of defect operators \cite{Correa:2012at, Cavaglia:2022yvv}.

For simplicity, we consider a deformation in the internal space along a direction orthogonal to the polarization of the insertions, for instance $\delta n^I= \theta \, \delta_3^I$. 
Then the small-angle expansion allows us to relate   the generalized anomalous dimension to the integrated correlators of the displacement and tilt operators with the  insertions  of~$Z^L(0)$ and~$\bar{Z}^L(\infty)$ as in
\begin{eqnarray}\label{deformation}
   \G_L(\lambda,\phi,0)&\cong& \phi^2\int d \tau_1 d \tau_2 \, \tau_1 \tau_2 \frac{\langle\!\langle Z^L(0) \mathbb{D}_2\left(\tau_1\right) \mathbb{D}_2\left(\tau_2\right)\bar{Z}^L(\infty)\rangle\!\rangle_c}{\langle\!\langle Z^L(0) \bar{Z}^L(\infty)\rangle\!\rangle}+\text{perm.} +O\left(\phi^3\right)\,, \\
   \G_L(\lambda,0,\theta)&\cong& \theta^2\int d \tau_1 d \tau_2 \, \frac{\langle\!\langle Z^L(0) t_3\left(\tau_1\right) t_3\left(\tau_2\right)\bar{Z}^L(\infty)\rangle\!\rangle_c}{\langle\!\langle Z^L(0) \bar{Z}^L(\infty)\rangle\!\rangle}+\text{perm.} +O\left(\theta^3\right)\,.
\end{eqnarray}
Above, $\text{perm.}$ stands for contributions from cyclic permutations of the insertions and we have omitted correlators that vanish by symmetry%
\footnote{For example, the linear order term $
    \int d\tau \frac{\langle\!\langle Z^L(0)\, t_3(\tau)\,\bar{Z}^L(\infty)\rangle\!\rangle_c}{\langle\!\langle Z^L(0) \bar{Z}^L(\infty)\rangle\!\rangle}
$ vanishes because $t_3 = \Phi_3$ carries an $R$-symmetry polarization which is  by construction orthogonal to that of $Z^L$. The three-point function, whose structure involves scalar products of the $R$-symmetry polarizations, is thus identically zero.}. 
Integrated correlators of tilt and displacement operators with external insertions have recently been studied in~\cite{Gabai:2025zcs, Drukker:2025dfm}. These expansions provide a defect CFT interpretation of the small-angle limit of ~\eqref{EBPS} and~\eqref{ElargeL} in terms of integrated correlators of defect insertions.
For instance\footnote{We focus on the tilt correlators, since the tilt and displacement operators belong to the same supermultiplet and the tilt is the superprimary \cite{Agmon:2020pde}. Furthermore, we always consider the $\phi^2 \sim \theta^2$ limit in order to preserve supersymmetry.}, setting $\phi=0$, from~\eqref{BCdef} we can read the small-$\theta$ expansion of the cusp anomalous dimension,
\begin{equation}
    \G_L(\lambda,0,\theta) = \mathbf{B}_L \,\theta^2 -\left(\frac{\mathbf{B}_L}{12}-\frac{\mathbf{C}_L}{4}\right)\theta^4+\mathcal{O}(\theta^6)\,,
\end{equation}
where  $\mathbf{B}_L \equiv B_L (0)$ and $\mathbf{C}_L=C_L(0)$ are respectively the generalized Bremsstrahlung and  Curvature functions defined in~\eqref{BCdef}. Their explicit value at $\L<\!\!<1$ is given in \eqref{Brem}, \eqref{curvature} and \eqref{BCsmall}, while the $\L>\!\!>1$ result can be read from \eqref{Blarge1} and \eqref{Clarge1}. Comparing with \eqref{deformation}, we read
\begin{equation}\label{intcorr}
    \mathbf{B}_L \cong \int d \tau_1 d \tau_2 \, \frac{\langle\!\langle Z^L(0) t_3\left(\tau_1\right) t_3\left(\tau_2\right)\bar{Z}^L(\infty)\rangle\!\rangle_c}{\langle\!\langle Z^L(0) \bar{Z}^L(\infty)\rangle\!\rangle}+\text{perm.}\,.
\end{equation}
Conformal symmetry of the four-point function implies
\begin{equation}\label{4pt1d}
    \langle\!\langle \mathcal{O}(\tau_1)\mathcal{O}(\tau_2) \mathcal{O}(\tau_3) \mathcal{O}(\tau_4)\rangle\!\rangle = \langle\!\langle \mathcal{O}(\tau_1)\mathcal{O}(\tau_2)\rangle\!\rangle\langle\!\langle \mathcal{O}(\tau_3) \mathcal{O}(\tau_4)\rangle\!\rangle \, \mathcal{G}(z)\,, \quad z=\frac{\tau_{12} \tau_{34}}{\tau_{13} \tau_{24}}\,, \quad \tau_{ij}=\tau_i-\tau_j\,,
\end{equation}
where $\mathcal{G}(z)$ is function of the cross-ratio $z$.
By performing an appropriate conformal transformation, we can set $\tau_1 = 0$, $\tau_3 = 1$ and $\tau_4 = \infty$. In this frame, the cross-ratio reduces to $z=\tau_2$. With these simplifications, the integrated correlator \eqref{intcorr} takes the form
\begin{equation}
   \mathbf{B}_L \cong \int d z \frac{\langle\!\langle Z^L (0) t_{3}(z) t_{3}(1)  \bar{Z}^L(\infty)\rangle\!\rangle_c}{\langle\!\langle Z^L (0)  \bar{Z}^L(\infty)\rangle\!\rangle}+\text{perm.}\,.
\end{equation}
To regulate potential divergences, we follow the prescription of \cite{Friedan:2012hi} and use a hard-sphere regularization scheme, which amounts to replacing the domain of integration with
\begin{equation}
    \Sigma_{\varepsilon}\left(\tau_1, \tau_2\right)=\left\{\left(\tau_1, \tau_2\right)\left|\varepsilon^{-1}>\left|\tau_1\right|>\varepsilon, \varepsilon^{-1}>\left|\tau_2\right|>\varepsilon,\left|\tau_1-\tau_2\right|>\varepsilon\right\}\right.\,.
\end{equation}
The integrated correlator then simplifies to \cite{Kong:2025sbk}
\begin{eqnarray}\label{intreg}
\mathbf{B}_L &=& \int d \tau F_{\varepsilon}(\tau) \frac{\langle\!\langle Z^L (0) t_{3}(\tau) t_{3}(1)  \bar{Z}^L(\infty)\rangle\!\rangle_c}{\langle\!\langle Z^L (0)  \bar{Z}^L(\infty)\rangle\!\rangle}+\text{perm.} \nonumber \\
 &=& \int d \tau \log |\tau| \frac{\langle\!\langle Z^L (0) t_{3}(\tau) t_{3}(1)  \bar{Z}^L(\infty)\rangle\!\rangle_c}{\langle\!\langle Z^L (0)  \bar{Z}^L(\infty)\rangle\!\rangle} +\text{perm.}\,,
\end{eqnarray}
\label{renorR}
where one uses \cite{Friedan:2012hi}
\be 
    F_{\varepsilon}(\tau)=\left(\int_{ \Sigma_{\varepsilon}(x,x\tau)} -\int_{ \Sigma_{\varepsilon}(x\tau,x)} \right)\frac{dx}{x}\,, \qquad\qquad
   \lim\limits_{\varepsilon \rightarrow 0} F_{\varepsilon}(\tau)=\log|\tau|\,.
\ee
In principle, higher orders in the small-angle expansion of the near-BPS generalized cusp anomalous dimension~\eqref{EBPS} correspond to integrated higher-point\footnote{Higher orders in the small-angle expansion generate integrated correlators involving operators beyond the tilt, $t_I = \Phi_I$ with $I=1,\ldots5$, including, for example, $\Phi_6$.} defect correlators involving tilts/displacements and $Z^L$. For example, expanding the generalized cusp anomalous dimension~\eqref{EBPS} at order $\theta^4$, we obtain a constraint on an integrated six-point function. We can write it as
\begin{equation}
    -\frac{\mathbf{B}_L}{12}+\frac{\mathbf{C}_L}{4}\cong\int d\tau_{1}d\tau_{2}d\tau_{3}d\tau_{4} \,\frac{\langle\!\langle  Z^L (0) \, t_3 (\tau_1) t_3 (\tau_2) t_3 (\tau_3) t_3(\tau_4) \bar{Z}^L (\infty)\rangle\!\rangle_c}{\langle\!\langle Z^L (0) \bar{Z}^L (\infty) \rangle\!\rangle}+\text{perm.}\,, 
\end{equation}
where $\mathbf{C}_L$ is the curvature function defined in \eqref{BCdef}. In principle, one can reduce the six-point function to a function of three cross-ratios and then regularize the corresponding integral, similarly to what was done in \eqref{4pt1d} and \eqref{intreg} for the four-point function.

\subsection{The antiparallel lines limit and defect fusion}
We now turn to the interpretation of the $\phi \rightarrow\pi$ limit of the cusp, which corresponds to two antiparallel Wilson lines, in terms of defect fusion. When the two Wilson lines approach each other, they fuse and undergo an RG flow that terminates at a new conformal line defect. The dynamics on the fused defect are governed by an effective action of the form \cite{Kravchuk:2024qoh}
\begin{equation}\label{effus}
    S_{\mathrm{eff}}=\int d \tau \sqrt{\gamma}\Bigg(\lambda_1 \mathbf{1}+\sum_{\mathcal{O}} \lambda_{\mathcal{O}} \mathcal{O}\Bigg)\,, \quad \lambda_{\mathcal{O}} \sim (\pi-\phi)^{\Delta_{\mathcal{O}}-1} \,,
\end{equation}
where $\gamma$ is the induced metric on the defect and $\lambda_\mathcal{O}$ are the Wilson coefficients associated to the irrelevant operators $\mathcal{O}$ that live on the fused conformal defect. The cusp anomalous dimension encodes information about these coefficients. As discussed in \cite{Kravchuk:2024qoh, Cuomo:2024psk}, in this limit the energy or the cusp anomalous dimension admit an expansion of the type
\begin{equation}
    \Gamma_{L}(\lambda,\phi,\theta)=E=\frac{E_\text{Cas}}{\pi-\phi}+\Delta_{L}+b\,(\pi-\phi)+ O(\pi-\phi)^2\,,
\end{equation}
where each term can be interpreted as a (higher-derivative) correction to the Wilson coefficient of the identity or of an irrelevant operator. The divergent term represents the Casimir energy (or potential) that must be overcome to fuse the two line defects, and can be interpreted as arising from $\lambda_1$. The constant term corresponds to the scaling dimension of the defect-creating operator and in particular it vanishes precisely when the fused defect is trivial. Higher-order terms arise from higher-derivative corrections to the Wilson coefficient of the identity as well as from the coefficients of irrelevant operators generated by the RG flow on the fused defect. For example, the linear term is related to $\lambda_1$ as \cite{Kravchuk:2024qoh},
\begin{equation}\label{lambda1}
    \lambda_1=-\frac{E_\text{Cas}}{\pi-\phi}-3 b\,(\pi-\phi)+\cdots
\end{equation}
Thus, we see that our results~\eqref{Einterm}  and~\eqref{EnergyB} provide substantial information about the effective action of the fused defect. By expanding it to a sufficiently high order at $\phi \rightarrow\pi$, one can in principle extract all Wilson coefficients appearing in the defect effective action.

In contrast to the case without insertions~\cite{DF}, at strong coupling $g$, we find two qualitatively distinct behaviors of the cusp anomalous dimension depending on the value of $\mathcal{L}$.

For $\mathcal{L}<\mathcal{L}_c$, the energy of the string \eqref{Einterm} takes the form
\begin{equation}
    \frac{\Gamma_{L}(\lambda,\phi,\theta)}{4g}=\mathcal{E}^{(A)}=\frac{\mathcal{E}^{(A)}_{Cas}}{\pi-\phi}+b^{(A)}(\pi-\phi)+O((\pi-\phi)^2)\,,
\end{equation}
where $\mathcal{E}^{(A)}_{Cas}$ and $b^{(A)}$ are defined in~\eqref{cL}. The first term, proportional to $\mathcal{E}^{(A)}_{Cas}$, is the potential energy between the generalized quark-antiquark pair in the excited state generated by $Z^L$, or in other words,  the Casimir energy between the two line defects. The linear term gives a prediction for the  Wilson coefficient $\lambda_1$ at strong coupling, see \eqref{lambda1}. The absence of a constant term indicates that, in this regime, the fused defect is trivial (its creating operator is the identity). This is precisely the same behavior observed in the case without insertions. Physically, this suggests that the local insertion carries only a small charge and therefore has a negligible effect on the fusion process.

For $\mathcal{L}>\mathcal{L}_c$, the structure of the cusp anomalous dimension changes dramatically,
\begin{equation}
  \frac{\Gamma_{L}(\lambda,\phi,\theta)}{4g}= \mathcal{E}^{(B)}=\Delta_L^{(B)}+\beta\,(\pi-\phi)^2+O((\pi-\phi)^4)\,,
\end{equation}
with $\Delta^{(B)}$ defined in~\eqref{DeltaB}. In this regime, the constant term is non zero but the potential term is absent, suggesting that the large charge insertion screens the interaction between the two Wilson lines.

\bigskip
We can also extract information about the behavior of correlators of bulk operators (i.e. operators not inserted on the defect) in the presence of a circular\footnote{Since a circle can be mapped to a straight line by a conformal transformation, correlators of local operators on a circular defect coincide with those on a straight line, up to a change of variables.} Wilson loop from the structure of the potential energy. In particular, the value of the potential determines the asymptotic behavior of one-point functions of bulk operators with large scaling dimension in the presence of a circular Wilson loop~\cite{Kravchuk:2024qoh}. To see this, we need to consider the two-point function of two circular Wilson loops with opposite orientation, $\langle W  \bar {W}\rangle$, and expand it as an infinite sum of correlators of local bulk operators $\mathcal{O}_n$~\cite{Berenstein:1998ij,Arutyunov:2001hs,Gadde:2016fbj}.

If we take the first Wilson loop to lie on a unit circle in some plane, and describe the second loop by its radius $r$ and relative angular orientation $\vec{\theta}=(\theta_1,\theta_2)$, the two-point function can be written as 
\begin{equation}
    \langle \mathcal{W}  \bar{\mathcal{W}} \rangle=\left\langle \mathcal{W}\right| e^{\log r D-\vec{\theta} \cdot \vec{M}}\left|\mathcal{W}\right\rangle,
\end{equation}
where $D$ and $\vec{M}$ are the generators of dilatations and rotations. Following \cite{Kravchuk:2024qoh}, we insert a complete basis of bulk operators $\mathcal{O}_n$ and rewrite the two-point function as
\begin{equation}
   \langle \mathcal{W} \bar{\mathcal{W}}\rangle=\sum_n\left\langle \mathcal{W}\right|\mathcal{O}_n\rangle \langle O_n| e^{\log r D-\vec{\theta} \cdot \vec{M}}\left|\mathcal{W} \right\rangle= \sum_n\left|\langle \mathcal{W}| \mathcal{O}_n\rangle\right|^2 r^{\Delta_n} e^{-i  \vec{J}_n \cdot \vec{\theta}}\,,
\end{equation}
 where $\Delta_n$ and the $\vec{J}$ are the scaling dimensions and spins of $\mathcal{O}_n$,
 \begin{eqnarray}
D|\mathcal{O}_n\rangle & =&\Delta_n|\mathcal{O}_n\rangle\,, \\
\vec{M}|\mathcal{O}_n\rangle & =&i \vec{J}_n|\mathcal{O}_n\rangle.
 \end{eqnarray}
 Introducing the one-point function density $\r (\Delta,\vec{J})$, the equation above becomes
\begin{eqnarray}
    \langle \mathcal{W}  \bar{\mathcal{W}}\rangle&=&\int d\Delta d\vec{J}\,\r(\Delta,\vec{J})\, r^{\Delta}e^{-i \vec{J} \cdot \vec{\theta}}\,, \\
    \r (\Delta,\vec{J})&=& \sum_n |\langle \mathcal{W}|\mathcal O_n\rangle|^2 \delta(\Delta-\Delta_n)\delta(\vec{J}-\vec{J}_n)\,.
\end{eqnarray}
Comparing the above decomposition with the predictions of the effective theory \eqref{effus} in the regime where the two defects approach each other and fuse, $r \rightarrow1$, the authors of \cite{Kravchuk:2024qoh} obtained the asymptotic expansion of $\r (\Delta,\vec{J})$ at large $\Delta$. For scalars operators $\vec{J}=0$, they found \cite{Kravchuk:2024qoh}
\begin{equation}
    \r(\Delta,\vec{J}=0) \sim \Delta^{-3 / 4} \exp \left[2 \sqrt{2 \pi E_{\text{Cas}}}\, \Delta^{1 / 2}\right]\,, \quad \Delta >>1\,,
\end{equation}
where $E_{Cas}$ is the potential energy for the fusion of the two Wilson lines, which can be read from the $\phi \rightarrow\pi$ limit of the cusp anomalous dimension with $L=0$\footnote{The potential or Casimir energy for two circular loops is exactly the same as the one for two antiparallel lines \cite{Arutyunov:2001hs}. Intuitively, this follows from the fact that the Casimir energy is a short-distance effect that does not know about the global structure of the defects.}.

We can extend the analysis of~\cite{Kravchuk:2024qoh} to Wilson lines containing insertions of local operators $Z^L$. Repeating the same steps, the correlator of two such Wilson lines becomes
\begin{equation}
    \langle \mathcal{W}_L  \bar{\mathcal{W}}_L\rangle=\int d\Delta d\vec{J}\,\r_L(\Delta,\vec{J})\, r^{\Delta}e^{-i \vec{J} \cdot \vec{\theta}}
\end{equation}
where $\mathcal{W}_L$ here denotes the circular Wilson loop with insertions $Z^L$ and $\bar{Z}^L$ and
\begin{equation}
    \r_L (\Delta,\vec{J}) = \sum_n |\langle \mathcal{W}_L|\mathcal O_n\rangle|^2 \delta(\Delta-\Delta_n)\delta(\vec{J}-\vec{J}_n)\,.
\end{equation}
In terms of the original defect without insertions, the overlaps become the squares of defect three-point functions involving two defect operators and one bulk operator
\begin{equation}\label{1to3pt}
    |\langle \mathcal{W}_L|\mathcal O_n\rangle|^2 = |\langle\!\langle Z^L \bar{Z}^L\mathcal O_n\rangle\!\rangle|^2\,,
\end{equation}
From the analysis of~\cite{Kravchuk:2024qoh}, the large $\Delta$ asymptotic behavior of $\r_L (\Delta,\vec{J})$ for scalar operators is then
\begin{equation}\label{three-point-pred}
     \r_L(\Delta,\vec{J}=0) \sim \Delta^{-3 / 4} \exp \left[2 \sqrt{2 \pi \,E_{\text{Cas}}}\, \Delta^{1 / 2}\right]\,, \quad \Delta >>1\,,
\end{equation}
where $E_{\text{Cas}}\equiv -4 \,g\,\mathcal{E}^{(A)}_{\text{Cas}}$  in~\eqref{cL} if $\mathcal{L}<\mathcal{L}_c^\theta$. Thus, in this regime, our result for the potential energy leads to a prediction for the asymptotic spectrum of defect three-point functions involving heavy bulk operators in the presence of a straight Wilson line defect with $Z^L$ insertions. If $\mathcal{L}>\mathcal{L}_c^\theta$,  $E_{\text{Cas}}=0$ and the fusion limit is governed by subleading contributions in the effective action \eqref{effus}. Therefore, we expect that the density of defect three-point functions will exhibit a qualitatively different behavior for $\mathcal{L}>\mathcal{L}_c^\theta$.
\bigskip

\newpage
\appendix  
\section{Classical setup}
\label{app-classical}

For the reader convenience, we collect here a more detailed description of the classical string solution~\cite{Correa:2012at, Drukker:2006xg,GS} presented in Section~\ref{sec-classical}. In terms of  embedding coordinates, $\text{AdS}_5$ and $S^5$ satisfy respectively the hyperboloid and sphere constraints 
(here with unit radius)  
\begin{equation}
\begin{aligned}
 \text{AdS}_5:&\qquad -X_1^2-X_2^2+X_3^2+X_4^2+X_5^2+X_6^2= -1\,, \\
 S^5:&\qquad  Y_1^2+Y_2^2+Y_3^2+Y_4^2+Y_5^2+Y_6^2=  1\,.
\end{aligned}
\label{embedding}
\end{equation}
One of the parametrizations for $X_i$ and $Y_i$  used in this paper reads
\begin{equation}
\begin{aligned}
\!\!\!\text{AdS}_5\!:&X_1\!+\!i X_2\!= \!\!\cosh\!{\rho}\,e^{i t},~X_3\!+\!i X_4\!=\!  {\textstyle  \!\frac{2 x_1 }{x_1^2+y_1^2+1}}\sinh\!\rho\,  e^{i \varphi } , ~ X_5\!+\! i X_6\!= \!  \textstyle\frac{2 y_1 + i (1 - x_1^2 - y_1^2)}{1 + x_1^2 + y_1^2}\sinh\!\rho\,,\\
\!\!\!\!\!\!\textstyle S^5\!:&Y_1\!+i Y_2\!= \!\!\cos\psi\,e^{i\nu },~Y_3\!+i Y_4\!= \!
{\textstyle\frac{2 x_2 }{x_2^2+y_2^2+1}}\sin\psi \,e^{i \vartheta}, ~ Y_5\!+i Y_6\!=\!   {\textstyle\frac{2 y_2 + i (1 - x_2^2 - y_2^2)}{1 + x_2^2 + y_2^2}}\sin\psi\,, 
\end{aligned}
\label{parametrization1}
\end{equation}
with corresponding  metric $ds^2= (ds^2)_{\text{AdS}_5}+ (ds^2)_{S^5}$ given by
\begin{equation}
\begin{aligned}
 (ds^2)_{\text{AdS}_5}&=\textstyle-\cosh^2\!\rho\, \,dt^2+d\rho^2+\frac{4\sinh^2\rho}{(1+x_1^2+y_1^2)^2}\,(x_1^2 \,d\varphi^2+dx_1^2+dy_1^2)\,,
  \\
 (ds^2)_{S^5}&=\textstyle\cos^2\!\psi \, d\nu^2+d\psi^2
 +\frac{4\sin^2\psi}{(1+x_2^2+y_2^2)^2}\,(x_2^2 \,d\vartheta^2+dx_2^2+dy_2^2)\,.
\label{metric1}
\end{aligned}
\end{equation}
In $\text{AdS}_5$, we single out an $\text{AdS}_3$ subspace parametrized by the coordinates $(\rho,t,\varphi)$, with $\rho>0$, decompactified time $t \in (-\infty,\infty)$, and angular coordinate $\varphi \in [0,2\pi)$. Likewise, inside $S^5$ we select an $S^3$ spanned by $(\psi,\nu,\vartheta)$, with $\psi \in \left[0,\tfrac{\pi}{2}\right)$, $\nu \in [0,2\pi)$, and $\vartheta \in [0,2\pi)$. Real stereographic coordinates are adopted for the directions transverse to the $\text{AdS}_3\times S^3$ subspace probed by the classical string solution: $(x_1, y_1)$ for the two-sphere $S^2 \subset \text{AdS}_5$ and $(x_2, y_2)$ for the two-sphere $S^2 \subset S^5$. In these coordinates, the ansatz for the classical string solution is
\begin{equation}
\begin{aligned}
&t=\kappa\, \tau, \qquad \rho=\rho(\sigma),\qquad~ \varphi=\varphi(\sigma),\qquad x_1=1,\qquad y_1=0\,,\\
&\nu=\gamma\,\tau,\qquad \psi=\psi(\sigma)\,,\qquad \vartheta=\vartheta(\sigma),\qquad x_2=1,\qquad y_2=0\,. 
\end{aligned}
\label{ansatz2app}
\end{equation}

\noindent An alternative parametrization for the embedding coordinates is  
\begin{equation}
\begin{aligned}
\!\!\!\!\!\!\!\! \text{AdS}_5:&X_1+ X_2= e^{i t}\sqrt{1+r^2},
~~~~ X_3+i X_4= r \,e^{i\varphi}\cos\zeta_1,~~~~X_5\!+ i X_6= r \,\sin \zeta_1 \,e^{i \zeta_2} \,, \\
\!\!\!\!\!\!\!\!S^5:& Y_1+ Y_2=  e^{i\nu }\textstyle\small\sqrt{1-\tilde\rho^2},~~~~  Y_3\!+i Y_4= \tilde\rho\, e^{i \vartheta}\cos\chi_1,~~~~ Y_5+i Y_6=\tilde\rho\,\sin\chi_1\, e^{i \chi_2}\,, 
\end{aligned}
\label{parametrization2}
\end{equation}
with  metric
\begin{equation}
\begin{aligned}
\!\!\!\!\!\! (ds^2)_{\text{AdS}_5}&=- \textstyle\left(r^2+1\right)dt^2+\frac{dr^2}{r^2+1}+r^2 \big(\cos
   ^2 \zeta_1 \,{d\varphi }^2   + d\zeta_1^2 + \sin^2\zeta_1 d\zeta_2^2\big)  \,, \\
\!\!\!\!\!\! (ds^2)_{S^5}&= \textstyle\left(1-\tilde\rho^2\right) d\nu^2 +\frac{d\tilde\rho^2}{1-\tilde\rho^2} +   \tilde\rho^2 \big(\cos ^2 \chi_1 \,d\vartheta^2
+{d\chi_1 }^2 +\sin ^2\chi_1 d\chi_2^2\big)\,. 
\end{aligned}
\label{metric2}
\end{equation}
This is the choice adopted in~\cite{GS} and, to avoid confusion with the radial AdS coordinate $\rho$ in~\eqref{metric1}–\eqref{parametrization1}, we label the radial coordinate on $S^5$ as $\tilde\rho$. The AdS$_5$ radial coordinate is $r>0$, the time $t$ is decompactified, $t\in(-\infty,\infty)$, and the angular coordinates range as $\varphi\in[0,2\pi)$, $\zeta_1\in[0,\pi)$, and $\zeta_2\in[0,2\pi)$. On $S^5$ we take $0<\tilde\rho<1$, $\nu\in[0,2\pi)$, $\vartheta\in[0,2\pi)$, $\chi_1\in[0,\pi)$, and $\chi_2\in[0,2\pi)$. The ansatz for the classical solution is again $t=\kappa\, \tau,~r=r(\sigma), ~\varphi=\varphi(\sigma), ~\nu=\gamma\,\tau,~\tilde\rho=\tilde\rho(\sigma),~\vartheta=\vartheta(\sigma)$ with $\zeta_1=\zeta_2=\chi_1=\chi_2=0$.
The metrics \eqref{metric1}-\eqref{metric2} are related, for the relevant $\text{AdS}_3\times S^3$ subspace,  via  $\sinh \rho = r$, $\sin \psi = \tilde{\rho}$.
The induced metric on the worldsheet $\Sigma$ reads
\be\label{induced}
\begin{split}
ds^2_\Sigma&\equiv\partial_\alpha X^\mu \partial_\beta X^\nu\,G_{\mu\nu}\equiv h_{\alpha\beta}\,d\sigma^\alpha d\sigma^\beta\\
&=-[\kappa^2-\gamma^2+\kappa^2 r^2  +\gamma^2 \tilde\rho^2] d\tau^2+\Big[\!\frac{r^{\prime 2} }{1+r^2 }+r^2 \varphi^{\prime 2}+\frac{\tilde\rho^{\prime 2}}{1-\tilde\rho^2}+\tilde\rho^2\vartheta^{\prime 2} 
 \!\Big] d\sigma^2\,. 
 \end{split}
\ee
When Virasoro constraints are satisfied, $h_{\alpha\beta}$ becomes conformally the flat one~\eqref{inducedflat}.  
The metric~\eqref{metric2} admits 15 Killing vectors in AdS$_5$ and 15 in $S^5$, of which the (commuting) ones associated with the cyclic coordinates for this problem are 
$K_t=\partial_t$, $K_\varphi=\partial_\varphi$, $K_{\zeta_2}=\partial_{\zeta_2}$, $K_\nu=\partial_\nu$,  $K_\vartheta=\partial_\theta$, $K_{\chi_2}=\partial_{\chi_2}$.
The corresponding conserved currents are the conjugate momenta
$
\Pi_\mu^\alpha = 2 g \,\eta^{\alpha\beta} G_{\mu\nu} \,\partial_\beta X^\nu \,,
$
with $\mu = t,\varphi,\zeta_2,\nu,\vartheta,\chi_2$. For $K_{\zeta_2}$ and $K_{\chi_2}$ the associated currents are trivial (they vanish identically). In contrast, for $K_\varphi$ (in AdS$_5$) and $K_\vartheta$ (in $S^5$), current conservation is equivalent to requiring that the quantities

\begin{equation}
\Pi_\varphi^\sigma = 2 g   r^2 \varphi'(\sigma)\,,\qquad\qquad
\Pi_\vartheta^\sigma = 2 g \rho^2 \vartheta'(\sigma)\,
\end{equation}
are $\sigma-$independent.
The invariance under global time translations in AdS$_5$ ($K_t$) and rotations along the $\nu$ direction in $S^5$ ($K_\nu$) implies the existence of two conserved charges in the usual sense, the energy $E$ and the angular momentum $L$:
\begin{eqnarray}\label{Energy2}
E &=& \int_{-s/2}^{s/2} d\sigma  \,\Pi_t^0 = 2 g \kappa \int_{-s/2}^{s/2} d\sigma (1 + r^2(\sigma))\,,
\\\label{Lconserved2}
L&=& - \int_{-s/2}^{s/2} d\sigma \, \Pi_\nu^0 = 2 g \gamma \int_{-s/2}^{s/2} d\sigma (1 - \tilde\rho^2(\sigma))\,.
\end{eqnarray}
Their values can be computed explicitly, as in~\eqref{E} and~\eqref{L}, by rewriting the integrals in terms of the radial coordinates using $d\sigma = dr/r'$ and $d\sigma = d\tilde\rho/\tilde\rho'$, and then employing the equations of motion~\eqref{req},~\eqref{rhoeq} to eliminate the dependence on $r'$ and $\tilde\rho'$. In terms of the auxiliary parameters introduced in~\eqref{nm} and~\eqref{nLmL}, the radial derivatives are given by
\begin{eqnarray}\label{rprime}
r^\prime&=&\pm
\frac{\sqrt{1+r^2} \sqrt{r^2(n-m)+n} \sqrt{
   r^2(n-m)-1+n}}{r \sqrt{n^2-m}\sqrt{1-n-m}}\,,\\\label{rhotildeprime}
\tilde\rho^\prime&=&\pm
\frac{\sqrt{1-\tilde\rho^2} \sqrt{\tilde\rho^2+n_{\text{\tiny $L$}}-1} \sqrt{ \tilde\rho^2m_{\text{\tiny $L$}}
   +n_{\text{\tiny $L$}}-m_{\text{\tiny $L$}}}}{r \sqrt{m_{\text{\tiny $L$}}  n_{\text{\tiny $L$}} +n_{\text{\tiny $L$}}-m_{\text{\tiny $L$}}}}.
\end{eqnarray}
We can easily solve \eqref{rprime} and \eqref{rhotildeprime}  in terms of trigonometric elliptic functions:
\begin{eqnarray}\label{rsol}
r(\sigma)&=&\sqrt{\frac{1-m}{(n-m) \text{cn}\left(\left.\frac{\sigma
   }{\sqrt{1-m-n}}\right|m\right)^2}-1}\,,
   \\\label{rhotildesol}
  \tilde\rho(\sigma)&=& \sqrt{1-\nL\, \text{cd}\left(\left.\frac{\sqrt{\nL}
   \sigma }{\sqrt{\mL
   (\nL-1)+\nL}}\right|\mL\right)^2}\,.
\end{eqnarray}
  
\subsection{Relation to parameters used in literature}
\label{app:notation}

In this appendix we summarize the various convention used in literature, some of the expressions below appear also in the main body, however  for reader's convenience we collect all the formulae here. In this paper we use the classical string solution of~\cite{GS}. The parameters used there to describe the phase space of the classical string configuration are related to our parameters according to the following maps for AdS$_5$ (cf. section \ref{sec:classical-ads}):
\begin{subequations}
\bea
\text{Region A$_L$}: &\quad  \kappa^2 = {n-m \over 1-n-m}\,, \qquad \ell_\varphi^2= {n(1-n)\over (n-m)(1-m-n)}\,,
\\ 
\text{Region B$_L$}: & \quad \kappa^2 ={ n'\over 1+ m' - n'}\,, \qquad \ell_\varphi^2= {(n'- m' )(1-n')\over n'(1+m' -n')}\,,
\eea
\end{subequations}
where the two sets of parameters are given by
\begin{subequations}
\bea
\text{Region A$_L$}: & \quad 1\ge n \ge 0\,, \qquad {1\over 2} \ge m\ge 0\,, \qquad 1\ge n\ge m\ge 0\,, 
\\ 
\text{Region B$_L$}: & \quad 1\ge n^\prime \ge 0\,, \qquad {1} \ge m^\prime\ge 0\,, \qquad 1\ge n^\prime\ge m^\prime\ge 0\,. 
\eea 
\end{subequations}
As explained in section \ref{sec:classical-ads}, we can cover the two regions A$_L$ and B$_L$ simply using $(n, m)$ and extending $m$ to negative values. Sometimes for convenience we use the set $(n^\prime, m^\prime)$  \eqref{mprimenprime} to have a compact phase space. The modular parameters are related by 
\be\label{mprime-nprime-app}
n^\prime=\frac{n-m}{1-m}\,, \qquad  m^\prime=\frac{m}{m-1}\, .
\ee
For S$^5$ (cf. section \ref{sec:classical-s}), the parameters used in \cite{GS} are related to our $(\nL, \mL)$ by 
\bea
 \gamma^2 = \frac{m_{\text{\tiny L}}}{n_{\text{\tiny L}}-m_{\text{\tiny L}}(1-n_{\text{\tiny L}})}\,, \qquad 
 \ell_\vartheta^2= \frac{(1-n_{\text{\tiny L}})(n_{\text{\tiny L}}-m_{\text{\tiny L}})}{n_{\text{\tiny L}}-m_{\text{\tiny L}}(1-n_{\text{\tiny L}})}\,.
\eea
Moreover,  in \cite{GS} the angular momenta $ \ell_\varphi$ and $\ell_\vartheta$ \eqref{conservedmomenta} are related to the above parameters with 
 \be
 \ell_\varphi^2= {k_\phi^4-(1-\kappa^2)^2\over 4\kappa^2}\,, \qquad
  \ell_\vartheta^2= {k_\theta^4-(1-\gamma^2)^2\over 4\gamma^2}\,,
 \ee
and $k_\phi$ and $k_\theta$ in \cite{GS} in terms of our notation read
\be
k_\phi^4= \Delta, \qquad k_\theta^4= \Delta_L, \qquad 
\ee
where $\Delta$ and $\Delta_L$ are defined in \eqref{nm} and \eqref{nLmL} respectively.

\subsubsection{The straight line limit}

The solution for $L$ orthogonal scalars inserted on the straight line, that is $\theta=\phi=0$ has been discussed in \cite{Drukker:2006xg, Giombi:2021zfb, Giombi:2022anm}. 
In our set-up, this corresponds to
\bea
\label{straight-line-lim}
\ell_\vartheta=\ell_\varphi=0\,, \qquad \kappa=\gamma >1\,,  
\eea
or equivalently 
\be
n=1\,, \quad n_L= m_L\,, \quad m={m_L\over m_L-1} \qquad \Rightarrow \qquad  \kappa=\gamma={1\over\sqrt{m_L}} >1\,, 
\ee
since $m_L\in (0, 1)$, see  equation~\eqref{rangenLkL}. 
In particular, 
In this limit in \cite{Giombi:2021zfb, Giombi:2022anm} a parameter $c$ is introduced to parametrize the minimal radius in S$^3$ reached by the string at the center of the worldsheet $\s\to 0$. In our notation $\kappa=\gamma={1\over\sqrt{m_L}} ={1\over c}$, and for $\mL=c\to 1$  the solutions of \cite{Drukker:2006xg} are recovered.

  \subsubsection{The generalized cusp limit}
 
In \cite{DF}, the solution is time independent on the compact space, and the string spans $\text{AdS}_3\times S_1$. This indeed corresponds to set $L=0$ while keeping the two opening angles $(\phi, \theta)$. Looking at the classical configuration \eqref{ansatz2}, we should set 
$$
\psi= {\pi\over 2} \,, ~~\text{or equivalently} ~~ \tilde\rho=1\,, \qquad \kappa \tau \to t\,.
$$
In \cite{DF}  the two opening angles $(\phi,\theta)$ of the dual operator enter the elliptic functions of the classical solution via two parameters,   $(p,q)$, or equivalently $(b,k)$. The relation between these two equivalent sets of parameters are~\cite{DF}
  \begin{equation}
\begin{aligned}
p^{2} &= \frac{b^{4}(1 - k^{2})}{b^{2} + k^{2}}, 
&\qquad 
q^{2} &= \frac{b^{2}(1 - 2k^{2} - k^{2}b^{2})}{b^{2} + k^{2}}\,,\\[6pt]
k^{2} &= \frac{b^{2}(b^{2} - p^{2})}{b^{4} + p^{2}}, 
&\qquad 
b^{2} &= \frac{1}{2} \left( p^{2} - q^{2} + \sqrt{(p^{2} - q^{2})^{2} + 4p^{2}} \right).
\end{aligned}
\end{equation}
From the discussion around equation \eqref{xi},  in order to have $L=0$, we need to impose 
\be\label{DF-lim1}
\ell_\vartheta^2=1\,, ~~\text{or equivalently} ~~ \nL=0\,.
\ee
The relations among the parameters introduced in \cite{GS} and \cite{DF} read then ($\kappa\to \pm 1/J$)
\begin{subequations}
\bea
\label{pq}
&& p=-\frac{\kappa}{\ell_\varphi}\,,\qquad  q=\frac{1}{\ell_\varphi}\,,
\\ \label{DF-lim2}
&& \kappa^2=\frac{b^2 p^2}{\left(b^2+1\right) p^2-b^4}\,, \qquad \ell_\varphi= {1\over q}=\frac{b}{\sqrt{\left(b^2+1\right) p^2-b^4}}\,.
\eea
\end{subequations}
For notational economy we work with the modular parameters~\eqref{nm}, related to~\cite{DF} via 
 \be\label{nmtobp}
n= \frac{b^4}{b^4+p^2}\,,\qquad \qquad m = k^2\,. 
 \ee
Notice that this identification is meaningful only in the so-called Region A$_L$, as explained in section \ref{sec:angle-s5}. 
It is also worth noticing that we need to use the relation $s_{\text{AdS}}=s_{S}$ in this limit, see  \eqref{L0}, in order to remove $\gamma$, since $\gamma$ is not constrained by the ansatz in~\cite{DF}. 


\section{Energy expansions in different regimes}
\label{app:B}
\subsection{The antiparallel limit}
In the case {\bf 1} ($n\to m$), which can only be realized in the Region~{\bf$A_L$},
the first step to obtain the expansion of the energy in the antiparallel lines limit is to write the factor $\sqrt{n-m}$ in~\eqref{E} as an expansion in $\pi-\phi$ via equation~\eqref{phi01}. Using~\eqref{Picompact}, one first write
\begin{align}\label{piminusphi}
\!\!\!
\pi\! -\! \phi =  \frac{2 \left(\mathds{E}\! -\! \left(1-m\right) \mathds{K}\right)}{\sqrt{m(1-m)}} \sqrt{n-m}  - \frac{\left(\left(1-2m\right) \mathds{E}\! -\! \left(1\!-\!m\right) \mathds{K}\right)}{3 \left(m(1-m)\right)^{\frac{3}{2}}} \left(n-m\right)^{\frac{3}{2}}\! +\!O((n\!-\!m)^2),
\end{align}
where \( \mathds{K} \equiv K(m) \), \( \mathds{E} \equiv E(m) \).  Inverting, we get the desired expansion:
\begin{equation}
\label{nminusm}
\!\!\!
\sqrt{n\!-\!m} = \frac{\sqrt{m(1\!-\!m)} \left(\pi\! -\! \phi\right)}{2 \left(\mathds{E} \!-\! \left(1\!-\!m\right) \mathds{K}\right)} 
+ \frac{\sqrt{m(1\!-\!m)} \left(\left(1\!-\!2m\right) \mathds{E} \!-\! \left(1\!-\!m\right) \mathds{K}\right)}{48 \left(\mathds{E}\! -\! \left(1\!-\!m\right) \mathds{K}\right)^4} \left(\pi\! -\! \phi\right)^3 \!+\! O( (\pi\!-\!\phi)^5)\,,
\end{equation}
and uses it in~\eqref{E} to obtain~\eqref{Einterm}
\begin{equation}
\label{energyA}
\!\!\!\mathcal{E} = -\frac{1}{ \left(\pi - \phi\right)}\frac{2 \left(\mathds{E} - \left(1-m\right) \mathds{K}\right)^2}{\sqrt{m\left(1-m\right)}} + \frac{\left(\pi - \phi\right) \left(\left(m-1\right) \mathds{K} + \left(1-2m\right) \mathds{E}\right)}{12 \sqrt{m \left(1-m\right)} \left(\mathds{E} - \left(1-m\right) \mathds{K}\right)} + O(\pi - \phi)^3)\,.
\end{equation}
As a second step, one needs then to express $m$ in terms of a solution of the consistency condition~\eqref{thetaeff} in the same $n\to m$ limit. This is done assuming  the expansion  $\displaystyle m =\mbar+a_1 (\pi-\phi)+a_2 (\pi-\phi)^2+ O((\pi-\phi)^3)$, with~$\mbar$ solution of~\eqref{consistencybarA}, finding 
\be\label{maroundmbar}
m=\mbar+\frac{\mbar^2 \left(1-\mbar\right)^2
\overline{  \mathds{K}}  \left(\pi -\phi\right)^2}{4 \left(\overline{  \mathds{E}}-\left(1-\mbar\right)
\overline{  \mathds{K}} \right){}^2 \left(\left(1-2 \mbar\right)
   \overline{  \mathds{E}} -\left(1-\mbar\right)
\overline{  \mathds{K}}\right)}+O(\left(\pi -\phi\right)^4)\,,
\ee
where $\overline{  \mathds{E}} \equiv  E(\mbar)$ and $\overline{\mathds{K}} \equiv K(\mbar)$.  
Plugging this into~\eqref{energyA} one obtains~\eqref{Einterm}. The dependence on the angular momentum $\mathcal{L}$ is encoded in $\mbar$. In the small–$\mathcal{L}$ regime, this dependence can be made explicit by expressing it as $O(\mathcal{L})$ corrections to the results of~\cite{Cuomo:2024psk}, expanding $\mbar$ around $\mbar_0$, where $\mbar_0$ is determined by the $\mathcal{L}\to 0$ limit~\eqref{consistency0} of~\eqref{consistencybarA}. As explained in the main body, the limit $\L \to 0$ corresponds to taking $\nL \to 0,  \mL \to 0$, with their ratio  finite and equal to~\eqref{xi}.  
Assuming the expansions $\mL = \sum_{j=1}^\infty a_j \mathcal{L}^j$ and $\nL = \sum_{j=1}^\infty b_j \mathcal{L}^j$, one can solve Eqs.~\eqref{theta} and~\eqref{L} order by order in $\mathcal{L}$, obtaining
\begin{align}
\label{nLmLexp}
  \nL=& \frac{4\,\mathcal{L}}{\sqrt{\pi^2-\theta^2}}\!+\! \frac{2\bigl(\theta^2\!-\!3\pi^2\bigr)\mathcal{L}^2}{\pi^2(\pi^2-\theta^2)}
  \!+\! \frac{\bigl(2\theta^4\!-\!7\pi^2\theta^2\!+\!3\pi^4\bigr)\mathcal{L}^3}{\pi^4\bigl(\pi^2-\theta^2\bigr)^{3/2}}\!+\! \frac{\bigl(5\pi^4\!+\!18\pi^2\theta^2\!-\!7\theta^4\bigr)\mathcal{L}^4}{4\pi^6\bigl(\pi^2-\theta^2\bigr)}+O(\mathcal{L}^5)\,,
\notag\\
\mL=& \frac{4\mathcal{L}\sqrt{\pi^2-\theta^2}}
        {\pi^2}\
   - \frac{2\left(3\pi^2 - 5\theta^2\right)\mathcal{L}^2}{\pi^4} + \frac{\left(20 \theta ^4-17 \pi ^2 \theta ^2+3 \pi ^4\right) \mathcal{L}^3}{4 \pi ^6
   \sqrt{\pi ^2-\theta ^2}}\\
  &+ \frac{\left(139\theta^6 - 105\pi^2\theta^4 - 7\pi^4\theta^2 + 5\pi^6\right)\mathcal{L}^4}
         {4\pi^8\!\left(\pi^2 - \theta^2\right)}+O(\mathcal{L}^5)\notag\,,
   \end{align}
from which 
\begin{align}\label{theffsmallL}
\Theff=&\theta +\frac{2 \sqrt{\pi ^2-\theta ^2} \mathcal{L}}{\theta }+\frac{\left(3 \theta ^2-2 \pi ^2\right) \mathcal{L}^2}{\theta ^3}+\frac{\left(\theta ^6+13 \pi ^2 \theta ^4-20 \pi ^4 \theta
   ^2+8 \pi ^6\right) \mathcal{L}^3}{2 \pi ^2 \theta ^5 \sqrt{\pi ^2-\theta ^2}}+\nonumber\\
   &+\frac{\left(3 \theta ^{10}+12 \pi ^2 \theta ^8+133 \pi ^4 \theta ^6-348 \pi ^6 \theta ^4+288 \pi ^8 \theta
   ^2-80 \pi ^{10}\right) \mathcal{L}^4}{8 \pi ^4 \theta ^7 \left(\pi ^2-\theta ^2\right)}+O(\mathcal{L}^5)\,.
\end{align}
Using this result in~\eqref{consistencybarA}, one obtains the power series in $\mathcal{L}$ for the modular parameter $\mbar$ in the antiparallel lines limit
\begin{align}\label{mbararoundm0}
\mbar\!=&\mbar_0\!-\!\frac{2 \mbar_0(1-\mbar_0) \sqrt{1-2 \mbar_0} \sqrt{\pi^2-\theta^2}\mathcal{L}}{ \left((1-\mbar_0) \mathds{K}_0+\left(2 \mbar_0-1\right)
 \mathds{E}_0\right)\theta}
 \!+\!\frac{\mbar_0 \left(1-\mbar_0\right)\,(\pi-\theta^2)^{3/2}\, \mathcal{L}^2 }{\left(\left(\mbar_0-1\right)  \mathds{K}_0+\left(1-2 \mbar_0\right)
 \mathds{E}_0\right)^3\theta^3\sqrt{\pi^2-\theta^2}}\notag\\
 &\times \Biggl[\left(\mbar_0-1\right)  \mathds{K}_0 \Big(\sqrt{1-2 \mbar_0}
   \left(\mbar_0-1\right) \frac{ \left(3 \theta^2-2 \pi ^2\right)}{\pi ^2-\theta^2} \mathds{K}_0+ \left(4 k_0^4-7
   \mbar_0+2\right)  \theta\Big)-\notag\\ &\left.-2 \left(2 \mbar_0-1\right) \mathds{E}_0
   \left(\sqrt{1-2 \mbar_0} \left(\mbar_0-1\right) \frac{ \left(3 \theta^2-2 \pi ^2\right)}{\pi ^2-\theta^2}
    \mathds{K}_0+ \left(2 \mbar_0^2-2 \mbar_0+1\right) \sqrt{\xi -1}\right)\right.\notag\\ &\left.+\left(1-2
   \mbar_0\right){}^{5/2} \frac{\left(3 \theta^2-2 \pi ^2\right)}{\pi ^2-\theta^2}  \mathds{E}_0{}^2\right]+O\left(\mathcal{L}^3\right)\,,
 \end{align}
 where $\mathds{E}_0\equiv  E(\mbar_0)$ and $\mathds{K}_0\equiv K(\mbar_0)$. 
 The expansion above is used  to obtain from~\eqref{Einterm} the expressions~\eqref{cL}.
 
 The antiparallel line limit of the energy can be also obtained in  Region~{\bf$B_L$}. One considers the $n\to0$ limit and negative values of $m$. 
 Proceeding in a very similar way, here one uses that
 \be\label{nexp}
 \frac{\sqrt{n}}{\sqrt{\!-m}}=\frac{(\pi-\phi) }{2 (\mathds{E} -\mathds{K} )}+\frac{(m-1) \mathds{K} +(m+1) \mathds{E}}{48
   (E(m)-K(m))^4}  (\pi-\phi)^3+O ((\pi-\phi)^3)\,,
 \ee
 and obtains
 \be\label{EnergyintermB}
\mathcal{E}= \frac{(1-m) \mathds{K}-\mathds{E}}{\sqrt{-m}}-\frac{(1-m)
   \mathds{K}-\mathds{E}}{8 \sqrt{-m} (\mathds{E}-\mathds{K})^2}\,(\pi-\phi)^2+O((\pi-\phi)^4)~.
 \ee
The modulus $m$ is then expressed in terms of a solution of the consistency condition~\eqref{thetaeff} in the same $n\to0$ limit, equation~\eqref{consistencyhatB}, finding 
\be\label{maroundmhat}
m=\widehat{m}+\frac{\widehat{m}^2\,\widehat{K}}{4 (\widehat{K}-\widehat{E})^3}\left(\pi -\phi\right){}^2+O\left(\pi -\phi\right){}^4\,,
\ee
where $\widehat{K}=K(\widehat m)$, $\widehat{E}=E(\widehat m)$ and $\widehat{m}$ solves~\eqref{consistencyhatB}. Plugging this into~\eqref{EnergyintermB}, one obtains~\eqref{EnergyB}. The dependence on $\mathcal{L}$ is again encoded in the parameter $\widehat m$. In this region, however, the standard small-$\mathcal{L}$ expansion is not meaningful, since the region collapses to a point in the limit $\mathcal{L} \to 0$.

\subsubsection{Expansion around the critical point $\L_c^\theta$}

We remain in the antiparallel lines limit and  explore the behavior of the energy in a neighborhood of $\mathcal{L}_c$. To obtain its expansion  around the transition point $\mathcal{L}_c$, we keep $\theta$ fixed and expand in the small parameter $\mathcal{L} - \mathcal{L}_c$. The parameters $\mL$ and $\nL$ must then solve
  \be\label{aroundcr}
  \mathcal{L}=\frac{\sqrt{\nL}}{\sqrt{\mL}}(K(\mL)-E(\mL))\,,
\qquad\mathrm{and}\qquad
  \theta=2 \sqrt{\frac{\left(1-\nL\right) \left(\nL-\mL\right)}{\nL}}\Pi \left(\left. \nL\right|\mL\right)\,.
   \ee
At the lowest order in $\mathcal{L}-\mathcal{L}_c$, the solution is~\footnote{One determines $b_1$ in terms of $a_1$ by imposing that the width of the cusp is still $\theta$ at the first order in the expansion in $\mathcal{L}-\mathcal{L}_c$, $a_1$ is then fixed by solving the first equation in~\eqref{aroundcr}.}
   \be
   \begin{split}
   \nL=\nL^c+&a_1 (\mathcal{L}-\mathcal{L}_c)+O( (\mathcal{L}-\mathcal{L}_c)^2)\,,\qquad
  \mL=\mL^c+b_1 (\mathcal{L}-\mathcal{L}_c)+O( (\mathcal{L}-\mathcal{L}_c)^2)\,,\\ 
a_1=&\frac{2{\mL^c}^{3/2}  (1-{\mL^c}) (1-{\nL^c}) \sqrt{{\nL^c}} E({\mL^c})}{({\nL^c}-{\mL^c})
   ((1-{\mL^c}) K({\mL^c})-E({\mL^c}))^2+{\mL^c}^2 (1-{\nL^c}) E({\mL^c})^2}\,, \\
b_1=& \frac{2 (1-{\mL^c})^2 {\mL^c}^{3/2} (K({\mL^c}) ({\mL^c}-{\nL^c})+{\nL^c}
   E({\mL^c}))}{\sqrt{{\nL^c}} \left(({\nL^c}-{\mL^c}) ((1-{\mL^c})
   K({\mL^c})-E({\mL^c}))^2+{\mL^c}^2 (1-{\nL^c}) E({\mL^c})^2\right)}\,,
\end{split}
   \ee
   where $\mL^c$ and $\nL^c$ solve \eqref{mncritical}.
 Then the expansion of $\Theff(\theta, \mathcal{L})$ around $\mathcal{L}=\mathcal{L}_c$ reads
  \be
  \label{ThetaLc}
\Theff(\theta, \mathcal{L})=\pi-\frac{2\sqrt{\mL^c} (\mathcal{L}_c-\mathcal{L}) }{\sqrt{\nL^c} \sqrt{\mL^c\left(1-\frac{1}{\nL^c}\right)+1}}+O((\mathcal{L}_c-\mathcal{L})^2)\,.
  \ee
The final step is to find the $\text{AdS}_5$ modulus $\overline{m}_c$ satisfying the consistency condition~\eqref{consistencybarA} around the critical point. We must solve 
 \be
 2 \sqrt{1 - 2 \overline{m}_c} K\left(\overline{m}_c\right)=\Theff(\theta, \mathcal{L})=\pi-\frac{2\sqrt{\mL^c} (\mathcal{L}_c-\mathcal{L}) }{\sqrt{\nL^c} 
 \sqrt{\mL^c \left(1-\frac{1}{\nL^c}\right)+1}}+O((\mathcal{L}_c-\mathcal{L})^2).
 \ee
 The solution  is 
 \be\label{mcriticalA}
\overline{m}_c= \frac{8 \sqrt{\mL^c }(\mathcal{L}_c-\mathcal{L})}{3 \pi  \sqrt{-\mL^c +\mL^c  \nL^c+\nL^c}}+O((\mathcal{L}-\mathcal{L}_c)^2)\,,
 \ee
 such that the coefficient  $\mathcal{E}^{(A)}_\text{Cas}(\theta,\mathcal{L})$ of the pole in $\pi-\phi$ scales, next to $\mathcal{L}_c$, as in~\eqref{EAcritical}. 
 
 Above the transition, the analysis in $S^5$ is unvaried and equation~\eqref{ThetaLc} still holds.  However, this time we have to solve~\eqref{consistencyhatB} to determine the appropriate  $\widehat{m}_c$, namely
\be
2 \sqrt{1-\widehat{m}_c}\,\mathds{K}(\widehat{m}_c)\,=\pi+\frac{2\sqrt{\mL^c} (\mathcal{L}-\mathcal{L}_c) }{\sqrt{\nL^c} \sqrt{\mL^c\left(1-\frac{1}{\nL^c}\right)+1}}+O\left((\mathcal{L}-\mathcal{L}_c)^2\right)\,.
\ee
At the lowest order, this equation is solved by
\be\label{mcriticalB}
\widehat m_c=-\frac{8 \sqrt{\mL^c}}{\pi  \sqrt{-\mL^c+\mL^c \nL^c+\nL^c}} (\mathcal{L}-\mathcal{L}_c)+O\left((\mathcal{L}-\mathcal{L}_c)^2\right)\,,
\ee
such that the correction to the energy in~\eqref{EnergyB} reads as in~\eqref{EBcritical}.

\subsection{Large-$\L$ expansion}
\label{app:largeL}

To understand the origin of the large-\(\mathcal L\) expansion, recall the definition of \(\mathcal L\) in \eqref{L} together with the parameter ordering
\(
0 \le m_{\mbox{\tiny $L$}} \le n_{\mbox{\tiny $L$}} \le 1 .
\)
This ordering implies the bounds
\begin{equation}\label{rangeinequality}
K(m_{\mbox{\tiny $L$}})-E(m_{\mbox{\tiny $L$}})
\;\le\;
\mathcal L
\;\le\;
\frac{K(m_{\mbox{\tiny $L$}})-E(m_{\mbox{\tiny $L$}})}{\sqrt{m_{\mbox{\tiny $L$}}}},
\end{equation}
where the lower (upper) bound corresponds to \(n_{\mbox{\tiny $L$}}=m_{\mbox{\tiny $L$}}\) (\(n_{\mbox{\tiny $L$}}=1\)). On the interval \(0\le m_{\mbox{\tiny $L$}}\le 1\), the elliptic integrals $K$ and $E$ satisfy the inequality
\begin{equation}\label{ellipticbound}
\frac{K(m_{\mbox{\tiny $L$}})-E(m_{\mbox{\tiny $L$}})}{\sqrt{m_{\mbox{\tiny $L$}}}}
\;\le\;
\frac12 \log\!\left(\frac{1+\sqrt{m_{\mbox{\tiny $L$}}}}{1-\sqrt{m_{\mbox{\tiny $L$}}}}\right).
\end{equation}
This follows directly from the standard trigonometric representation of \(K\) and \(E\)~\footnote{From the usual trigonometric integral representation for elliptic integrals, we find immediately \[\scalebox{0.90}{$\displaystyle K(m_{\mbox{\tiny $L$}})-E(m_{\mbox{\tiny $L$}})=\int_0^{\pi/2}\!\!\!\! \!\! \!\!d\theta~ \frac{m_{\mbox{\tiny $L$}}\sin^2\theta}{\sqrt{1-m_{\mbox{\tiny $L$}} \sin^2\theta}}=\int_0^1\!\!\!\! dt \frac{m_{\mbox{\tiny $L$}} \sqrt{1-t^2}}{\sqrt{m_{\mbox{\tiny $L$}} t^2-m_{\mbox{\tiny $L$}}+1}}\leqslant \int_0^1\!\!\! dt \frac{m_{\mbox{\tiny $L$}} }{\sqrt{m_{\mbox{\tiny $L$}} t^2+1-m_{\mbox{\tiny $L$}}}}=\sqrt{m_{\mbox{\tiny $L$}}}\log\!\Bigl(\frac{1+\sqrt{m_{\mbox{\tiny $L$}}}}{1-\sqrt{ m_{\mbox{\tiny $L$}}}}\Bigr).$} \] Above, we have used the change of variable $\theta=\arccos(t)$.}.
Combining \eqref{rangeinequality} and \eqref{ellipticbound}, we obtain
\[
\mathcal L
\;\le\;
\frac12 \log\!\left(\frac{1+\sqrt{m_{\mbox{\tiny $L$}}}}{1-\sqrt{m_{\mbox{\tiny $L$}}}}\right),
\ \mbox{
which can be inverted to give}\ 
\sqrt{m_{\mbox{\tiny $L$}}}
\;\ge\;
\frac{1-e^{-2\mathcal L}}{1+e^{-2\mathcal L}}
\;\ge\;
1-2e^{-2\mathcal L}.
\]
Hence,
\begin{equation}\label{inequalityexp}
1-2e^{-2\mathcal L}
\;\le\;
\sqrt{m_{\mbox{\tiny $L$}}}
\;\le\;
1.
\end{equation}
Therefore, in the large-\(\mathcal L\) limit, one has \(m_{\mbox{\tiny $L$}}\to 1\) with exponentially small corrections. Since \(m_{\mbox{\tiny $L$}}\le n_{\mbox{\tiny $L$}}\le 1\), the parameter \(n_{\mbox{\tiny $L$}}\) must also approach \(1\) at the same exponential rate. Guided by this behavior, we introduce the expansions
\begin{equation}\label{mLnLlargeL}
m_{\mbox{\tiny $L$}} = 1-\eL+\sum_{k=2}^{\infty} a_k\,\eL^k,
\qquad
n_{\mbox{\tiny $L$}} = 1-\sum_{k=1}^{\infty} b_k\,\eL^k,
\end{equation}
where \(\eL\) is exponentially small in \(\mathcal L\). Its dependence on \(\mathcal L\), as well as the coefficients \(a_k\) and \(b_k\), will be determined by solving \eqref{theta} and \eqref{L} perturbatively. The choice of unit coefficient for the linear term in \(m_{\mbox{\tiny $L$}}\) fixes the overall normalization of \(\eL\).

\noindent
We proceed by first solving \eqref{theta} for \(\theta\), using the expansion of \(\Pi(m,n)\) derived in Appendix~\ref{app:expansionPi}, and then solving \eqref{L} for \(\mathcal L\). At leading order, \eqref{L} fixes the expansion parameter \(\eL\) as in~\eqref{epsilonL}.  
Proceeding order by order in the solutions, we get the following expansions for $n_{\mbox{\tiny $L$}}$ and $m_{\mbox{\tiny $L$}}$ up to $\eL^4$:
\begin{align}\label{mLlargeL}
m_{\mbox{\tiny $L$}}=&1-\eL+\frac{ \eL^2}{2} (1+\mathcal{L} \cos \theta )-\frac{\eL^3}{64}  \big(20 \mathcal{L}^2
   \cos 2 \theta +4 \mathcal{L}^2+32 \mathcal{L} \cos \theta +6 \mathcal{L} \cos 2 \theta -4
   \mathcal{L}+15\big)+\nonumber\\
   &+\frac{\eL^4}{1536} \big (120 \mathcal{L}^3 \cos \theta +392 \mathcal{L}^3 \cos 3
   \theta -132 \mathcal{L}^2 \cos \theta +576 \mathcal{L}^2 \cos 2 \theta +228 \mathcal{L}^2 \cos
   3 \theta +\nonumber\\
   &+192 \mathcal{L}^2+525 \mathcal{L} \cos \theta +144 \mathcal{L} \cos 2 \theta +39 \mathcal{L}
   \cos 3 \theta -96 \mathcal{L}+168\big)+O(\eL^5)\,,
   \\\label{nLlargeL}
  n_{\mbox{\tiny $L$}}=& 1-\eL  \cos ^2\frac{\theta }{2}+\frac{\eL ^2}{4} 
   \cos ^2\frac{\theta }{2} ((4 \mathcal{L}+3) \cos \theta
   -2 \mathcal{L}-1)-\frac{\eL ^3}{128}  \cos
   ^2\frac{\theta }{2} \big(64 \mathcal{L}^2-2 (56 \mathcal{L}^2+\nonumber\\
   &+52
   \mathcal{L}+15) \cos \theta +(16 \mathcal{L}
   (6 \mathcal{L}+7)+39) \cos 2 \theta +60
   \mathcal{L}+21\big)+\nonumber\\
   &+\frac{\eL ^4}{6144} \big(6
   (\mathcal{L} (40 \mathcal{L}^2+44
   \mathcal{L}+43)+17) \cos \theta +4 (6 \mathcal{L} (4
   \mathcal{L}+11)+25) \cos 2 \theta +\nonumber\\
   &+2 (\mathcal{L} (4 \mathcal{L} (98
   \mathcal{L}+183)+483)+117) \cos (3 \theta )+(8 \mathcal{L} (64
   \mathcal{L} (2 \mathcal{L}+3)+111)+\nonumber\\
   &+197) \cos 4 \theta +24 \mathcal{L}
   (4 \mathcal{L}+3)+39)+O(\eL ^5)\,.
\end{align}
Given the expansion for  $n_{\mbox{\tiny $L$}}$ and $m_{\mbox{\tiny $L$}}$, we can compute the expansion of $\Theta_{\text{eff}}(\theta,\mathcal{L})$ in power of $\eL$:
\begin{align}\label{thetaeff-largeL}
\Theta_{\text{eff}}(\theta,\mathcal{L})=&2 (\mathcal{L}+1)-\frac{\eL}{2}   \cos \theta +\frac{ \eL ^2}{32}
   ((4 \mathcal{L}+1) \cos 2 \theta -2)+\frac{ \eL ^3}{256}
   \cos \theta  (4 \mathcal{L} (4 \mathcal{L}+5)+1-\nonumber\\
   &-(16 \mathcal{L} (2 \mathcal{L}+1)+5) \cos
   2 \theta )+\frac{\eL ^4
  }{24576} (-8 (96 \mathcal{L}^2+7) \cos 2 \theta +\nonumber\\
  &+(8
   \mathcal{L} (32 \mathcal{L} (4 \mathcal{L}+3)+45)+59) \cos 4 \theta
   +96 (1-2 \mathcal{L}) \mathcal{L}-66)+O(\eL
   ^5)\,.
\end{align}
We now determine the parameters \(m\) and \(n\) in \(\text{AdS}_5\) in the large-\(\mathcal L\) regime. 
Because \(\Theta_{\mathrm{eff}}(\theta,\mathcal L)\) grows linearly with \(\mathcal L\), the solution lies in region B$_L$, characterized by \(m<0\). 
Imposing the consistency condition \eqref{thetaeff} then requires \(m\to -\infty\) as \(\mathcal L\to\infty\). 
Moreover, matching the leading \(\mathcal L\)-dependence of \(\Theta_{\mathrm{eff}}\) fixes the scaling of $m$ to be
\(
 -\frac{1}{\eL}+O(1).
\)
By contrast, \(n\) remains finite in this limit.  In order to apply the expansion of Appendix~\ref{app:expansionPi}, it is convenient to perform a modular transformation that maps the parameters $m$ and $n$ to the interval $[0,1]$, see equation \eqref{mprime-nprime-app}. 
The complete elliptic integrals then transform according to
\begin{equation}
K(m)=\sqrt{1-m'}\,K(m'), \qquad
\Pi(n,m)=\frac{\sqrt{1-m'}}{n'}\left(m'\,K(m')+n\,(1-m')\,\Pi(n',m')\right).
\end{equation}
With this reparametrization, the equation \eqref{phi01} determining $\phi$ is unchanged in form, upon replacing $(n,m)$ by $(n',m')$. The consistency condition is correspondingly modified to
\be
\label{cons1}
\Theta_{\text{eff}}(\theta,\mathcal L)=2\sqrt{1-n'+m'}\,K(m').
\ee
As $\mathcal L\to\infty$, the transformed parameters satisfy $m'\to1$ and $n'\to1$ with exponentially small deviations. Accordingly, the following ansatz is adopted:
\begin{equation}\label{mhatnhatargeL}
m'=1-\eL+\sum_{k=2}^{\infty} c_k\,\eL^k,
\qquad
n'=1-\sum_{k=1}^{\infty} d_k\,\eL^k,
\end{equation}
where the normalization of $\eL$ is fixed by imposing the asymptotic behavior
\(
m=-\eL^{-1}+O(1).
\)
Equations \eqref{phi01} and \eqref{cons1} are solved perturbatively in powers of $\eL$, using the expansion presented in Appendix~\ref{app:expansionPi}, together with the standard asymptotic expansions of the elliptic integrals $K$ and $E$ in the limit $m\to1$. The results are given up to, and including, terms of order $O(\eL^4)$:
\begin{align}
m'=& 
1-\eL +\frac{\eL ^2}{2}  (\cos\phi-\cos\theta+\mathcal{L} \cos\phi+1)+\frac{\eL ^3 }{64}  (-4
   \mathcal{L}^2+8 \cos\theta ((4 \mathcal{L}+3) \cos\phi+4)+\notag\\ &+(8
   \mathcal{L}-2) \cos 2 \theta -32 (\mathcal{L}+1) \cos\phi-10
   (\mathcal{L}+1) (2 \mathcal{L}+1) \cos 2 \varphi -12
   \mathcal{L}-27 )+\notag\\ &+\frac{\eL ^4 }{1536} (12
   \cos 2 \theta   ( (5-16 \mathcal{L}^2 ) \cos \phi
   -16 \mathcal{L}+12 )-11 \cos 3 \theta -3 \cos\theta  (4  (68
   \mathcal{L}^2+86 \mathcal{L}+\notag\\ &+23 ) \cos 2 \phi +64 (6
   \mathcal{L}+5) \cos\phi+8 \mathcal{L} (2
   \mathcal{L}+3)+195 )-96 \mathcal{L}^2 \cos 3 \theta +192
   \mathcal{L}^2+\notag\\ &+3 (\mathcal{L} (20 \mathcal{L} (2
   \mathcal{L}+7)+371)+231) \cos\phi+48 (\mathcal{L}+1) (12
   \mathcal{L}+7) \cos 2 \phi +\notag\\ &+(\mathcal{L}+1) (4 \mathcal{L} (98
   \mathcal{L}+109)+119) \cos 3 \phi )+480
   \mathcal{L}+648 )+O(\eL ^5)\,,
\end{align}
\begin{align}
n'=&1-\frac{ \eL}{2}\sin^2\frac\phi2  +\frac{ \eL ^2}{4} \sin
   ^2\frac{\phi }{2} ((4 \mathcal{L}+3)
   \cos \phi -2 \cos \theta +2 \mathcal{L}+3)-\frac{\eL ^3 }{128}   \sin
   ^2\frac{\phi }{2}  (64 \mathcal{L}^2-  \notag\\&-16 \cos
   \theta  (2 (5 \mathcal{L}+3)   \cos\phi+6 \mathcal{L}+7)+(4-16
   \mathcal{L}) \cos 2 \theta +2 (4 \mathcal{L} (14 \mathcal{L}+33)+  \notag\\&+71)
     \cos\phi+3 (16 \mathcal{L} (2 \mathcal{L}+3)+17) \cos 2 \phi
   +140 \mathcal{L}+101  )  +  \notag\\&+\frac{\eL ^4   }{1536}(\frac{1}{2}
     (96 \mathcal{L}^2+11  ) \cos 3 \theta  (\cos \phi
   -1)+\sin ^2  \frac{\phi }{2}     (-12 \cos (2 \theta )
     (32 \mathcal{L}^2+  \notag\\&+2 (\mathcal{L} (24 \mathcal{L}-5)-9) \cos
   (\phi )+6 \mathcal{L}-25  )+(8 \mathcal{L} (2 \mathcal{L} (94
   \mathcal{L}+309)+795)+2455)   \cos\phi+  \notag\\&+2 (\mathcal{L} (4
   \mathcal{L} (158 \mathcal{L}+507)+2013)+623) \cos 2 \phi +(4+
   \mathcal{L} (64 \mathcal{L} (4 \mathcal{L}+9)+423)+  \notag\\&+409) \cos 3 \phi
   +2 (\mathcal{L}+1) (4 \mathcal{L} (94 \mathcal{L}+245)+895)  )-3
     \cos\theta \sin ^2  \frac{\phi }{2}  (4 (4 \mathcal{L}
   (66 \mathcal{L}+  \notag\\&+133)+243)   \cos\phi+2 (128 \mathcal{L} (3
   \mathcal{L}+4)+161) \cos 2 \phi +64 \mathcal{L} (9
   \mathcal{L}+17)+617)  )+O  (\eL ^5  )\,.
\end{align}
At this stage, all necessary ingredients are in place to evaluate the energy. In terms of the transformed parameters, it takes the form
\be
\mathcal{E}=\frac{1}{\sqrt{n'}}\bigl(K(m')-E(m')\bigr)\,,
\ee
leading to a  compact expression for the large $\L$ expansion, equation~\eqref{ElargeL}.

 \subsection{Near BPS expansion}

To approach the BPS point, we fix $\mathcal{L}$ and $\theta$ and consider the limit in which $\phi$ approaches $\theta$. To isolate how a nonzero $\mathcal{L}$ modifies the result, compared to the standard generalized cusp of \cite{DF}, we expand in small $\mathcal{L}$. In this small–$\mathcal{L}$ limit, the quantities $\nL$ and $\mL$ take the form shown in \eqref{nLmLexp}, and the effective angle $\Theta_{\text{eff}}(\theta,\mathcal{L})$ is given by \eqref{theffsmallL}.

For the $\text{AdS}_5$ moduli  solving the conditions~\eqref{thetaeff} and \eqref{phi01}, one postulates  a double series-expansion in $\hL$ and $\epsilon=\cos\varphi-\cos\theta$
\be
m=\sum_{k,j=0}^\infty m_{kj}(\htheta)\epsilon^k \hL^J\,, \ \qquad \  n=\sum_{k,j=0}^\infty n_{kj}(\htheta)\epsilon^k \hL^j\,.
\ee
Here we have defined for future convenience $\hL=\L/\pi$ and $\htheta=\theta/\pi$.
At the BPS point \((\epsilon=0,\, \mathcal{L}=0)\) one has \(\mL=0\), which because of~\eqref{newparameters} implies \(m=0\), or \(m_{00}=0\). The expansions read
\begin{align}\label{mBPS}
m
= &\biggl[-4 \sqrt{1\!-\!\hat \theta ^2} \hat{\mathcal{L}}\!+\!(6 \hat\theta ^2\!-\!10) \hat{\mathcal{L}^2}\!-\!\frac{(4 \htheta ^4\!-\!17 \htheta ^2\!+\!19 )
   \hL}{ \sqrt{1\! -\!\htheta^2 }}
   -\frac{(9\! -\!5 \htheta ^2) (\htheta ^4\!-\!6 \htheta ^2\!+\!13 ) \hL^4}{4  (1 \!-\!\htheta^2 ) }\!+\!O(\hL^5)\biggr]\!+\!\nonumber\\
  &+\frac{ \htheta }{\pi \sin\pi\htheta}\biggl[-2+\frac{2   (3 \htheta ^2-4 ) \hL }{ \sqrt{1 -\htheta^2 }}+3 
   (2 \htheta ^2-5 ) \hL^2   +\frac{ (5 \htheta ^6-33  \htheta ^4+68  \htheta ^2-26 ) \hL^3  }{2 (1 -\htheta ^2)^{3/2}}+\biggr.\nonumber\\ &\biggl.+\frac{ (-15 \htheta ^8+130  \htheta ^6-444  \htheta ^4+558  \htheta ^2+91) \hL^4
  }{16 (1-\htheta ^2)^2}+O(\hL^5)\biggr]\epsilon\nonumber
  \\
   & \!+\!\frac{\csc (\pi  \hat{\theta) }}{2 \pi ^2}\Biggl[\frac{\pi  (3 \hat{\theta }^4 \!-\!6 \hat{\theta }^2 \!+\!2 \pi   (\hat{\theta }^2 \!-\!1 ) \hat{\theta } \cot (\pi 
   \hat{\theta }) \!+\!2)}{\hat{\theta } (1 \!-\!\hat{\theta }^2)} 
    \!-\!\frac{\pi  \hat{\mathcal{L}}  }{\hat{\theta }  (1 \!-\!\hat{\theta }^2 )^{3/2}}(6
   \hat{\theta }^6 \!-\!24 \hat{\theta }^4 \!+\!21 \hat{\theta }^2\!+\!\nonumber\\ & \!+\!2 \pi   (3 \hat{\theta }^4 \!-\!7 \hat{\theta }^2 \!+\!4 )
   \hat{\theta } \cot  (\pi  \hat{\theta } ) \!-\!8 )\!+\!\frac{3
   \pi  \hat{\mathcal{L}}^2 }{4 \hat{\theta }  (\hat{\theta }^2 \!-\!1 )^2} (5 \hat{\theta }^8 \!-\!36 \hat{\theta }^6 \!+\!65 \hat{\theta }^4 \!-\!42 \hat{\theta }^2 \!+\!\nonumber\\& \!+\! 4 \pi 
    (\hat{\theta }^2 \!-\!1 )^2  (2 \hat{\theta }^2 \!-\!5 ) \hat{\theta } \cot  (\pi  \hat{\theta
   } ) \!+\!20 ) \!-\!\frac{\hat{\mathcal{L}}^3}{8  \hat{\theta }  (1 \!-\!\hat{\theta }^2 )^{5/2} }   (\pi   (15
   \hat{\theta }^{10} \!-\!138 \hat{\theta }^8 \!+\!428 \hat{\theta }^6 \!-\!444 \hat{\theta }^4 \!+\!\nonumber\\ &  \!+\!359 \hat{\theta }^2 \!+\!4 \pi   (5
   \hat{\theta }^8 \!-\!38 \hat{\theta }^6 \!+\!101 \hat{\theta }^4 \!-\!94 \hat{\theta }^2 \!+\!26 ) \hat{\theta } \cot  (\pi 
   \hat{\theta } ) \!-\!104 ) )\!+\! \nonumber\\ &\!+\! \frac{\pi 
   \hat{\mathcal{L}}^4 }{64  (\hat{\theta } \!-\!1 )^3 \hat{\theta }  (\hat{\theta
   } \!+\!1 )^4}  (\hat{\theta }  (\hat{\theta }  (\hat{\theta }  (896100 \!-\!\hat{\theta }
    (\hat{\theta } \!+\!1 )  (218295 \hat{\theta }^8 \!-\!1116072 \hat{\theta }^6 \!+\!2317014 \hat{\theta }^4 \!-\!\nonumber\\
    &\!-\!2493676
   \hat{\theta }^2 \!+\!1520151 ) ) \!+\!9060 ) \!+\!364 ) \!-\!4 \pi   (\hat{\theta } \!-\!1 ) \hat{\theta }  (15
   \hat{\theta }^8 \!-\!130 \hat{\theta }^6 \!+\!444 \hat{\theta }^4 \!-\!558 \hat{\theta }^2\!-\!\nonumber\\ & \!-\!91 )  (\hat{\theta } \!+\!1 )^2 \cot
    (\pi  \hat{\theta } ) \!+\!364 ) \!+\!O (\hat{\mathcal{L}}^5)\Biggr]\epsilon^2
 \!+\! O(\epsilon^{2}) \,,  \\\nonumber
n =& \biggl[
    (1  \!-\! \htheta^{2})
     \!-\! 2\,\htheta^{2}\sqrt{1 \!-\!\htheta^2}\,\hL
     \!+\! \frac{3\,\htheta^{2}\,(\htheta^{2} \!-\!1)\,\hL^{2}}{2}
     \!-\! \frac{\bigl(\htheta^{6}  \!-\! \htheta^{4}  \!-\! 2\htheta^{2}\bigr)\hL^{3}}{2\,\sqrt{1 \!-\!\htheta^2}}\\\nonumber
    &
     \!+\! \frac{\bigl(5\htheta^{8} \!-\!15\htheta^{6} \!-\!33\htheta^{4} \!+\!107 \htheta^{2}\bigr)\hL^{4}}{32\,\,(1 \!-\!\htheta^{2})}
     \!+\! O(\hL^{5})
\biggr]  \!+\!\,\frac{ \htheta }{\pi \sin\pi\htheta}\biggl[
    2 \!-\!\htheta^{2}
     \!+\! \frac{(3\htheta^{4} \!-\!9\htheta^{2} \!+\!8)\,\hL}{2\sqrt{1 \!-\!\htheta^2}}\\\nonumber
& 
     \!+\! \frac{3\,(\htheta^{4} \!-\!3\htheta^{2} \!+\!4)\,\hL^{2}}{4}    \!+\! \frac{\htheta\,\bigl( 5\htheta^{8} \!-\!25\htheta^{6} \!+\!51\htheta^{4} \!-\!55\htheta^{2} \!-\!32\bigr)\,\hL^{3}}{16(1 \!-\!\htheta^2)^{3/2}}   
    \\\label{nBPS}
&  \!+\! \frac{\htheta\,\bigl( 6\htheta^{8} \!-\!47\htheta^{6} \!+\!129\htheta^{4} \!-\!141\htheta^{2} \!-\!107\bigr)\,\hL^{4}}{16\,(1 \!-\!\htheta^{2})^{2}}
     \!+\! O(\hL^{5})\biggr] \epsilon +\nonumber\\
     &\!+\! \frac{\csc ^2(\pi  \hat{\theta })}{8 \pi ^2}\biggl[ \frac{1}{1\!-\!\hat{\theta }}\!+\!\frac{1}{\hat{\theta }\!+\!1}\!+\!\hat{\theta }  (8 \pi  \cot  (\pi  \hat{\theta
   })\!+\!\hat{\theta }  (\!-\!3 \hat{\theta }^2\!-\!4 \pi  \hat{\theta } \cot  (\pi  \hat{\theta
   })\!+\!18))- 10 - \nonumber\\
   &\!-\!\frac{\hat{\mathcal{L}}  (\hat{\theta }  (\hat{\theta }  (\hat{\theta }
    (34 \pi  \cot  (\pi  \hat{\theta })\!+\!3 \hat{\theta }  (\hat{\theta }^2\!-\!4)  (\hat{\theta }^2\!+\!2
   \pi  \hat{\theta } \cot  (\pi  \hat{\theta })\!-\!5))\!-\!42)\!-\!16 \pi  \cot  (\pi  \hat{\theta
   }))\!+\!16)}{ (1\!-\!\hat{\theta }^2)^{3/2}}\!+\!\nonumber\\
   &\frac{3}{8} \hat{\mathcal{L}}^2  (5 \hat{\theta
   }^6\!-\!37 \hat{\theta }^4\!+\!64 \hat{\theta }^2\!-\!\frac{8  (\hat{\theta }^4\!+\!\hat{\theta }^2\!+\!4)}{ (\hat{\theta
   }^2\!-\!1)^2}\!+\!8 \pi  \hat{\theta }^5 \cot  (\pi  \hat{\theta })\!-\!24 \pi  \hat{\theta }^3 \cot  (\pi 
   \hat{\theta })\!+\!32 \pi  \hat{\theta } \cot  (\pi  \hat{\theta }))\!+\!\nonumber\\ 
   &+\frac{\hat{\mathcal{L}}^3  }{8  (1\!-\!\hat{\theta }^2)^{5/2}}(39
   \hat{\theta }^{10}\!-\!289 \hat{\theta }^8\!+\!827 \hat{\theta }^6\!-\!1023 \hat{\theta }^4\!+\!614 \hat{\theta }^2\!-\!2 \pi   (5
   \hat{\theta }^{10}\!-\!30 \hat{\theta }^8\!+\!76 \hat{\theta }^6\!-\!106 \hat{\theta }^4\!+\!\nonumber\\ 
   &\!+\!23 \hat{\theta }^2\!+\!32) \hat{\theta }
   \cot  (\pi  \hat{\theta })\!+\!64)\!+\!\frac{\hat{\mathcal{L}}^4
}{32  (1\!-\!\hat{\theta
   })^3  (\hat{\theta }\!+\!1)^4}    (\hat{\theta }  (\hat{\theta }  (\hat{\theta }  (\hat{\theta }  (\hat{\theta }  (\hat{\theta }
    (\hat{\theta }\!+\!1)  (218295 \hat{\theta }^8\!-\! \nonumber\\ &\!-\!1552098 \hat{\theta }^6\!+\!4544430 \hat{\theta }^4\!-\!7112988
   \hat{\theta }^2\!+\!6486403)\!-\!3917490)\!-\!3030450)\!+\!1783344)\!+\! \nonumber\\ &\!+\!9264)\!+\!856)\!+\!8 \pi 
    (\hat{\theta }\!-\!1) \hat{\theta }  (\!-\!6 \hat{\theta }^8\!+\!47 \hat{\theta }^6\!-\!129 \hat{\theta }^4\!+\!141 \hat{\theta
   }^2\!+\!107)  (\hat{\theta }\!+\!1)^2 \cot  (\pi  \hat{\theta })\!+\!\nonumber\\ &\!+\!856)\!+\!O(\hat{\mathcal{L}}^5) \biggr]\epsilon^2\!+\!O(\epsilon^3).
\end{align}
The expansion of the energy close to the BPS points and at small $\mathcal{L}$ reads
\begin{align}\nonumber
\mathcal{E}=&\hL + \frac {\htheta  }{ \sin\htheta\sqrt{\pi^2 -\htheta^2}} \biggl[\frac{1}{2}-\frac{3    \hL}{2 \sqrt{1-\htheta^2}}-\frac{3  (\htheta ^2-3 ) \hL^2}{8  (1 -\htheta^2 )}+\frac{   (5 +6  \htheta ^2-3 \htheta
   ^4 ) \hL^3}{8   (1-\htheta ^2 )^{\frac{3}{2}}}\biggr. 
   \\\nonumber
 &-\frac{15   (3 \htheta ^6-3  \htheta ^4-11  \htheta ^2+3 
    ) \hL^4}{128  (1 -\htheta^2)^{2}}+O(\hL^5)\biggr]\,\epsilon\\\nonumber
   &+ \frac{\csc^2\pi\htheta}{8\pi(1 -\!\htheta^2 )^{\frac{3}{2}}} \biggl [    \!-\!3 \hat{\theta }^2  \!-\!2 \pi  \hat{\theta }^3 \cot   (\pi  \hat{\theta } ) +2 \pi  \hat{\theta } \cot   (\pi 
   \hat{\theta }  )  \!-\!2 \!-\!\nonumber\\
   &   \!-\!\frac{3 \hat{\mathcal{L}}   (\hat{\theta }^4  \!-\!13 \hat{\theta }^2  \!-\!4 \pi  \hat{\theta }^3
   \cot   (\pi  \hat{\theta }  )+4 \pi  \hat{\theta } \cot   (\pi  \hat{\theta }  )  \!-\!4  )}{2
   \sqrt{1  \!-\!\hat{\theta }^2}}  \!-\!\frac{3 \hat{\mathcal{L}}^2   }{4   (\hat{\theta }^2  \!-\!1  )}(  \!-\!3 \hat{\theta }^6+15 \hat{\theta }^4  \!-\!28 \hat{\theta }^2\!+\!\nonumber\\ &\!+\!2 \pi
    \hat{\theta }^5 \cot   (\pi  \hat{\theta }  )  \!-\!8 \pi  \hat{\theta }^3 \cot   (\pi  \hat{\theta }  )+6 \pi 
   \hat{\theta } \cot   (\pi  \hat{\theta }  )  \!-\!6  )+\frac{\hat{\mathcal{L}}^3
}{16
     (1  \!-\!\hat{\theta }^2  )^{3/2}}      (  \!-\!45 \hat{\theta }^8+183 \hat{\theta }^6  \!-\!\nonumber\\ &\!-\!159 \hat{\theta }^4  \!-\!387 \hat{\theta }^2+24 \pi  \hat{\theta }^7 \cot
     (\pi  \hat{\theta }  )  \!-\!72 \pi  \hat{\theta }^5 \cot   (\pi  \hat{\theta }  )+8 \pi  \hat{\theta }^3 \cot
     (\pi  \hat{\theta }  )+40 \pi  \hat{\theta } \cot   (\pi  \hat{\theta }  )  \!-\!40  )\!+\!\nonumber\\ 
     & \!+\!\frac{15}{64} \hat{\mathcal{L}}^4 (\!-\!14553 \hat{\theta }^8+45264 \hat{\theta }^6\!-\!49200 \hat{\theta }^4\!+\!22257 \hat{\theta
   }^2\!+\!\nonumber\\
   &\!+\!\frac{6\!-\!2 \hat{\theta } (\hat{\theta } (\hat{\theta } (3737 \hat{\theta } (\hat{\theta
   }\!+\!1)\!-\!18450)+11118)\!-\!3)}{(\hat{\theta }\!-\!1)^2 (\hat{\theta }\!+\!1)^3}+6 \pi 
   \hat{\theta }^5 \cot (\pi  \hat{\theta })\!-\!22 \pi  \hat{\theta } \cot (\pi  \hat{\theta })\!-\!\nonumber\\\label{EBPSlong}
   &\!-\!\frac{16
   \pi  \hat{\theta } \cot (\pi  \hat{\theta })}{\hat{\theta }^2\!-\!1})\!+\!O(\hat{\mathcal{L}}^5)\biggr]\,\epsilon ^2\!+\!O(\epsilon ^3)\,. 
      \end{align}
The expansion of~\eqref{Brem} at small angles, with the substitution $\L=L/(4g)$ reproduces of course all the results in~\cite{GS}.

\section{Quantum fluctuations}
\label{app-quantum}

In this appendix, we report some technical details used for the computations of Section~\ref{sec:fluctuations}.

\subsection{Bosonic fluctuations}
\label{app:bosons-oneloop}

In order to compute the Lagrangian for the bosonic fluctuations using the algorithm developed in~\cite{Forini:2015mca}, we need to specify a basis $N_i^A$ (with $i=1, \dots, 8$ and $A=0,\dots, 9$) in the normal bundle as well as the tangent vectors  along the tangent bundle, i.e. $t^A_\alpha = \pa_\alpha X^\mu E^A_\mu$ where $E^A_\mu$ are the target space vielbein. We choose the following basis:
\bea
\label{def-t}
t_\tau &=& (\sqrt{\XX^2+1},0,0,0,0, \sqrt{1-\YY^2} ,0,0,0,0)\,, 
\\ \nn
t_\s &=& \Big(0, \frac{\XX \XX'}{\sqrt{(\XX^2+1)(\XX^2+1 -\kappa ^2)}} ,\frac{\kappa \ell_\varphi  }{\sqrt{\XX^2+1-\kappa ^2}} ,0,0,
\\ \nn 
&& ~~ 0, \frac{\YY \YY'}{\sqrt{(1-\YY^2) (\YY^2+\gamma ^2-1)}},  -\frac{\gamma  \ell_\vartheta}{\sqrt{\YY^2-1+\gamma ^2}} ,0,0\Big)\,,
\eea
\bea
\label{N-vectors}
&\text{AdS}_5:& N_1= (0, 0, 0, 1, 0, 0, 0, 0, 0, 0) \,,  \qquad  N_2=(0, 0, 0, 0, 1, 0, 0, 0, 0, 0) \,, \\ \nn
&& N_7=\Big(0,\frac{\kappa  \ell_\varphi}{\XX \sqrt{\XX^2+1-\kappa ^2}},-\frac{\XX'}{\sqrt{(\XX^2+1) (\XX^2+1-\kappa ^2)}},0,0,0,0,0,0,0\Big) \,, 
\\ \nn
\\[2.5 pt]
&S^5:& N_3= (0, 0, 0, 0, 0, 0, 0, 0, 1, 0)\,, \qquad N_4 = (0,0,0,0,0,0,0,0,0,1) \,, \\ \nn
&& N_6=\Big( 0,0,0,0,0,0, \frac{\gamma  \ell_\vartheta}{\YY \sqrt{\YY^2-1+\gamma ^2}}\,, \frac{\YY'}{\sqrt{(1-\YY^2) (\YY^2-1+\gamma ^2)}}, 0,0\Big)\,, \\ \nn 
\\[2.5 pt]
&\text{Mixed}:& N_5= e^{-\Lambda}(\sqrt{1-\YY^2}, 0,0,0,0, \sqrt{1+\XX^2}\,, 0,0,0,0)\,,
\\ \nn 
&& N_8=  e^{-\Lambda}  \Big(0, -\frac{\YY \XX'}{\sqrt{(\XX^2+1)(\XX^2+1-\kappa ^2)}}\,,-\frac{\kappa  \ell_\varphi \YY}{\XX \sqrt{\XX^2+1-\kappa ^2}}\,, 0, 0, 
\\ \nn
&& \qquad \qquad \quad  0, \frac{\XX \YY'}{\sqrt{(1-\YY^2) (\YY^2-1+\gamma ^2)}}\,, -\frac{\gamma  \ell_\vartheta \XX}{\YY \sqrt{\YY^2-1+\gamma ^2}}, 0, 0\Big)\,. 
\eea

\subsubsection{The straight line limit}

In \cite{Giombi:2021zfb, Giombi:2022anm}, the Lagrangian for the quadratic bosonic fluctuations was obtained for the case of $L$ orthogonal insertions on the straight line with $\phi=\theta=0$. 
We remind the reader that for us this implies \eqref{straight-line-lim}. 
In this limit the solutions \eqref{rsol}-\eqref{rhotildesol} satisfy the relation
\be
\gamma \cos\psi={1\over \cosh\rho}\,,
\ee
which in terms of the functions $\XX, \YY$ \eqref{def-XandY} this means 
\bea
&& \YY  \to \tanh\rho\,,\qquad \YY^2={\XX^2+1-\kappa^2\over \XX^2+1}=1 -\frac{1}{\mL(1+\XX^2)}\,,
\\\nn
&& e^{2\Lambda}= 1+\XX^2-\frac{\kappa^2}{\XX^2+1} = 1+\XX^2-\frac{1}{\mL(\XX^2+1)}\,. 
\eea
In this limit \eqref{straight-line-lim} the mass matrix \eqref{massmatrix-last4} becomes diagonal, and particularly simple~\footnote{In~\eqref{massesstraight}, a mixed used of the modular parameter  ($\mL$) and the parameters of the classical solution is made to ensure compactness of the expression.}:
\begin{subequations}\label{massesstraight}
\bea
 -\sqrt g\mathcal M_{ii} &=&  1+2\XX^2\,,  \qquad i=1,2, 7\,,
 \\
-\sqrt g \mathcal M_{ii} &=& -1+2\frac{\kappa^2}{1+\XX^2} \,, \qquad i=3,4,6\,,
\\ 
-\sqrt g \mathcal M_{ii} &=& -1+2\frac{1}{\mL(1+\XX^2)} \,, \qquad i=3,4,6\,,
\\ 
 -\sqrt g \mathcal M_{ii} &=&1+ \frac{2 \kappa  (\kappa -1)^2 \left(\XX^2+1\right)}{\left(-\kappa +\XX^2+1\right)^2}-\frac{2 \kappa  (\kappa +1)^2 \left(\XX^2+1\right)}{\left(\kappa +\XX^2+1\right)^2}\,, \qquad i=5, 8\,. 
\eea
\end{subequations}
The connections along the $\sigma$-direction \eqref{Asigma} vanish identically, and along the $\tau$-direction \eqref{Atau} only one component survives
\be
A_\tau^{58}\to \frac{\kappa  \left(\kappa ^2+\XX^4-1\right)}{\left(-\kappa +\XX^2+1\right) \left(\kappa +\XX^2+1\right)}\,.
\ee
In this limit our quadratic bosonic Lagrangian agree with the one computed in \cite{Giombi:2022anm} (setting $c={1\over\kappa}=\sqrt{\mL}$).
Finally the fermionic masses \eqref{L-fer-flux-fin} become
\be
\mathcal L_{\rm flux} \to  2 i  e^{-2\Lambda}\,  \bar\Theta \big(\XX \sqrt{1+\XX^2}\, \Gamma_{\hat 2\hat 3\hat 4}
+\frac{\kappa  \sqrt{\XX^2+1-\kappa ^2}}{\XX^2+1}\, \Gamma_{\hat 6\hat 8\hat 9} \big)\Theta\,.
\ee
Finally in the limit $\kappa\to 1^+$ ($\mL\to 1^-$) \cite{Drukker:2006xg}, then we have further simplifications as in can be seen from the above formulae, and notice that $\XX\to \sinh\rho$ in this case. 

\subsubsection{The generalised cusp limit}

We have checked that our one-loop Lagrangian for the bosonic and fermionic modes \eqref{L-bos-oneloop}, \eqref{L-fer-fin} reproduces the results of \cite{DF}, once taken the appropriate limits \eqref{DF-lim1}-\eqref{DF-lim2}. 
In comparing our results with \cite{DF}, it should be kept in mind that the worldsheet coordinate $\s$ spans between $(-K(k^2), K(k^2))$ in \cite{DF}, while we have
$$-\frac{\sqrt{-b^4+b^2 p^2+p^2}}{\sqrt{b^4+p^2}} K(k^2)\le \s \le  \frac{\sqrt{-b^4+b^2 p^2+p^2}}{\sqrt{b^4+p^2}} K(k^2)\,.$$

\subsection{Fermionic Lagrangian}
\label{app:fermions}

In this section we compute the quadratic part of the fermionic Lagrangian~\cite{Cvetic:1999zs}, which for type IIB GS superstring (in Lorentzian signature) is \be
\label{Lf-general}
\mathcal L_{\rm f}=\bar \Psi^I \left(\sqrt h \,h^{\a\b} \, \delta^{IJ}- \varepsilon^{\a\b} \tau_3^{IJ}\right) \Gamma_\a \left(\delta^{JK} D_\b+\frac 18 \tau_3^{JK} \pa_\b X^\mu H_{\mu\nu\lambda} \G^{\nu\lambda}+{e^{\Phi}\over 8}\mathcal F^{JK} \Gamma_\beta\right)\Psi^K,
\ee
where $\Gamma_\a$ is the pull-back of the ten-dimensional Gamma matrices onto the worldsheet, $D_\a$ is the pulled back covariant derivative $D_\mu= \pa_\mu +\frac 14 \Omega_\mu^{\a\b}\Gamma_{\a\b}$, with $\Omega$ the ten-dimensional spin connection (we follow here the notation of~\cite{Chen-Lin:2017pay}). 
The ten-dimensional Majorana-Weyl fermions $\Psi^I$ ($I=1,2$)  satisfy the chiral condition $\Gamma_{11}\Psi^I=\Psi^I$, and the $2\times 2$ Pauli matrices $\tau_1, \tau_2, \tau_3$ act on the indices~$I, J$. 
In the case of \text{AdS}$_5\times S^5$, there is no contribution from the $B$-field, and thus to $H_{(3)}$.  The  ten-dimensional dilaton $\Phi$ is simply related to the string coupling by $e^\Phi=g_s$, and 
the only contribution to the flux term is given by (self-dual) five-form 
$
{e^{\Phi} \over 8 }\mathcal F^{JK} =-{e^\Phi\over 8}  \slashed{F}_{(5)} \tau_2^{JK}
$\,.
In type IIB GS superstrings, it is useful to set $\Psi :\, = \Psi^1=\Psi^2$ to fix the $\kappa$-symmetry. For the background we are interested in, the general result in~\cite{Drukker:2000ep,Forini:2015mca} gives
\begin{subequations}
\bea
\label{L-fer-fin}
\mathcal L_{\rm f} &=& \mathcal L_{\rm kin}+\mathcal L_{\rm flux}\, ,
\\ \label{L-fer-kin-fin}
\mathcal L_{\rm kin} &=& 2i \sqrt{g} g^{\alpha\beta} \bar\Theta \Gamma_{a} e^{\varrho} \delta_\alpha^a \Big(\pa_\beta+\frac 14 \omega_\beta^{bc}\Gamma_{bc}-\frac 14 A_{\beta}^{ij} \Gamma_{ij} \Big)\Theta\,,
\\ 
 \label{L-fer-flux-fin}
\mathcal L_{\rm flux} &=& -2 i  e^{-2\varrho}\,  \bar\Theta \big(\XX \sqrt{1+\XX^2}\, \Gamma_{127}
+\YY \sqrt{1-\YY^2}\, \Gamma_{346} \big)\Theta\,,
\eea
\end{subequations}
where the $\Gamma-$matrices in \eqref{L-fer-flux-fin} are labelled with the $\grSO(8)$ normal indices. Moreover, the ten-dimensional spinor $\Theta$ is obtained from $\Psi$ via a local $\grSO(1,9)$ rotation which  allows one to  transform the quadratic fermionic term in the Green-Schwarz action into the action for a set of two-dimensional fermions~\cite{Drukker:2000ep}. 
The connections $A_{\a}^{ij}$ are defined in~\eqref{Atau}, \eqref{Asigma} and $\omega_\beta$ is the two-dimensional spin connection
$$\omega_\tau^{01}=\pa_\s \Lambda=e^{-2\Lambda}(\XX \XX'+\YY\YY')\,,$$
with all other components vanishing.  In writing~  
\eqref{L-fer-flux-fin}   the chirality condition $\Gamma_{11}\Theta=\Theta$ was used.


The trace relation for fermionic and bosonic mass matrices \cite{Drukker:2000ep, Forini:2015mca} which controls the universality of UV logarithmic divergences~\cite{Giombi:2020mhz}  is obeyed,
\be
\Tr\, \mathcal M+R^{(2)}= -8 \Big(\XX^2(1+\XX^2)+\YY^2(1-\YY^2)\Big)= \frac12 \Tr\left(\mathcal M_F \Gamma_\alpha \mathcal M_F \Gamma^\alpha\right)\,,
\ee
providing a consistency check of our result.

\section{Frequencies}
\label{app:D}
\subsection{Normal modes frequencies for the uncoupled fluctuations}
\label{app:Lame}

In this section we compute in a closed form the normal mode frequencies for the uncoupled fluctuations $\xi^i$, with $i=1,2,3,4$, discussed in section \ref{sec:fluctuations}, by recasting the equations of motion into an integrable Lam\'e differential equation.

\subsubsection{Lam\'e differential equations}

Let us consider the equation \eqref{K11} for the transverse modes $i=1,2$. We first change the spatial $\s$ coordinate as follows
\bea
\s \to x={\s\over \sqrt{1-m-n}}-K(m)-i K(1-m)\,,
\eea
and, after Fourier transforming $\xi(\s,\tau)\to e^{i\omega \tau} \hat\xi(\s)$, we scale the Fourier frequency $\omega$ as 
\bea
 \omega \to {\Omega\over \sqrt{1-n-m}}\,. 
\eea
Then, using the above transformations, as well as the explicit form of the solution \eqref{rsol}, the equation \eqref{K11} becomes 
\be
 -\hat\xi_i''(x)+2 m \, {\rm sn}^2(x\vert m) \hat\xi_i(x)=h \,\hat\xi_i(x)\,, \quad h=1+m-n+\Omega^2\, ,\quad i=1,2\,,
\ee
where $h$ is the effective  eigenvalue in the Lam\'e equation. 
For the single-gap Lam\'e problem, two independent solutions can be constructed~\cite{Braden1985}
$$
\hat\xi_{\pm}=\frac{H(x\pm \alpha)}{\Theta(x)} e^{\mp x Z(\alpha)}\,,
$$
where $H, \Theta, Z$ are Jacobi functions defined in Appendix~\ref{app:elliptic}, and the parameter $\alpha$ is determined by the condition
\be\label{cond-os-1}
{\mathrm{sn}}(\alpha\vert m)= \sqrt{{1+m-h\over m}} = \sqrt{{n-\Omega^2\over m}}\,. 
\ee
Imposing homogeneous Dirichlet conditions at $\s=\pm s/2$ on the general solution gives the quantization condition on the phase:
\be\label{cond-quant-lame-1}
4 K( m) \, Z(\alpha) = 2\pi i\ell\,, \qquad \ell =1, 2, \dots \qquad \text{i.e.} \quad \ell \in \mathbb N\,. 
\ee
The two equations \eqref{cond-os-1} and \eqref{cond-quant-lame-1} relate the eigenvalues $h$, and thus the original frequencies $\omega$, to the quantum number $\ell$. 

We can repeat the above analysis in complete analogy  for the transverse modes in the compact space, that is $i=3,4$. 
Indeed, the equation \eqref{K33} can be brought into a standard Lam\'e form by transforming the spatial $\s$ coordinate as follows
\be
\s \to x= \frac{\sqrt{\nL}}{\sqrt{\nL-\mL (1-\nL)}} \s -K(\mL) \,,
\ee
and by rescaling the Fourier frequencies as
\be
\omega \to \omega=\Omega {\sqrt{\nL}\over \sqrt{\nL-\mL (1-\nL)}}\,.
\ee
Then, using the above transformations, as well as the explicit form of the solution \eqref{rhotildesol}, the equation of motion \eqref{K33} becomes 
\be\label{lame-s}
-\hat\xi_i''(x)+2 \mL \, {\rm sn}^2(x\vert \mL) \hat\xi_i(x)=h \,\hat\xi_i(x)\,, \quad h=1+\mL-{\mL\over \nL}+\Omega^2\,, \qquad i=3, 4\,,
\ee
where the condition on the effective eigenvalue $h$ reads now
\be
\label{cond-os-2}
{\mathrm{sn}}(\alpha\vert \mL) = {\sqrt{1+\mL-h}\over \sqrt{\mL}} =\sqrt{\frac 1{\nL}-\frac{\Omega^2}{\mL}}\,. 
\ee
The Dirichlet boundary conditions then read formally as in \eqref{cond-quant-lame-1}, where now $m\to \mL$, that is 
\be
\label{cond-quant-lame-2}
4 K( \mL) \, Z(\alpha) = 2\pi i\ell\,, \qquad \ell =1, 2, \dots \qquad \text{i.e.} \quad \ell \in \mathbb N\,. 
\ee
and $\alpha$ satisfies the above condition \eqref{cond-os-2}.

  \subsubsection{Energy densities}
  \label{app:energydensity}

In order to clarify the implicit relation between the normal mode frequencies and the quantum numbers, as well as to compute the various asymptotic regimes for large and/or small $\mathcal L$, it is useful to introduce  ``density'' functions  of the frequencies $ \rho_\omega, \tilde \rho_\omega $, following \cite{Giombi:2022anm}. 

We start by considering the transverse modes in S$^5$, that is $i=3,4$. 
The quantization condition \eqref{cond-quant-lame-2} can be rewritten as~\cite{Giombi:2022anm}
\be\label{def-rho-density}
\ell \, ={1\over 2\pi i} \int_0^{\alpha_\ell} \tilde f'(\alpha) \dif \alpha ={1\over 2\pi i} \int_{\tilde{\omega}_0}^{\omega_\ell} \tilde f'(\alpha)  \dfrac{\dif \alpha}{\dif\omega} \dif \omega =\, : \int_{\tilde{\omega}_0}^{\omega_\ell} \tilde\rho_\omega(\omega)\dif \omega    \,, \qquad  \ell\in\mathbb N\,,
\ee
where the function $\tilde f$ is defined by 
\be\label{def-f-lame}
\tilde f(\alpha) :\, = 4 K( \mL) \, Z(\alpha)\,. 
\ee
The derivatives $\tilde f'(\alpha)$ and $\dfrac{\dif \alpha}{\dif\omega}$ can be computed using the relations \eqref{jacobi123}, as well as the on-shell condition \eqref{cond-os-2}.
We obtain 
\bea\label{density-s}
\Scale[0.90]{
\tilde\rho_\omega(\omega)=
\frac{2 \omega  \left(\mL
   \left(\nL-1\right)+\nL\right) \left(K\left(\mL\right) \left(\omega
   ^2 \left(\mL \left(\nL-1\right)+\nL\right)-\mL+\nL\right)-\nL
   E\left(\mL\right)\right)}{\pi  \sqrt{\nL} \sqrt{\left(\omega ^2
   \left(\mL \left(\nL-1\right)+\nL\right)-\mL\right) \left(\omega ^2
   \left(\mL \left(\nL-1\right)+\nL\right)+\mL
   \left(\nL-1\right)\right) \left(\omega ^2 \left(\mL
   \left(\nL-1\right)+\nL\right)-\mL+\nL\right)}}\,.\quad\quad
}
\eea
The extreme of integration $\tilde{\omega}_0$ is given by the condition $\tilde f(0)=0$, that implies $\alpha=0$ as lowest root, since $ \mL \in (0, 1)$ according to \eqref{rangenLkL}.%
\footnote{For $\mL\to 1$ then the elliptic function $K$ has a logarithmic divergence in $1-m$, while $Z(\alpha, 1)=\tanh\alpha$. }
Then, from the relation \eqref{cond-os-2}, we obtain 
\be\label{omega0s}
\tilde{\omega}_0=\sqrt{\frac{\mL}{\mL \left(\nL-1\right)+\nL}}\,. 
\ee
Notice that the normal mode frequencies $\omega_\ell$ with $\ell\in \mathbb N$ are strictly greater than $\tilde{\omega}_0$.

We can proceed in a similar manner for the transverse modes in AdS$_5$ ($i=1,2$) to compute the  density $\rho_\omega$. 
In this case, we can define an auxiliary function $f$ according to the quantization condition \eqref{cond-quant-lame-1}, that is 
\be\label{def-f-lame-2}
 f(\alpha) :\, = 4 K( m) \, Z(\alpha)\,,
\ee
and an extreme of integration $\omega_0$, such that 
\be
\ell \, ={1\over 2\pi i} \int_0^{\alpha_\ell} f'(\alpha) \dif \alpha =\, : \int_{{\omega}_0}^{\omega_\ell}\rho_\omega(\omega)\dif \omega    \,, \qquad  \ell\in\mathbb N\,.
\ee
We obtain for the energy density
\bea\label{density-ads}
\rho_\omega(\omega)=\frac{2 \omega \,  (1-m-n)\, \left(K(m) \left(\omega ^2 (1-m-n)-n+1\right)-E(m)\right)}{\pi 
   \sqrt{\left(\omega ^2 (1\!-\!m\!-\!n)\!-\!n\right) \left(\omega ^2 (1\!-\!m\!-\!n)\!-\!n\!+\!1\right) \left(\omega ^2 (1\!-\!m\!-\!n)\!+\!m\!-\!n\right)}}\,.\qquad
\eea
The only subtle point in this case is the extreme of integration $\omega_0$. 
Indeed, looking at the equation 
\be\label{f0}
f(\alpha)=0\,,
\ee
with $f$ given in \eqref{def-f-lame-2}, we need to recall that now $m \le  \tfrac 12$ \eqref{rangenm}, and in particular for $m\to -\infty$ the elliptic function $K$ vanishes. 
In order to correctly compute $\omega_0$, it is useful to split the range of the parameters $(n, m)$ according to the sign of $m$. 
For $m\in (0,1/2)$, the lowest solution to equation \eqref{f0} for any $m$, is given by $\alpha=0$, which implies 
\be\label{omega0ads-A}
\omega_0= \frac{\sqrt{n}}{\sqrt{1-m-n}}\,.
\ee
As discussed in section~\ref{subsec-antiparallel} this corresponds to the case {\bf 1}, that is the Region A$_L$, which plays a role once we examine the antiparallel lines limit in section~\ref{sec:omega-antiparallel}.

\noindent
For negative values of $m$,  it is useful to use the modular parameters~\eqref{mprimenprime}. 
In terms of $(n', m')$ the two conditions~\eqref{cond-os-1} and~\eqref{cond-quant-lame-1} read formally the same
\be
{\mathrm{sn}}(\alpha\vert m') = {\sqrt{1+m'-h}\over \sqrt{m'}} =\sqrt{{n'- \Omega^2\over m'}}\,,  \qquad 4 K(m') Z(\alpha, m')=2 \pi i \ell\,, \qquad \ell\in \mathbb N\,,
\ee
where now the rescaled frequencies are obtained by $\omega = {\Omega \over \sqrt{1+m'-n'}}$.%
\footnote{It is straightforward to see that by using the following transformations
$$
\sigma\to \sqrt{1+m'-n'} (x+i K(1-m')+K(m')\,, \qquad \omega\to {\Omega \over \sqrt{1+m'-n'}}\,,
$$
the equation of motion \eqref{K11} reduces to the following Lam\'e form
$$
 -\hat\xi_i''(x)+2 m' \, {\rm sn}^2(x\vert m') \hat\xi_i(x)=h \,\hat\xi_i(x)\,, \quad h=1+m'-n'+\Omega^2\, ,\quad i=1,2\,. 
$$
}
Then, the lowest zero of \eqref{f0} in terms of parameters $(n', m')$ is given by  
\be\label{omega0ads-B}
\hat \omega_0= \frac{\sqrt{n'}}{\sqrt{1+m'-n'}}=\frac{\sqrt{n-m}}{\sqrt{1-m-n}}\,. 
\ee
This is the correct extreme of integration for the region B$_L$ (see also case {\bf 2} in section~ \ref{subsec-antiparallel} in relation to the antiparallel lines limit), and thus for expanding in large $\mathcal L$ as discussed in section~ \ref{sec:largeL-frequencies}. 
Our functions $\rho_\omega, \tilde \rho_\omega$ \eqref{density-ads},  \eqref{density-s} reduce to the energy density computed in \cite{Giombi:2022anm} in the straight line limit \eqref{straight-line-lim} with rescaling our $\omega$  by $1/c$ and renaming it $E$, that is $\omega \to {E\over c}$, where $c$ in  \cite{Giombi:2022anm} corresponds to our parameter $\sqrt{\mL}$. It is straightforward to see that $\tilde{\omega}_0$ in \eqref{omega0s} reduces to ${1\over c}$ in this limit \eqref{straight-line-lim}, however, only $\hat \omega_0$ in \eqref{omega0ads-B} reduces to the same value in agreement with \cite{Giombi:2022anm}, while $\omega_0$ in \eqref{omega0ads-A} reduces to $\sqrt{1-c^2}/c$. 

\bigskip

It is not difficult to check that in the limit of large quantum number $\ell$, the exact analysis for the uncoupled scalar sector reproduces the WKB result. One starts from the quantization condition written in the form~\eqref{def-rho-density} for the $S^5$ case, and its analogue for the $\text{AdS}_5$ case. 
To single out the behavior for large $\ell$ it is sufficient to consider the  large $\omega$ expansions of the density~\eqref{density-s} and \eqref{density-ads}
\be\label{rhoexp}
\begin{split}
\!\!\!\!
\tilde\rho_\omega(\omega)&=\frac{2 K(\mL)
   \sqrt{-\frac{\mL}{\nL}+\mL+1}}{\pi }
   +\frac{1}{\omega ^2}\frac{K(\mL)
   \left(\frac{\mL}{\nL}-\mL+1\right)-2
   E(\mL)}{\pi  
   \sqrt{-\frac{\mL}{\nL}+\mL+1}}+  O(\omega^{-4}),\\
\!\!\!\!
\rho_\omega(\omega)&=\frac{2 K(m)
   \sqrt{1-m-n}}{\pi }+ \frac{1}{\omega^2}\frac{K(m) (1-m+n)-2 E(m)}{\pi \sqrt{1-m-n}}+  O(\omega^{-4}). \\
   \end{split}
\ee 
One proceeds substituting them respectively in~\eqref{def-rho-density} (and in its analog for $\text{AdS}_5$) and integrating term by term between the lower bounds 
($\tilde{\omega}_0$ in \eqref{omega0s} for $S^5$ and $\omega_0$ in \eqref{omega0ads-A} for $\text{AdS}$) and~$\omega_\ell$. In doing this, we have to be careful with possible constant term coming from the lower extreme of integration in the integrals $\int_{\omega_0}^{\omega_\ell}\omega^{-2k}$ in the expansion of the densities. 
To single out this  term one has to subtract from the integral of the density its leading behavior and take the limit of $\omega_\ell\to\infty$, 
 for example in the case of~$S^5$
\be\label{r0const}
r_0 \equiv \lim_{\omega_\ell^{S^5}\to\infty}\,\Big(\,\int_{\tilde{\omega}_0}^{\omega_\ell^{S^5}} \tilde\rho(\omega)d\omega-\frac{2 K(\mL)}{\pi}
   \sqrt{1-\frac{\mL}{\nL}+\mL}\,\omega_\ell^{S^5}\,\Big) \,.
\ee 
In both cases, the result is totally independent on the modular parameters and is given by $r_0=-1$. Therefore, its effect  is  simply to shift 
the integer $\ell$ of one unit. Inverting the expansion, one obtains 
\be
\label{Omegas}
\begin{split}
&\omega_\ell^{S^5}= \tfrac{\pi \,(\ell+1) }{2 K(\mL)  \sqrt{1-\frac{\mL}{\nL}+\mL}}
   + \tfrac{K(\mL) (\mL+\nL-\mL \nL)-2
   \nL E(\mL)}{\pi  \nL
   \sqrt{1-\frac{\mL}{\nL}+\mL}} \tfrac{1}{\ell+1}+O(\tfrac{1}{(\ell+1)^2})\,,
   \\
&\omega_\ell^{\text{AdS}_5}= \tfrac{\pi }{2 K(m) \sqrt{1-m-n}}\,(\ell+1) 
+\tfrac{K(m) (1-m+n)-2 E(m)}{\pi  \sqrt{1-m-n}}\tfrac{1}{\ell+1}+O(\tfrac{1}{(\ell+1)^2})\,.
\end{split}
\ee
Recalling the expressions for the range of the spatial worldsheet coordinate, equations~\eqref{sS5} and~\eqref{sAdS5}, as well as those of the angular momentum~\eqref{L} and of the energy~\eqref{E}, one can verify that the term linear in $\ell$ and the term of order $1/\ell$ above coincide with the corresponding expressions appearing in~\eqref{omega34-semicl} and~\eqref{omega12-semicl} respectively.

\subsection{WKB analysis of the normal mode frequencies}
\label{app:WKB}

In this section we elucidate the WKB analysis for the coupled transverse modes $\xi^i$, with $i=5, \dots, 8$. 
For these modes, the equation of motions \eqref{eom-schem} represent a system of coupled $4\times 4$ differential equations. 
After Fourier transforming, we write schematically the differential equation as 
\be\label{Schr-matrix}
\left( -\pa_\s^2-\omega^2-2 i \omega \mathbb A_\tau +2 \mathbb A_\s\pa_\s+ \pa_\s \mathbb A_\s +\mathbb A_\tau^2-\mathbb A_\s^2-\sqrt h\mathcal M \right)\hat\Xi(\s)=0\,,
\ee
where now $\hat\Xi$ is a $\mathbb C^4$ wave function. We have indicated $\mathbb A_\tau, \mathbb A_\s$ the $4\times 4$ antisymmetric matrices whose components are the connections \eqref{Atau}, \eqref{Asigma} respectively, computed in section \ref{sec:fluctuations}. 
We can always rescale the radial component $\hat\Xi$ to remove the connection $\mathbb A_\s$, however we find more convenient to not perform such a transformation. 
We assume that there exists a $4\times 4$ matrix $\mathrm P$, which is defined by the relation
\be
\label{def-P-Riccati}
\pa_\s \hat\Xi (\s)=i\,  \mathrm P(\s)\, \hat\Xi(\s)\,.
\ee
In general, $\mathrm P$ is a (smooth) function of the worldsheet coordinate $\s$. Using the above relation, the equation \eqref{Schr-matrix} becomes a Riccati matrix equation for $\mathrm P$:
\be
\label{eq-P-matrix}
 -i \pa_\s \mathrm P +\mathrm P^2 -\omega^2-2 i \omega \mathbb A_\tau +2 i \mathbb A_\s \mathrm P+ \pa_\s \mathbb A_\s+\mathbb A_\tau^2-\mathbb A_\s^2-\sqrt h \, \mathcal M=0\,. 
\ee
In complete analogy to the semiclassical expansion \eqref{xihat-exp}, we can assume the following large $\omega$ approximation for the matrix $\mathrm P$
\be\label{exp-P-matrix}
\mathrm P=\omega \left(\mathrm P_0+\frac1\omega  \mathrm P_1+\frac 1{\omega^2} \mathrm P_2+\dots\right)\,,
\ee
where now all the $\mathrm P_p$, with $p=0, 1\dots$ are $4\times 4$ matrices.
With the ansatz \eqref{exp-P-matrix}, the Riccati equation \eqref{eq-P-matrix} can be solved order by order in large $\omega$. At the leading order, this is nothing but
\be
\mathrm P_0^2-\mathrm I=0\,, \qquad \text{that is} \qquad \mathrm P_{0,\pm}=\pm \mathrm I\,,
\ee
where $\mathrm I$ is the $4\times4$ identity matrix. 
At the next leading order, we obtain an algebraic equation for $\mathrm P_1$
\be
\left\{ \mathrm P_{0}, \mathrm P_1\right\}- 2 i\mathbb A_\tau+2 i \mathbb A_\s \mathrm P_0=0\,, 
\qquad 
\text{which gives}
\qquad
\mathrm P_{1,\pm}=- i \left( \mathbb A_\s \mp  \mathbb A_\tau  \right)\,. 
\ee
At the next-to-next leading order, the algebraic equation for $\mathrm P_2$ reads
\be
-i\pa_\s \mathrm P_1 +\left\{ \mathrm P_0, \mathrm P_2\right\} +\mathrm P_1^2+2 i\mathbb A_\s \mathrm P_1+ \pa_\s \mathbb A_\s+\mathbb A_\tau^2-\mathbb A_\s^2-\sqrt h \, \mathcal M=0\,,
\ee
and a formal solution for $\mathrm P_2$ is 
\bea
\label{eq-P2-matrix}
\mathrm P_{2, \pm} &=&\pm{1\over 2} \left(i \pa_\s \mathrm P_{1, \pm}-\mathrm P_{1, \pm}^2-2 i\mathbb A_\s \mathrm P_{1, \pm}- \pa_\s \mathbb A_\s-\mathbb A_\tau^2+\mathbb A_\s^2+\sqrt h \, \mathcal M\right)
\\ \nn
&=& \frac i2 \left( [ \mathbb A_\s, \mathbb A_\tau] -\pa_\s \mathbb A_\tau \pm \sqrt h \mathcal M\right)\,. 
\eea

The above formal expansion can be used to reconstruct the wave function $\hat\Xi$ via \eqref{def-P-Riccati}. At the leading order, since $\mathrm P_0$ is a constant diagonal matrix, the Dirichlet boundary conditions at $\s=\pm s/2$, give the quantization condition
\be
\omega s= \ell \pi\,, \qquad \ell\in \mathbb N\,.
\ee
At the next leading order, the connections \eqref{Atau}, \eqref{Asigma} modify this relation. 
The solution $\hat \Xi$ \eqref{def-P-Riccati} can be written formally as path-ordered exponential matrix acting on a constant vector $v_\pm$:
\be
\hat\Xi_\pm (\s)=\left[ \mathcal P \exp \int_{-s/2}^\s \left(\pm i \omega \s' + \mathbb A_\s (\s')\mp \mathbb A_\tau(\s') \right) \dif \s' \right]v_\pm\,.
\ee
As before the Dirichlet boundary conditions at $\s=-s/2$ will give a relation among the trivial solutions, while at $\s=s/2$, we obtain
\be
e^{i \omega s} \left[  \mathcal P \exp \int_{-s/2}^{s/2} \left( \mathbb A_\s (\s)- \mathbb A_\tau(\s)\right)\dif \s \right]v_-=
e^{-i \omega s} \left[  \mathcal P \exp \int_{-s/2}^{s/2} \left( \mathbb A_\s (\s)+ \mathbb A_\tau(\s)\right)\dif \s \right]v_-\,.
\ee
Defining the holonomy matrix as 
\be\label{def-U1}
U^{(1)}_\pm(s/2) :\, = \mathcal P \exp \int_{-s/2}^{s/2} \left( \mathbb A_\s (\s) \mp \mathbb A_\tau(\s)\right)\dif \s\,,
\ee
this implies that in order to have a non-trivial solution satisfying the Dirichlet boundary conditions, we should have 
\be\label{det-qc}
\det \left( {U^{(1)}_+}^{-1}(s/2) U^{(1)}_-(s/2) - e^{2 i \omega s} \mathrm I\right)=0\,.
\ee
If the holonomies $U^{(1)}_\pm$ are $\grSO(4)$ matrices, then the above quantization relation can be written as 
\be\label{omega58-semicl}
\omega={\pi \ell \over s}+{\theta_I \over 2 s}\,, \qquad \ell\in \mathbb N\,,  \quad I=1,  \dots, 4\,,
\ee
where $\theta_I$ are the eigenvalues of $ {U^{(1)}_+}^{-1}(s/2) U^{(1)}_-(s/2)$. The above expression is what reported in equation \eqref{omega58-semicl-main} in section \ref{sec:freq}. 
The same argument can be applied to the next-to-next leading order in $\omega$, where formally the quantization condition is now given by \eqref{det-qc}, where the exponential matrices $U_\pm (s/2)$ contain also the matrix $\mathrm P_{2, \pm}$ \eqref{eq-P2-matrix}, that is
\be\label{def-U2}
U^{(2)}_\pm(s/2) :\, = \mathcal P \exp \int_{-s/2}^{s/2} \left( \mathbb A_\s (\s) \mp \mathbb A_\tau(\s) -\frac{1}{2\omega} \left([\mathbb A_\s(\s), \mathbb A_\tau(\s)] -\pa_\s \mathbb A_\tau(\s) \pm \sqrt h\mathcal M(\s)\right)\right)\dif \s\,.
\ee
It should be noticed that the above path-ordered exponential matrices, as \eqref{def-U1} or \eqref{def-U2}, should be expanded again in large $\omega$, and thus at the leading orders only few terms contribute to the wave function $\hat\Xi$ \eqref{def-P-Riccati}.

\section{Elliptic functions}
 \label{app:elliptic}
 
In this appendix we collect miscellaneous formulae for elliptic functions used throughout the main text.  
The incomplete elliptic integrals of the first, second and third kind are defined as 
\begin{align}
F(\varphi|m) &= \int_{0}^{\varphi} 
\frac{d\theta}{\sqrt{1 - m \sin^{2}\theta}},
\label{Fincomplete} \\
E(\varphi|m) &= \int_{0}^{\varphi}
\sqrt{1 - m \sin^{2}\theta}\, d\theta,
\label{Eincomplete} \\
\Pi(\varphi|n,m) &= \int_{0}^{\varphi}
\frac{d\theta}{(1 - n \sin^{2}\theta)\sqrt{1 - m \sin^{2}\theta}}.
\label{Piincomplete}
\end{align}
The related complete elliptic integrals are obtained by evaluating the above at $\varphi=\tfrac{\pi}{2}$
\be \label{KEP}
K(m) = F\!\left(\frac{\pi}{2}\,\middle|\,m\right)\,,\qquad
E(m) = E\!\left(\frac{\pi}{2}\,\middle|\,m\right)\,,\qquad
\Pi(n|m) = \Pi\!\left(\frac{\pi}{2}\,\middle|\,n,m\right)\,.
\ee

Defining the Jacobi amplitude as  
\be\label{jacobiam}
\alpha={\rm am}(z\vert m),~~~~{\rm where}~~~~z=\int_0^\alpha d\theta\  (1-m'\,\sin^2\theta)^{-1/2}\,,
\ee
with $m'=1-m$, the Jacobi elliptic functions  $\sn,\cn,\dn$  defined as
\be
\sn(z\,|\,m)=\sin \alpha,~~~~~~~\cn(z\vert m)=\cos \alpha,~~~~~~\dn(z\vert m)=(1-m\,\sin^2 \alpha)^{1/2}
\ee
are doubly-periodic functions of $z$, with real-valued periods that are either $2K(m)$ ($\dn$) or $4K(m)$ ($\sn$ and $\cn$) and purely imaginary periods that are either $2i\,K(m')$ ($\sn$) or $4i\,K(m')$ ($\cn$ and $\dn$). 

Other Jacobian elliptic functions useful for us are
\begin{eqnarray}
&&{\rm cd}(z\vert m)=\frac{\cn(z\vert m)}{\dn(z\vert m)},~~~~~~~~~~ {\rm sd}(z\vert m)=\frac{\sn(z\vert m)}{\dn(z\vert m)}\,,\\
&& {\rm ns}(z\vert m)=\frac{1}{\sn(z\vert m)},~~~~~~~~~~ {\rm nd}(z\vert m)=\frac{1}{\dn(z\vert m)}\,.
\end{eqnarray}

Useful relations   are
\be\label{jacobi123}
\begin{split}
&{\rm cn}^2(z\vert m)+{\rm sn}^2(z\vert m) = 1\,,\qquad {\rm dn}^2(z\vert m)+m\,  {\rm sn}^2(z\vert m) = 1\,,\\
&\dfrac{\dif}{\dif z} {\rm sn}(z\vert m) = {\rm cn}(z\vert m){\rm dn}(z\vert m)\,.
\end{split}
\ee

The Jacobi $H$, $\Theta$ and $Z$ functions are defined in terms of the Jacobi $\vartheta$ functions as
\be\label{jacobidef}
H(z\vert m) = \vartheta_1\left(\frac{\pi\,z}{2\,\mathbb{K}}, q\right), \qquad
\Theta(z\vert m) = \vartheta_4\left(\frac{\pi\,z}{2\,\mathbb{K}}, q\right),\qquad
Z(z\vert m) = \frac{\pi}{2\,\mathbb{K}}\,\frac{\vartheta_4'(\frac{\pi\,z}{2\,\mathbb{K}}, q)}
{\vartheta_4(\frac{\pi\,z}{2\,\mathbb{K}}, q)}\,,
\ee
where 
\be
q=q (m) =  \exp\Big(-\pi\frac{K(m')}{K(m)}\Big).
\ee
Useful periodicities for them are
\ba\label{periodicityH}
H(z+2K(m) \,|\,m)&=&-H(z \,|\,m),\\
\Theta(z+2K(m) \,|\,m)&=&\Theta(z \,|\,m),\\
Z(z+2K(m) \,|\,m)&=&Z(z\vert m)\,.
\ea
A useful representation for the Jacobi $Z$ function is  
\begin{equation}\label{ZetaJacobi}
{Z}(\alpha; m)=\int_0^\alpha dz \ {\rm dn}^2(z; m)  \  - \frac{E(m)}{K(m)}\, \alpha\, ,
\end{equation}
where $\alpha$ is the Jacobi amplitude in~\eqref{jacobiam}. It should be noticed that the input argument of $Z(\alpha; m)$ in~\eqref{ZetaJacobi} differs from the one of $Z(z\vert m)$ in~\eqref{jacobidef} and coincides with the one used by the  $\mathrm{Mathematica}$ function $\mathrm{JacobiZeta}[u,m]$. In other words, 
it is the Mathematica function JacobiZeta$[$JacobiAmplitude$[z, m], m]$ that reproduces the right-hand-side of~\eqref{jacobidef}.

\subsection*{Complete Elliptic integral of the third kind and its expansion}

The Taylor expansion of  the elliptic integral \(\Pi(n,m)\) with respect to the first argument, around a regular point \(n_0\), is given by
\begin{equation}
\label{Taylor-expans}
\Pi \left(n\left|m\right.\right) =
\frac{\pi}{2} \sum_{j=0}^\infty \frac{1}{j!} \left(\frac{1}{2}\right)_j
F_1\left(\frac{1}{2} + j; 1 + j, \frac{1}{2}; 1 + j; n_0, m\right) \left(n - n_0\right)^j,
\end{equation}
where \(F_1\left(a; b_1, b_2; c; z_1, z_2\right)\) is the Appell function~\cite{NIST}. The latter  simplifies to a standard hypergeometric function when \(z_1 = z_2\) 
\begin{equation}
F_1\left(\frac{1}{2} + j; 1 + j, \frac{1}{2}; 1 + j; m, m\right) = \, {}_2F_1\left(j + \frac{1}{2}, j + \frac{3}{2}; j + 1; m\right), 
\end{equation}
and a further simplification occurs  using the properties of the Gaussian hypergeometric function under differentiation, 
\begin{equation}
\frac{1}{j!} \left(\frac{1}{2}\right)_j \, {}_2F_1\left(j + \frac{1}{2}, j + \frac{3}{2}; j + 1; m\right)
= \frac{\sqrt{\pi}}{2 \Gamma \left(j + \frac{3}{2}\right)} \left. \frac{\partial^j}{\partial x^j} \left({}_2F_1\left(\frac{1}{2}, \frac{3}{2}; 1; x\right)\right)\right|_{x = m}.
\end{equation}
Using this and the identity
\begin{equation}
{}_2F_1\left(\frac{1}{2}, \frac{3}{2}; 1; x\right) = \frac{2E(x)}{\pi(1 - x)}
\end{equation}
 into~\eqref{Taylor-expans} for \(n_0 = m\), one obtains the compact form
\begin{equation}\label{Picompact}
\Pi \left(n\left|m\right.\right) = \frac{\sqrt{\pi}}{2} \sum_{j=0}^\infty \left. \frac{\partial^j}{\partial x^j} \left(\frac{E(x)}{1 - x}\right)\right|_{x = m} \frac{\left(n - m\right)^j}{\Gamma \left(j + \frac{3}{2}\right)}.
\end{equation}
 In this paper we also use  the modular transformations
\be
\label{modtransfKE}
K(z)=\frac{K\left(\frac{z}{z-1}\right)}{\sqrt{1-z}}~\text{/;}~| \arg (1-z)| <\pi\qquad
E(z)=\sqrt{1-z} E\left(\frac{z}{z-1}\right)~\text{/;}~| \arg (1-z)| <\pi\,,
\ee
and
  \be\label{Pimodul} 
  \Pi (n|m)=\frac{\Pi \left(\frac{n}{n-1}|\frac{m}{m-1}\right)}{\sqrt{1-m} (1-n)}\text{/;}\,\,| \arg (1-m)| <\pi \land | \arg (1-n)| <\pi\,,
  \ee
as well as the identity
   \be\label{Pitransf}
  \begin{split}
  \Pi (n|m)=&K(m)+\frac{i \pi }{2 \sqrt{\frac{m}{n}-m+n-1}}-\Pi \left(\!\left.\frac{m}{n}\right|m\!\right)\\
  &\text{/;}(0<n<1\land 0<m<1)\lor
   (0<n<1\land 0<\arg (m)\leq \pi ).
   \end{split}
  \ee

\subsubsection*{A useful integral representation of $\Pi(m,n)$}
\label{app:expansionPi}

In this appendix we derive an integral representation for $\Pi(m,n)$ that is
well suited to analyze the joint limit
\begin{equation}\label{limmnPi-1}
  m \to 1, \qquad n \to 1,
\end{equation}
with the ratio
\begin{equation}
  \frac{1-m}{1-n} = \alpha
  \label{eq:param-alpha}
\end{equation}
kept finite. Since we always assume
\(
  0 \leqslant m \leqslant n \leqslant 1,
\)
we have $\alpha \geqslant 1$.
We start from the canonical integral representation
\begin{equation}
  \Pi(m,n)
  =
  \int_0^{\pi/2}
  \frac{d\varphi}{\bigl(1-n\sin^2\varphi\bigr)\sqrt{1-m\sin^2\varphi}}\,.
  \label{eq:Pi-standard}
\end{equation}
The potential singularity in the limit $m,n\to 1$ comes from the region near
the upper endpoint $\varphi \simeq \pi/2$, where $\sin^2\varphi \simeq 1$.
We now perform the change of variables
\(
  \varphi = \mathrm{arccot} (t) ,
\)
which maps the interval $(0,\pi/2)$ to the positive half-line $(0,\infty)$ and
moves the singular region to $t=0$. In these variables the integral becomes
\begin{equation}
  \Pi(m,n)
  =
  \int_0^\infty dt\;
  \frac{\sqrt{1+t^2}}{\sqrt{1-m+t^2}\,\bigl(1-n+t^2\bigr)}\,.
  \label{eq:Pi-S2}
\end{equation}
Next, we make explicit the dependence on $n$ by rescaling
$t \to \sqrt{1-n}\,t$ and using the parametrization $m = 1 - \alpha(1-n)$ (see eq. \eqref{eq:param-alpha}).
This gives
\begin{equation}
  \Pi(m,n)
  =
  \int_0^\infty dt\;
  \frac{\sqrt{1+(1-n)t^2}}{(1-n)(1+t^2)\sqrt{t^2+\alpha}}.
  \label{eq:Pi-S4}
\end{equation}
The overall factor $1/(1-n)$ makes the leading divergence manifest as
$n \to 1$.  However, if one naively expands the square root in the numerator
in powers of $(1-n)$, all the resulting integrals over $t$ diverge,
except for the leading-order term. To obtain a
representation suitable for a systematic expansion in $(1-n)$, we rewrite the
square root using the integral identity
\begin{equation}
  \sqrt{1+(1-n)t^2}
  =
  \frac{2}{\pi}
  \int_0^\infty du\;
  \frac{1+(1-n)t^2}{u^2+1+(1-n)t^2},
  \label{eq:identity-root}
\end{equation}
which follows from
\(
\int_0^\infty \! du/(u^2+a^2) = \pi/(2a)
\)
with $a^2 = 1+(1-n)t^2$.
Inserting \eqref{eq:identity-root} into \eqref{eq:Pi-S4} we obtain the
double integral
\begin{equation}
  \Pi(m,n)
  =
  \int_0^\infty dt\int_0^\infty du\;
  \frac{2\bigl[1+(1-n)t^2\bigr]}{\pi(1-n)\,(1+t^2)\sqrt{t^2+\alpha}\;
  \bigl[u^2+1+(1-n)t^2\bigr]}\,.
\end{equation}
We now exchange the order of integration and first integrate over $t$. For
this purpose it is convenient to use the partial fraction decomposition
\begin{equation}
  \frac{2\bigl(1+(1-n)t^2\bigr)}{\pi (n-1)\bigl(t^2+1\bigr)
   \bigl(u^2+1+(1-n)t^2\bigr)}
  = \frac{2}{\pi (u^2+n)}
  \left[
    \underbrace{\frac{ n}{ (1-n)\bigl(t^2+1\bigr)}}_{\mathds{I}_1}
    +\underbrace{\frac{ u^2}{ u^2+1+(1-n)t^2}}_{\mathds{I}_2}
  \right].
\end{equation}
We first consider the contribution from $\mathds{I}_1$. 
In this case, both
integrations over $t$ and $u$ can be carried out explicitly, and we obtain
\begin{equation}
  \frac{\sqrt{n}\,\sec^{-1}\!\bigl(\sqrt{\alpha}\bigr)}{\sqrt{\alpha -1}\,(1-n)}.
\end{equation}
For the term associated with $\mathds{I}_2$, integrating over $t$ and
simplifying the result yields
\begin{equation}
  \int_0^\infty du\;
  \frac{u^2 \bigl[
   2 \log\!\bigl(\sqrt{1+u^2}+ \sqrt{
      1+u^2 - (1-n)\alpha}\bigr)-\log(1-n) - \log\alpha\bigr]}{\pi \,\sqrt{
 1+u^2}\,\bigl(n+u^2\bigr)\,\sqrt{1+u^2 -(1-n)\alpha}}\,.
\end{equation}
Collecting the two contributions, we arrive at the final representation
\begin{equation}
\label{eq:Pi-final-int}
  \Pi(m,n)
  =
  \frac{\sqrt{n}\,\sec^{-1}\!\bigl(\sqrt{\alpha}\bigr)}{\sqrt{\alpha \!-\!1}\,(1\!-\!n)}
  \!+\! \int_0^\infty du\;
  \frac{u^2 \bigl[
   2 \log\!\bigl(\sqrt{1\!+\!u^2}\!+\! \sqrt{
      1\!+\!u^2 \!-\! (1\!-\!n)\alpha}\bigr)\!-\!\log(1\!-\!n) \!-\! \log\alpha\bigr]}{\pi \,\sqrt{
 1\!+\!u^2}\,\bigl(n\!+\!u^2\bigr)\,\sqrt{1\!+\!u^2 \!-\!(1\!-\!n)\alpha}}\,.
\end{equation}
In terms of the original parameter $m$ and $n$,  the above formula reads
\begin{equation}
\label{eq:Pi-final-int2}
  \Pi(m,n)
  =
  \frac{\sqrt{n}\,\sec^{-1}\!\bigl(\sqrt{\frac{1-m}{1-n}}\bigr)}{\sqrt{(1-n)(n-m)}}
  \!+\! \int_0^\infty du\;
  \frac{u^2 \bigl[
   2 \log\!\bigl(\sqrt{1\!+\!u^2}\!+\! \sqrt{
      m\!+\!u^2 }\bigr)\!-\!\log(1\!-\!m) \bigr]}{\pi \,\sqrt{
 1\!+\!u^2}\,\bigl(n\!+\!u^2\bigr)\,\sqrt{m\!+\!u^2 }}\,.
\end{equation}
The advantage of \eqref{eq:Pi-final-int2} is that, under the limits \eqref{limmnPi-1}-\eqref{eq:param-alpha}, the divergent
contribution is completely factorized. In fact, the integrand
naturally splits into a term proportional to $-\log(1-m)$, which
gives the explicit logarithmic divergence, and a remainder that is
finite at $m,n=1$. Moreover, expanding the integrand in powers of
$(1-n)$ and $(1-m)$ generates only convergent integrals over $u$,
so that \eqref{eq:Pi-final-int} provides a convenient starting
point for a controlled asymptotic expansion around $m,n \to 1$
with fixed $\alpha = (1-m)/(1-n)$. Performing this double expansion
and integrating each term, we arrive at the following expansion:
\begin{align}
\Pi&(m,n)-\frac{\sqrt{n} \sec
   ^{-1}\left(\sqrt{\frac{1-m}{1-n}}\right)}{\sqrt{(1-n) (n-m)}}=\nonumber\\
&=\sum _{j,k=0}^{\infty} \Biggl[\log (1-m) \frac{(-1)^{j+1}   \Gamma \left(\frac{1}{2}+j+k\right)}{4 \Gamma
   \left(\frac{1}{2}-j\right) \Gamma (1+j) \Gamma (2+j+k)}(1-m)^j
   (1-n)^k+\nonumber\\ &+\sum _{l=1}^{\infty}  \frac{(-1)^{j+l+1}  \pi
   ^{1/2} \Gamma \left(\frac{1}{2}+j+k+l\right)}{4 l^2 \Gamma \left(\frac{1}{2}-j\right)
   \Gamma (1+j) \Gamma \left(\frac{1}{2}-l\right) \Gamma (l) \Gamma (2+j+k+l)} (1-m)^{j+l} (1-n)^k+\nonumber\\
   &+\ \frac{(-1)^{k} \pi 
   \left(H_{1+j+k}-H_{-\frac{1}{2}+j+k}+2\log2\right)}{4 \Gamma
   \left(\frac{1}{2}-j\right) \Gamma (1+j) \Gamma \left(\frac{1}{2}-j-k\right) \Gamma
   (2+j+k)}(1-m)^j (1-n)^k\Biggr]\,.
\end{align}
\newpage

\acknowledgments

We thank L.~Bianchi, A.~Cavaglià, N.~Drukker, C.~Meneghelli, M.~Preti,  N.~Primi, Z.~Komargodski, K.~Zarembo  for discussions. LG acknowledges particularly E.~ Armanini for many interesting discussions. LG and DS are partially supported by an INFN  grant--``GaST: Non-perturbative Dynamics in Gauge and String Theories".
DB is supported by the German Research Foundation DFG -- SFB 1624 -- “Higher structures, moduli spaces and integrability” -- 506632645.
DB and VF would like to thank the Isaac Newton Institute for Mathematical Sciences, Cambridge (UK), for support and hospitality during the programme ``Quantum Field Theories with Boundaries, Impurities and Defects'' where work on this paper was undertaken. This work was supported by EPSRC grant EP/Z000580/1.
VF is supported by the DFG via the Heisenberg Professorship program 506208580 and the Research Unit FOR 5582: Modern Foundations of Scattering Amplitudes, project number 508889767.  VGMP  is supported by the Icelandic Research Fund under grant 228952-053, and  partially supported by grants from the University of Iceland Research Fund.
VGMP would like to acknowledge the Simons Center for Geometry and Physics, Stony Brook University, and the program ``Black hole physics from strongly coupled thermal dynamics'', for their hospitality and partial support during the completion of this work.
VGMP would like to acknowledge the International Institute of Physics, Natal, and the organizers of the program ``Quantum Gravity, Holography and Quantum Information" for their hospitality and partial support during the preparation of this manuscript.
\newpage


\bibliographystyle{JHEP}
\bibliography{Ref-all-defects.bib}

\end{document}